\documentclass[traditabstract]{aa}
\usepackage{amsmath}
\usepackage{graphicx}
\usepackage{txfonts}
\usepackage{natbib}
\bibpunct{(}{)}{;}{a}{}{,}
\usepackage{ulem}
\begin{document}

\title{Structure and kinematics of the nearby dwarf galaxy \\ UGCA\,105}
\titlerunning{Structure and Kinematics of the Nearby Dwarf Galaxy UGCA\,105}
\subtitle{}
\author{P. Schmidt\inst{1}\thanks{Member of the International Max Planck Research School (IMPRS) for Astronomy and
Astrophysics at the Universities of Bonn and Cologne}
    \and
        G. I. G. J\'{o}zsa\inst{2,3}
    \and
        G. Gentile\inst{4,5}\thanks{Postdoctoral researcher of the FWO-Vlaanderen (Belgium)}
    \and
        S.-H. Oh\inst{6}\thanks{Member of the ARC Centre of Excellence for All-sky Astrophysics (CAASTRO)}
    \and
        Y. Schuberth\inst{3}
    \and
       \\ N. Ben Bekhti\inst{3}
    \and    
        B. Winkel\inst{1}        
    \and
        U. Klein\inst{3}
\institute{Max-Planck-Institut f\"ur Radioastronomie, Auf dem H\"ugel 69, 53121 Bonn, Germany\\
\email{[pschmidt;bwinkel]@mpifr-bonn.mpg.de}
\and Netherlands Institute for Radio Astronomy (ASTRON), Postbus 2, 7990 AA Dwingeloo, the Netherlands\\
\email{jozsa@astron.nl}
\and Argelander-Institut f\"ur Astronomie, Universit\"{a}t Bonn, Auf dem H\"ugel 71, 53121 Bonn, Germany\\
\email{[yschuber;nbekhti;uklein]@astro.uni-bonn.de} 
\and Sterrenkundig Observatorium, Universiteit Gent, Krijgslaan 281, B-9000 Gent, Belgium\\
\email{gianfranco.gentile@ugent.be}
\and Department of Physics and Astrophysics, Vrije Universiteit Brussel, Pleinlaan 2, 1050 Brussels, Belgium 
\and International Centre for Radio Astronomy Research (ICRAR), The University of Western Australia, 35 Stirling Hwy, Crawley, Western Australia, 6009 \\
\email{se-heon.oh@uwa.edu.au}
}
}
\date{Received 28 September 2011 / Accepted 15 October 2013}

\abstract{Owing to their shallow stellar potential, dwarf galaxies possess thick gas disks, which makes them good candidates for studies of the galactic vertical kinematical structure. We present 21 cm line observations of the isolated nearby dwarf irregular galaxy UGCA\,105, taken with the Westerbork Synthesis Radio Telescope (WSRT), and analyse the geometry of its neutral hydrogen ($\ion{H}{i}$) disk and its kinematics. The galaxy shows a fragmented $\ion{H}{i}$ distribution. It is more extended than the optical disk, and hence allows one to determine its kinematics out to very large galacto-centric distances. The \ion{H}{i} kinematics and morphology are well-ordered and symmetric for an irregular galaxy. The $\ion{H}{i}$ {is} sufficiently extended to observe a substantial amount of differential rotation. 
Moreover, UGCA\,105 shows strong signatures for the presence of a kinematically anomalous gas component. Performing tilted-ring modelling by use of the least-squares fitting routine \texttt{TiRiFiC}, we found that the $\ion{H}{i}$ disk of UGCA\,105 has a moderately warped and diffuse outermost part. Probing a wide range of parameter combinations, we succeeded in modelling the data cube as a disk with a strong vertical gradient in rotation velocity ($\approx -60\,\rm km\,s^{-1}\,kpc^{-1}$), as well as vertically increasing inwards motion ($\approx -70\,\rm km\,s^{-1}\,kpc^{-1}$) within the radius of the stellar disk. 
The inferred radial gas inflow amounts to $0.06\,\rm M_\odot \rm  yr^{-1}$, which is similar to the star formation rate of the galaxy. The observed kinematics are hence compatible with {direct or indirect accretion from the intergalactic medium, an} extreme backflow of material that has formerly been expelled from the disk{, or a combination of both}. 
}


\maketitle

\section{Introduction}
\label{sect:intro} 
Dwarf galaxies are by far the most frequent type of galaxy in the Universe. Studying their properties is essential for understanding galaxy formation and evolution, as well as star formation.

\begin{figure}
 \centering
 \includegraphics[width=0.5\textwidth,bb=0 0 595 842,clip=true,trim=50pt 137pt 50pt 179pt]{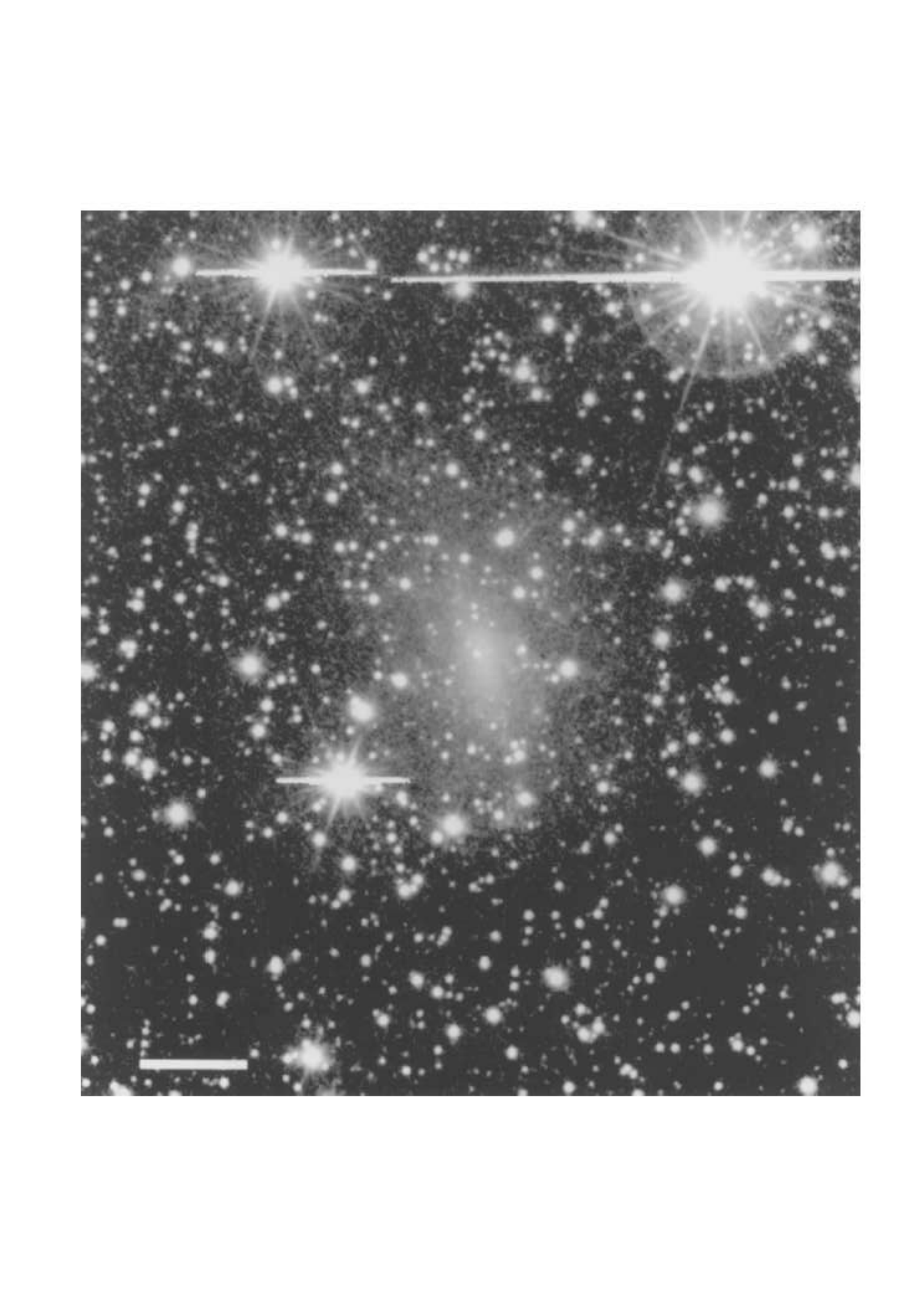}
 \caption{$I$-band image of UGCA\,105 \citep{buta99}. North is up, east is to the left. The scale bar in the lower left corner is 2\arcmin \, in length. 
 }
\label{fig:01}
\end{figure}

In the more massive dwarfs, dwarf irregulars (dIrrs) in particular, the $\ion{H}{i}$ disk is typically much more extended than its stellar counterpart. Previous observations of dwarf galaxies with an extremely large $\ion{H}{i}$ extent include DDO 154 \citep{carignan98,hoffman01} and NGC\,3741 \citep{begum05,gentile07}, which have $\ion{H}{i}$ disks extending out to radii larger than five times the Holmberg radius. The kinematics of these galaxies has been used to trace the gravitational potential out to the largest possible galacto-centric distances \citep{carignan98,gentile07}. 


Another important aspect in the study of $\ion{H}{i}$ in dwarf galaxies is the gas kinematics and distribution in the vertical direction with respect to their disks. They possess relatively thick disks compared to their radial extent because their gravitational potential is shallower {than that of} massive spiral galaxies (\citealt{Walter99}). 
{Because} any local structure in dwarf galaxies is conserved over long time-scales owing to their solid-body rotation, dwarf galaxies have been targeted to study star formation (\citealt{Walter99}). However, an important question regarding the evolution of dwarf galaxies that has not yet been addressed in depth is whether and how dwarf galaxies may accrete material from intergalactic space. {Because of the shallowness of the potential, and as a consequnce of the increased thickness of the \ion{H}{i} disk}, any interaction of the intergalactic medium (IGM) with the interstellar medium (ISM) should start at a considerable height above the plane, and hence impose its observable kinematical signature on the gas structure. {For this reason, dwarf galaxies should be very good candidates for studying the IGM-ISM interface.}

Accretion of low-metallicity gas from the IGM is required to maintain the observed star formation rates in massive galaxies and to explain their stellar metallicity budget (e.g. \citealt{Larson80, Bergh62, Sargent72, Pagel75}). For most systems, this still holds down to $z=0$ when taking into account the star formation and accretion history instead of the current star formation rate \citep{Fraternali12}. Models of galactic evolution in the mass range of spiral galaxies suggest that such accretion happens at all cosmic times in a "cold mode" (\citealt{Keres05}), that is, by relatively low-temperature gas ($<10^5\,\rm K$) streaming onto the galaxy disk. While at lower mass ranges accretion is expected to be suppressed for lower-mass DM halos \citep{Hoeft10}, dIrr galaxies fall in the transition range ($M_{\rm DM} = 10^9-10^{10}\,M_\odot$) where cold gas accretion is suppressed but is still expected to occur at $z=0$. To observationally constrain this transition mass 
range, a search for signatures of accretion in low-mass galaxies is hence important from an evolutionary point of view. High-velocity-clouds \citep{Woerden04} and Quasar absorption-line systems \citep{Richter12} have been suggested as evidence for cold gas accretion. {Studies of the vertical structure of gas disks and s}earches for evidence of cold accretion employing the $\ion{H}{i}$ emission line in external galaxies (e.g. \citealt{Swaters97, Schaap00, Lee01, Fraternali02, Barbieri05, westmeier05, Boomsma08, Oosterloo07, Zschaechner11, Kamphuis11b, Zschaechner12}, see \citealt{Heald11, Sancisi08} for a review) have disclosed in some cases the existence of a vertically extended gas layer with a strong vertical velocity gradient \citep{Fraternali02, Swaters97, Oosterloo07}. {The best-studied cases to date are NGC~2403} \citep{Fraternali02, Fraternali06, Fraternali08} {and NGC~891} \citep{Swaters97, Fraternali06, Fraternali08, Oosterloo07}.

While star formation is expected to expel gas into the halo of spiral galaxies that will condense and reaccrete onto the galaxy (galactic fountain model, \citealt{Shapiro76, Bregman80, Norman89}), and this reaccreting gas is also expected to show a lagging tangential velocity \citep{Benjamin02, Barnabe06}, model calculations indicate that the observed decrease of rotation velocity with height is incompatible with a closed system, but requires the accretion of low angular momentum material from the environment \citep{Collins02, Fraternali06, Heald07, Fraternali08, Marinacci11}. Direct evidence for a radial velocity component of high-latitude gas has been found for NGC~2403 \citep{Fraternali02} and was interpreted as an additional indication for accretion of ambient low angular momentum material \citep[][]{Fraternali06,Fraternali08}: major kinematic features (strength of radial motion, absence of strong vertical velocity components) in the extraplanar layers of NGC~2403 can only be reproduced in a 
ballistic model by feeding (accreted) low angular momentum gas into the fountain flow. {While} \citet[][]{Fraternali08} {assumed that such gas is directly accreted from inflowing cold streams of intergalactic matter and not from a hot ambient medium such as a hot galactic corona} \citep[e.g. ][]{Keres05}, \citet[][]{Marinacci10} {argued that for Milky-Way type galaxies, coronal gas can cool through the interaction with the fountain material on a short enough time scale to add a trailing stream of freshly condensed low angular momentum gas to the neutral gas fountain. In that picure, the intergalatic medium would accrete onto the hot corona instead of directly onto the neutral gas disk and the fountain flow, to be then swept up from the corona by the fountain, hereby adding mass to and removing angular momentum from the fountain gas} (see \citealt{Marinacci11}). {This mechanism, however, should become more and more ineffective with decreasing galaxy mass, since the accretion is expected to be 
more and more dominated by the cold, direct inflow of intergalactic matter rather than accretion onto the hot corona because of the missing shock-heating of infalling gas} \citep[e.g. ][]{Birnboim03,Keres05}.

{Although this clearly addresses an important mass regime, l}ittle is known about the global extraplanar velocity structure of dwarf galaxies. Some evidence for lagging, extraplanar gas in low-mass galaxies (NGC~2541, NGC~5204, UGC~3580) has been found by \citet{Jozsa07b}. \citet{Kamphuis11b} stated in a more detailed study of the $\ion{H}{i}$ structure of the dIrr {UGC}~1281 that a definite conclusion is not possible for {UGC}~1281{ and that a face-on warp of the disk could reproduce the data as well as slowly rotating extraplanar gas, a lagging disk.}

Here we present \ion{H}{i} observations of UGCA\,105, a dIrr taken from a sample of four galaxies (see Sect.~\ref{sect:sample}) {selected} because of their exceptionally massive $\ion{H}{i}$ disks. Exploring a number of kinematical \ion{H}{i} models that we compared with the data, we {searched for} kinematical signatures that have been interpreted {as signatures} of recent gas accretion in more massive galaxies. 

This paper is structured as follows: Sect. 2 contains some basic information on UGCA\,105, in addition to a list of criteria according to which the galaxy sample was selected. Sect. 3 describes details of our $\ion{H}{i}$ observations and data reduction, the outcome of which is presented in Sect. 4. Sect. 5 focuses on the geometric and kinematic analysis by means of tilted-ring modelling. We {present} a model accounting for the kinematic structure of an extraplanar $\ion{H}{i}$ component, which is discussed in depth in Sect. 6. A summary of our results is given in Sect. 7.

\begin{table*}[htbp]
\caption{Basic properties of UGCA\,105.}
\begin{center}
\begin{tabular}{llrrrr}
\hline
\hline
       Description & Parameter &            UGCA 105 &  Ref. \\
\hline
Classification of galaxy & Type & SAB(s)m &   (2)  \\
Right Ascension (J2000)  & R.A. & 05h14m15.3s &   (4)  \\
Declination (J2000)                               & Dec. & +62d34m48s &   (4)  \\
Heliocentric systemic velocity ($\rm km\,s^{-1}$) & $V_{\rm sys}$ & $     111 \,\pm\,   5  $ &  (1)  \\
Distance of object ($\rm Mpc$) & $\rm D$ & $     3.48\,^{+\,0.58}_{-\,0.49}  $ &  (3)  \\
Scale between distance on sky and true distance ($\,{\rm pc}\,arcsec^{-1}$) & $sc$ & $     16.9  \,\pm\, 2.4    $ &  (1)   \\
Apparent $B$-band magnitude ($\rm mag$) & $m_{\rm B}$ & $     14.04 \,\pm\,  1.27  $ &  (4)   \\
Apparent $I$-band magnitude ($\rm mag$) & $m_{\rm I}$ & $     10.41 \,\pm\,  0.06  $ &  (2)   \\
\hline
Reddening (mag)& $E(B\!-\!V)$&$0.218$&(5)\\
Galactic extinction in $B$-band (mag)& $A_{B}$&$0.941$&(*)\\
Galactic extinction in $I$-band (mag)& $A_{I}$&$0.423$&(*)\\
Absolute $B$-band magnitude ($\rm mag$) & $M_{\rm B}$ & $-14.68$ &  ($\dag$)   \\
Absolute $I$-band magnitude ($\rm mag$) & $M_{\rm I}$ & $     -17.80  $ & ($\dag$)   \\
\hline
Total {\ion{H}{i}} flux (${\rm Jy}\,{\rm km}\,{\rm s}^{-1}$) & $F_{\rm {\ion{H}{i}}}$ & $     225 \,\pm\, 11   $ &  (1)   \\
{\ion{H}{i}} mass ($10^{9}\,{M}_\odot$) & $M_{\rm {\ion{H}{i}}}$ & $     0.64  \,\pm\,  0.03  $ &  (1)   \\
$B$-band optical radius ($\,^{\prime\prime}$) & $r_{25}$ & $      169  \,\pm\,  12   $ & (4)   \\
{\ion{H}{i}} radius ($\,^{\prime\prime}$) & $r_{\rm {\ion{H}{i}}}$ & $     476   \,\pm\,  14    $ &  (1)   \\
$B$-band optical radius ($\rm kpc$) & $R_{25}$ & $      2.9 \,\pm\,  0.7  $ &  (1)   \\
{\ion{H}{i}} radius ($\rm kpc$) & $R_{\rm {\ion{H}{i}}}$ & $     8.0  \,\pm\,  1.1   $ & (1)   \\
Ratio of {\ion{H}{i}} to optical radius & $R_{\rm{\ion{H}{i}}}/R_{25}$ & $     2.8  \,\pm\,  0.6   $ &  (1)   \\
Inclination of inner {\ion{H}{i}} disk ($^{\circ}$) & $i_{{\ion{H}{i}},\rm i}$ & $50\,\pm\,2$ &  (1)   \\
Inclination of {\ion{H}{i}} disk at outermost radius ($^{\circ}$) & $i_{{\ion{H}{i}},\rm o}$ & $60\,\pm\,4$ &  (1)   \\
Dynamical mass ($10^{9}\,{M}_\odot$) & $M_{\rm dyn}$ & $ >\, 12   \,\pm\,  3    $ &  (1)   \\
\hline
\end{tabular}
\end{center}
\label{tab:properties}
\note{ References: (1) This work; (2) \cite{buta99}; (3) \cite{jacobs09}; (4) HyperLeda; (5) \cite{georgiev04}. (*) The conversions from $E(B-V)$ to extinction are based on the values given in Table\,6 of \cite{schlegel98}. ($\dag$) For the  absolute magnitudes quoted here, only Galactic foreground extinction has been considered. No corrections for internal extinction have been applied. }
\end{table*}

\begin{table*}
\caption{Summary of observations and data reduction.}
\begin{center}
\begin{tabular}{l r r r r}
\hline
\hline
Name of galaxy & & \multicolumn{2}{c}{UGCA 105}     \\
Total on-source integration time ($\rm h$) & & \multicolumn{2}{c}{$2\times 12$}  \\
Observation dates & & \multicolumn{2}{c}{4-5\,\&\,8-9/10/07}    \\
\hline
Applied weighting scheme & & Robust 0 & Robust 0.4       \\
Pixel size ($\,^{\prime\prime}$) &  $dx$ & $ 4     $ & $ 7     $          \\
Channel width (${\rm km}\,{\rm s}^{-1}$) &  $dv$ & $ 2.06  $ & $ 4.12  $          \\
HPBW along the beam major axis ($\,^{\prime\prime}$) &  ${\rm HPBW}_{\rm maj}$ & $ 14.84  $ & $ 39.86  $         \\
HPBW along the beam minor axis ($\,^{\prime\prime}$) &  ${\rm HPBW}_{\rm min}$ & $ 13.14  $ & $ 38.36  $        \\
Beam position angle ($\deg$) & ${\rm PA}_{\rm beam}$ & $ 1.2   $ & $ 5.4   $          \\
rms noise (${\rm mJy}/{\rm beam}$) in the cubes &  $\sigma_{\rm rms}$ & $ 0.59  $ & $ 0.46  $         \\
\hline
\end{tabular}
\end{center}
\label{tab:2}
\end{table*}

\section{{Galaxy} selection}
\label{sect:sample}
We searched for nearby dwarf galaxies that have massive $\ion{H}{i}$ disks (with respect to the optical ones) and are sufficiently isolated to avoid direct interaction with other galaxies. The sample considered was selected from the HyperLeda catalogue\footnote{({http://leda.univ-lyon1.fr}\nocite{paturel03})} based on the following criteria:
\begin{itemize}
 \item An $\ion{H}{i}$ mass-to-light ratio ($M_{\rm{\ion{H}{i}}}$ /$L_{B}) > 2.0$ ensures that the galaxies have an unusually large amount of $\ion{H}{i}$, with respect to {their} optical luminosity. 
 \item 25 km s$^{-1} < \varv_{max} < 100$ km s$^{-1}$ selects rotationally supported dwarf galaxies. 
 \item Total $\ion{H}{i}$ flux $>$ 10 Jy km s$^{-1}$
 \item Declination $> 30^{\circ}$
 \item An inclination of 30$^{\circ} < i < 80^{\circ}$ allows one to derive reliable kinematics.
 \item $D_{25} > 32\arcsec$: the optical radius of the galaxy will be larger than 16$\arcsec$, to have at least one synthesized beam HPBW ($\approx15\arcsec$) to probe the optical part of the galaxy. 
 \item Distance $\lesssim15$ Mpc, to have a spatial resolution similar to that of typical $\Lambda$CDM numerical simulations ($\lesssim1$ kpc)
 \item A projected distance from the nearest galaxy $>$ 20 times the optical radius excludes interaction with other galaxies, and hence ensures a relatively regular velocity structure.
\end{itemize}

With these selection criteria, four galaxies were found in the HyperLeda database: UGCA\,105, UGC\,11891, UGC\,5846, and PGC 168300. The first is the subject of this paper. An optical image of UGCA\,105, taken from \cite{buta99}, is shown in Fig.~\ref{fig:01}. {The} basic properties of the galaxy, compiled from the literature or derived in this work, are listed in Table\,\ref{tab:properties}.

UGCA\,105 is a dwarf irregular galaxy located in the IC 342/Maffei Group, which is one of the closest groups of galaxies. \cite{buta99}, who performed $V$- and $I$-band photometry of UGCA\,105, classified it as an SAB(s)m galaxy (while in HyperLeda it is listed as irregular), that is, a late-type spiral, because they found that it shows a small bar and very weak spiral structure in its optical component. The galaxy was also observed in H$\alpha$ by \cite{kingsburgh98} and by \citet{Kennicutt08}. We adopted the most recently determined distance of 3.48\,Mpc \citep{jacobs09} because of the galaxy's low Galactic latitude.


\section{Observations and data reduction}

UGCA\,105 was observed in the $\ion{H}{i}$ 21 cm line with the Westerbork Synthesis Radio Telescope (WSRT). We used the interferometer in its maxi-short configuration, which provides optimal uv-coverage for extended sources. The standard correlator setup was used, that is, we measured two linear polarisations in 1024 channels over a bandwidth of 10 MHz (allowing for a spectral resolution of 4 km s$^{-1}$), centred on the systemic velocity of the target. Our target was observed for two full 12-hour tracks, with an additional half-hour before and after each track spent measuring standard flux calibrator sources. 

The data reduction was performed using standard techniques of the Miriad software package \citep{sault95}. First, the raw dataset was visually inspected, and autocorrelations as well as bad data resulting from shadowing or solar interference were flagged where necessary. Subsequently, the calibrator sources were used for the absolute flux calibration as well as for gain and bandpass corrections. Next, we performed a second-order continuum subtraction in the visibility domain, and stored the continuum emission as channel-0 data. Several iterations of CLEAN deconvolution and self-calibration were performed on the inverted continuum data to correct for time-variable phase- and amplitude errors of the antenna gain. The gain calibration was applied to the complete uv-dataset, and the continuum model was subtracted to obtain the amplitude- and phase-calibrated spectral line data. By Fourier-inverting the line data of UGCA\,105, we produced two data cubes, using different weighting schemes: one with a 
Robust 
parameter of 0 at full resolution, and, to increase the sensitivity to extended emission, a Robust-0.4 cube. For the latter, we additionally applied a Gaussian taper to the visibilities, with a Gaussian equivalent in the image domain of $\rm{FWHM}=28\arcsec$, as well as a decrease in spectral resolution by a factor of two. After Hanning smoothing we obtained velocity resolutions of 4.12 km s$^{-1}$ and 8.24 km s$^{-1}$, respectively. 
Both data cubes were iteratively deconvolved by means of the Clark CLEAN algorithm, using masks to define the appropriate CLEAN regions for each iteration.
The cutoff value for the algorithm was decreased step by step, in both cases down to a value equal to the rms noise in the channel maps. From the CLEANed data cubes, moment-0 maps (corresponding to the distribution of total intensity) were computed after blanking out all pixels outside the CLEAN regions specified in the final iteration. Total intensity maps were converted into column density maps, assuming optically thin gas. In Table\,\ref{tab:2} we list basic parameters of the final data cubes.

\begin{figure*}
 \centering
 \includegraphics[width=0.24\textwidth,angle=270,bb=0 0 612 792,keepaspectratio=true,clip=true,trim=33pt 92pt 112pt 132pt]{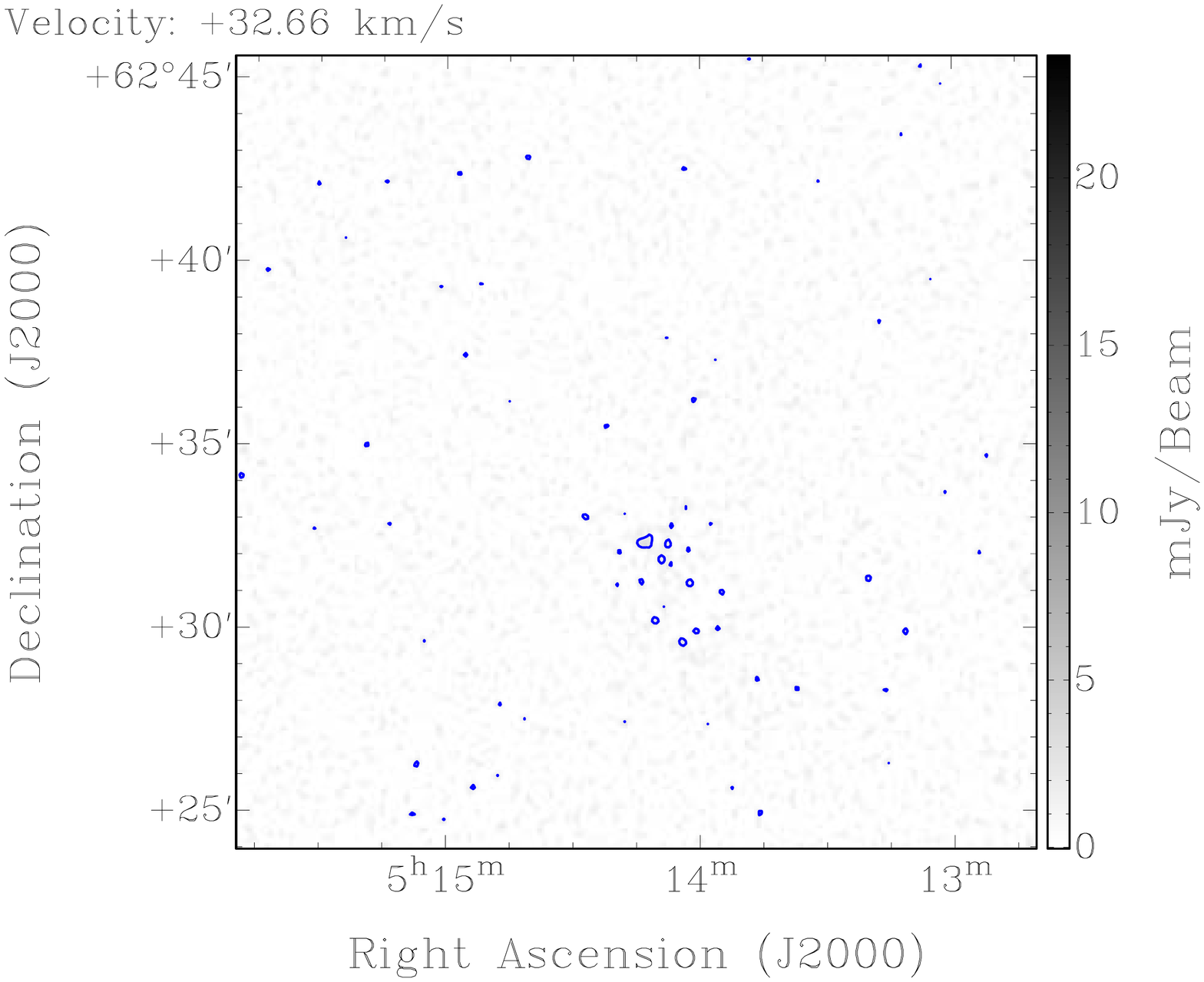}\includegraphics[width=0.24\textwidth,angle=270,bb=0 0 612 792,keepaspectratio=true,clip=true,trim=33pt 214pt 112pt 132pt]{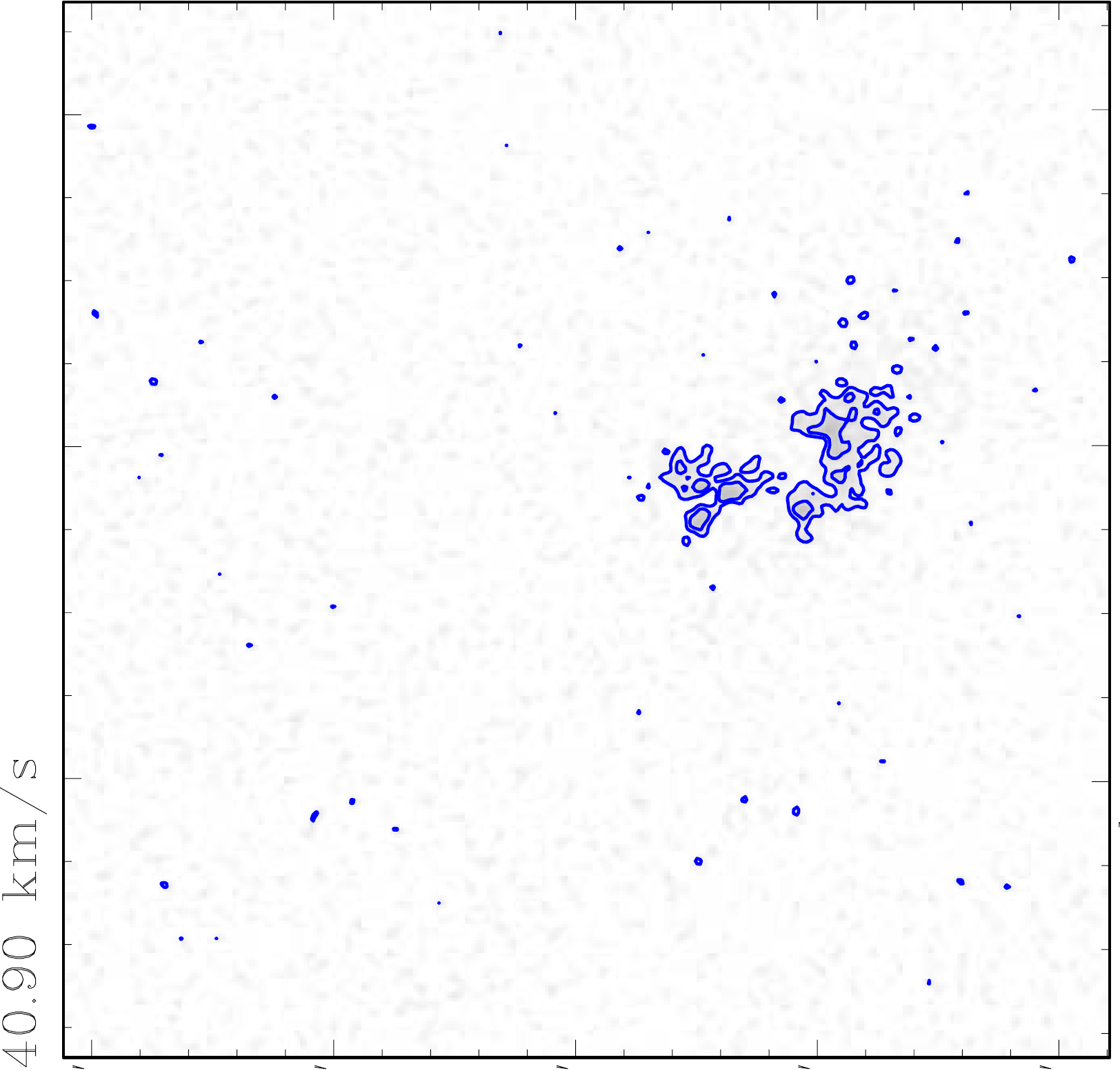}\includegraphics[width=0.24\textwidth,angle=270,bb=0 0 612 792,keepaspectratio=true,clip=true,trim=33pt 214pt 112pt 132pt]{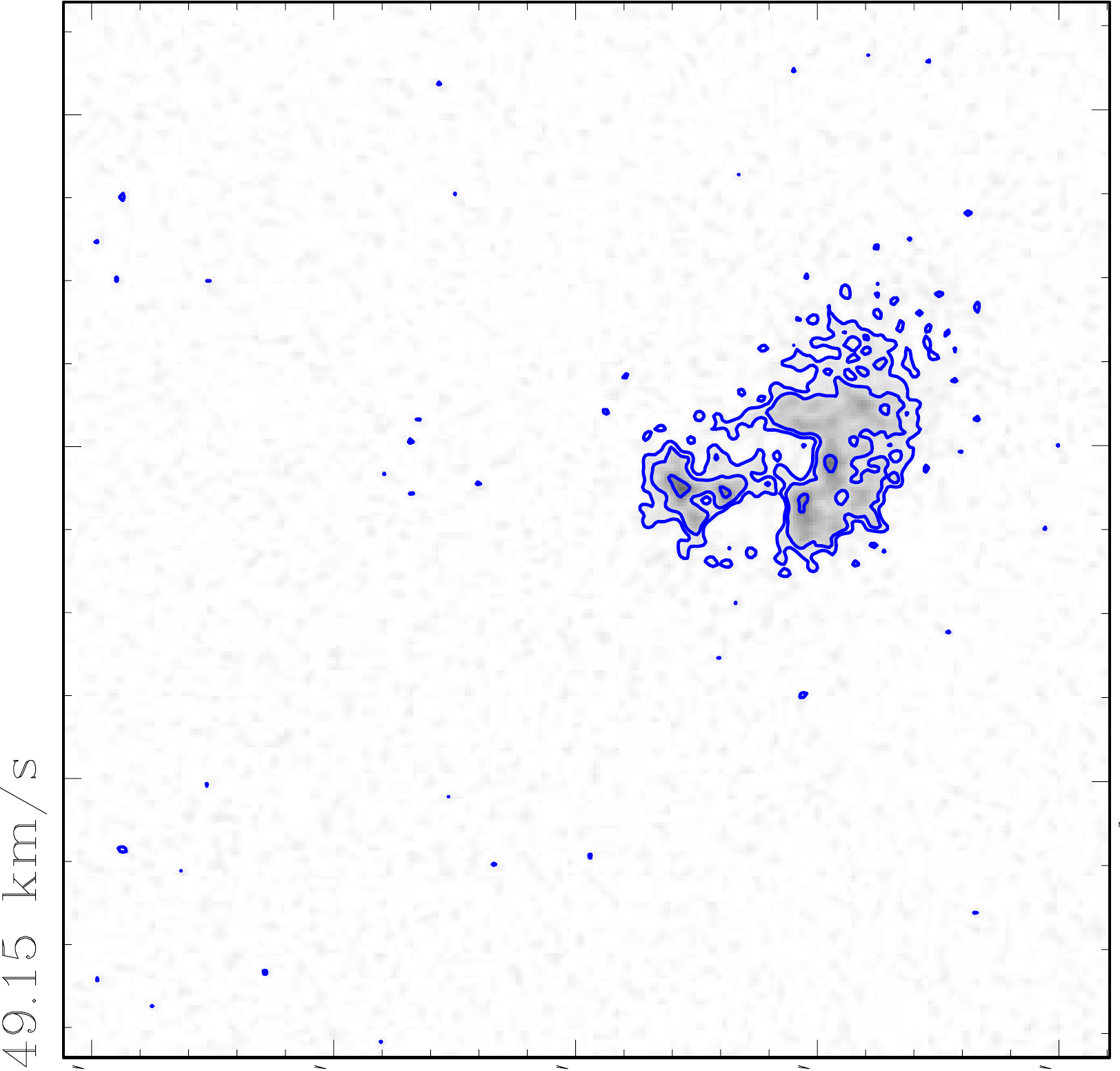}\includegraphics[width=0.24\textwidth,angle=270,bb=0 0 612 792,keepaspectratio=true,clip=true,trim=33pt 214pt 112pt 42pt]{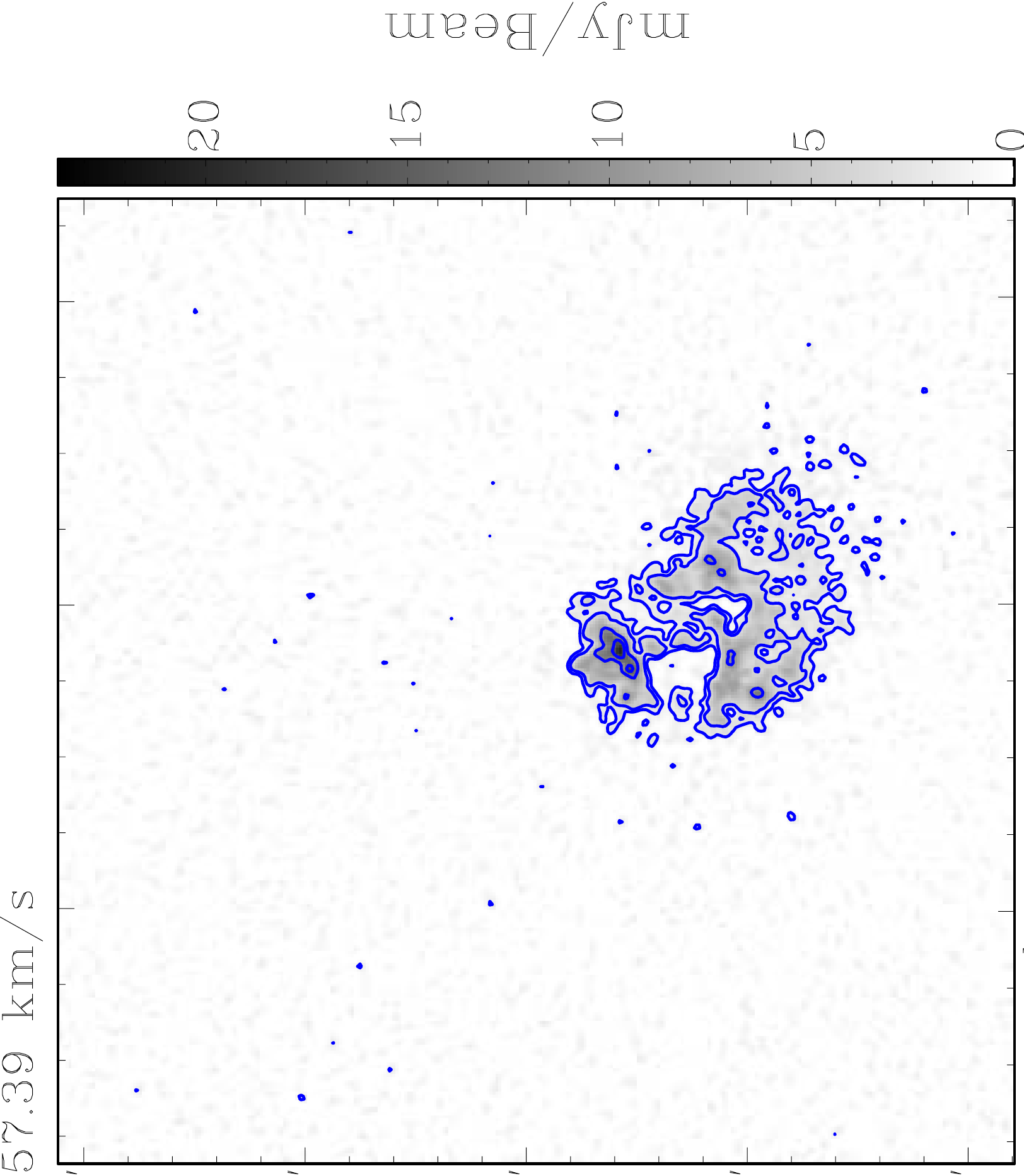}
\includegraphics[width=0.24\textwidth,angle=270,bb=0 0 612 792,keepaspectratio=true,clip=true,trim=33pt 92pt 112pt 132pt]{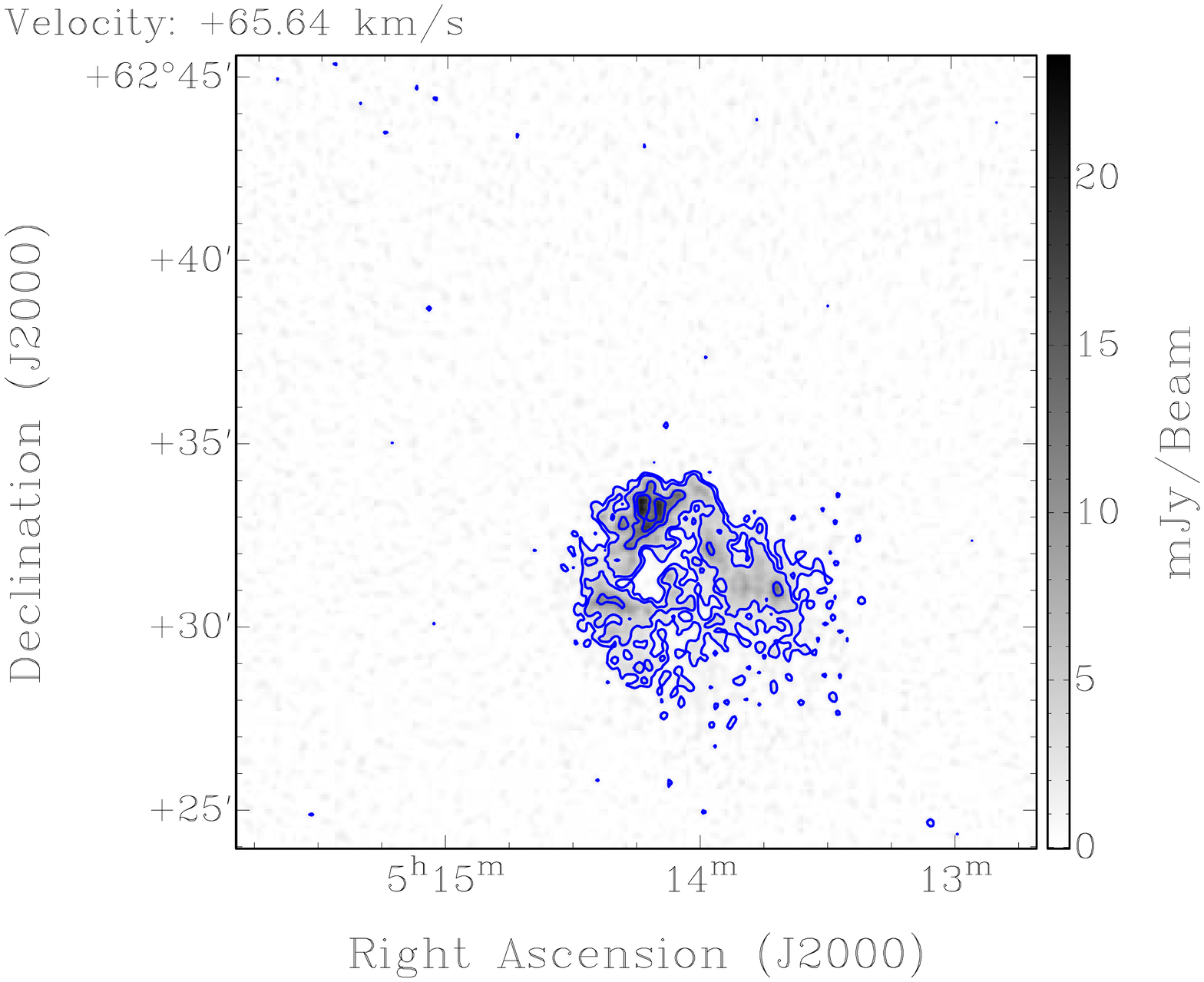}\includegraphics[width=0.24\textwidth,angle=270,bb=0 0 612 792,keepaspectratio=true,clip=true,trim=33pt 214pt 112pt 132pt]{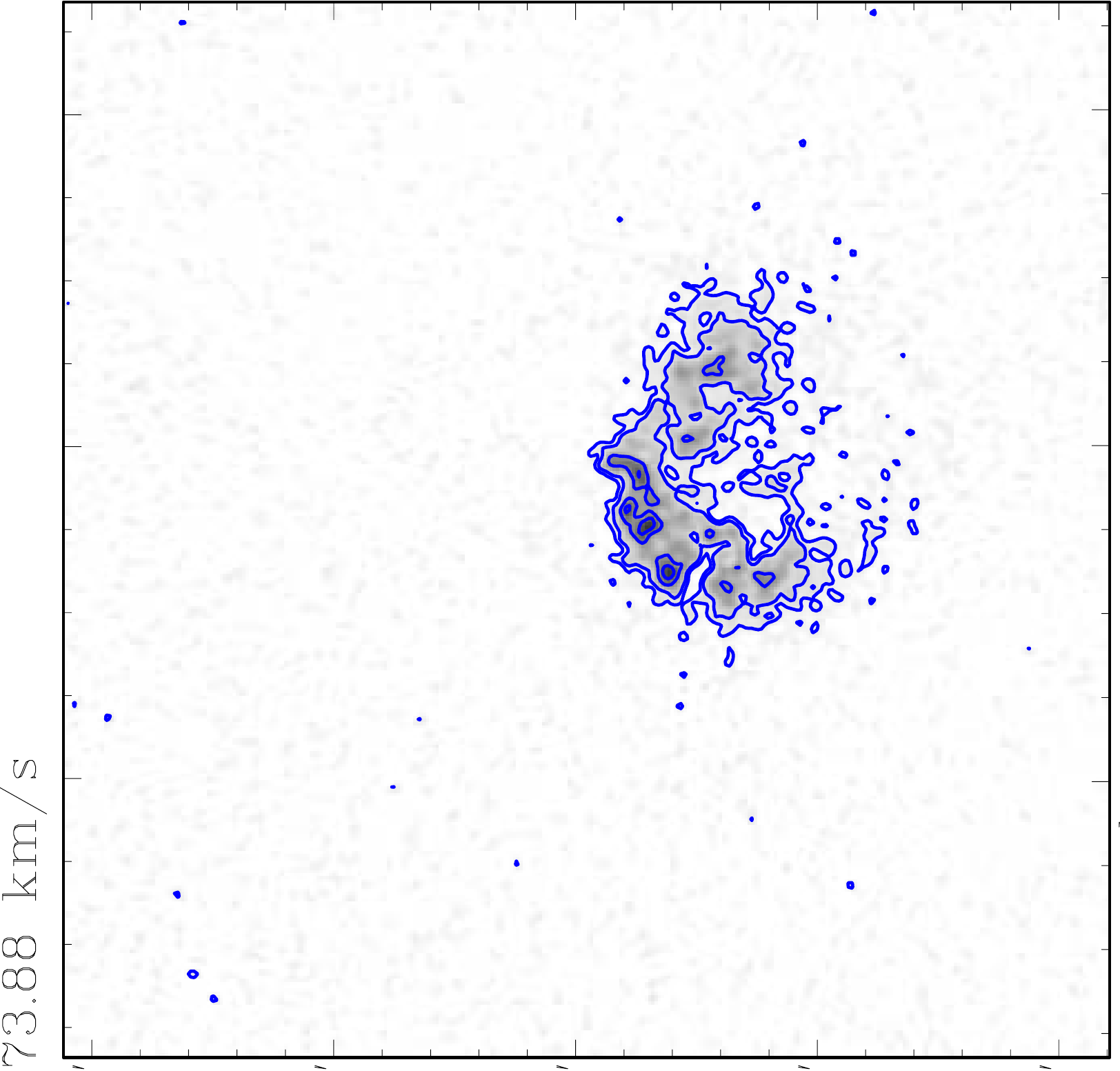}\includegraphics[width=0.24\textwidth,angle=270,bb=0 0 612 792,keepaspectratio=true,clip=true,trim=33pt 214pt 112pt 132pt]{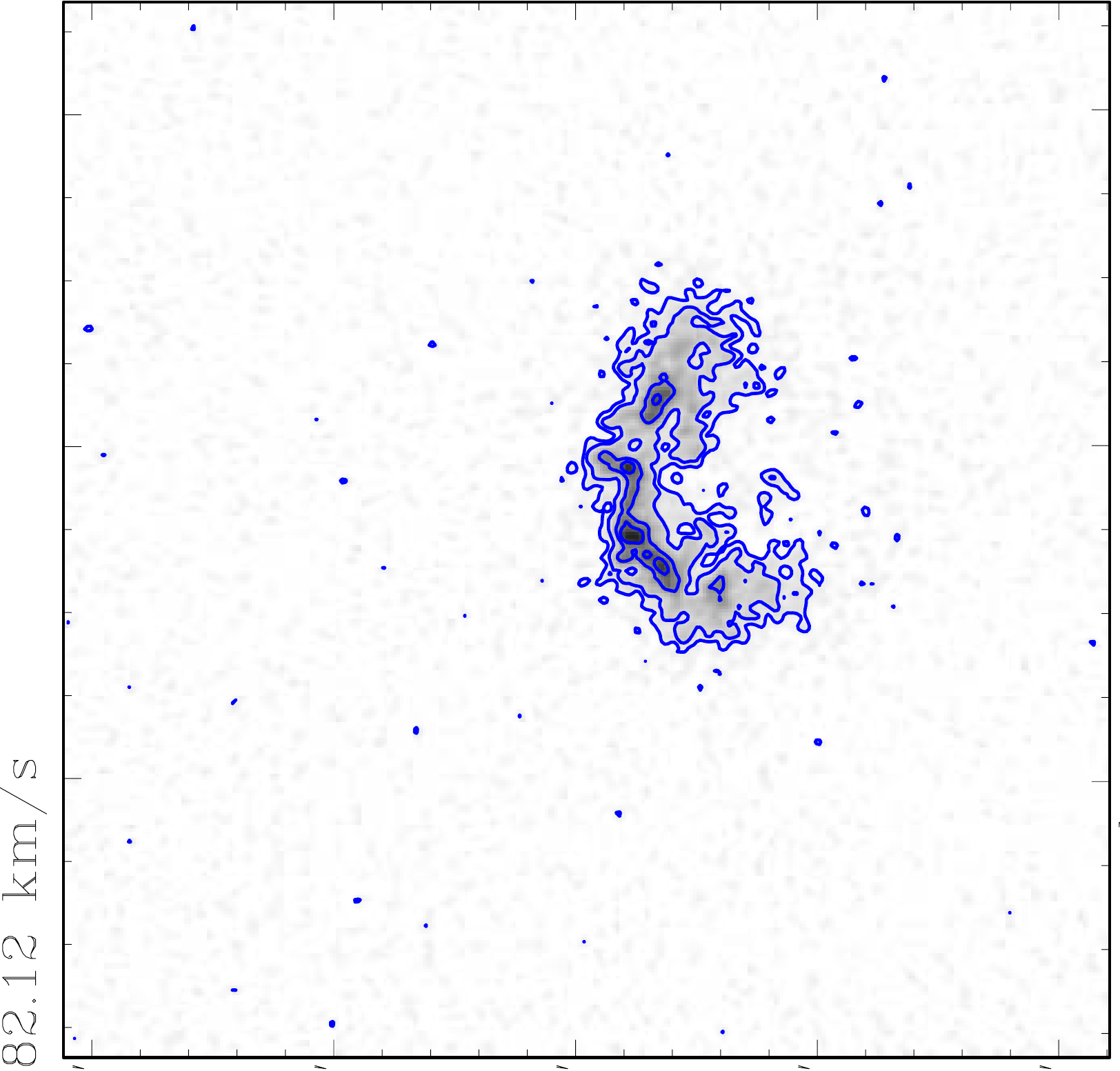}\includegraphics[width=0.24\textwidth,angle=270,bb=0 0 612 792,keepaspectratio=true,clip=true,trim=33pt 214pt 112pt 42pt]{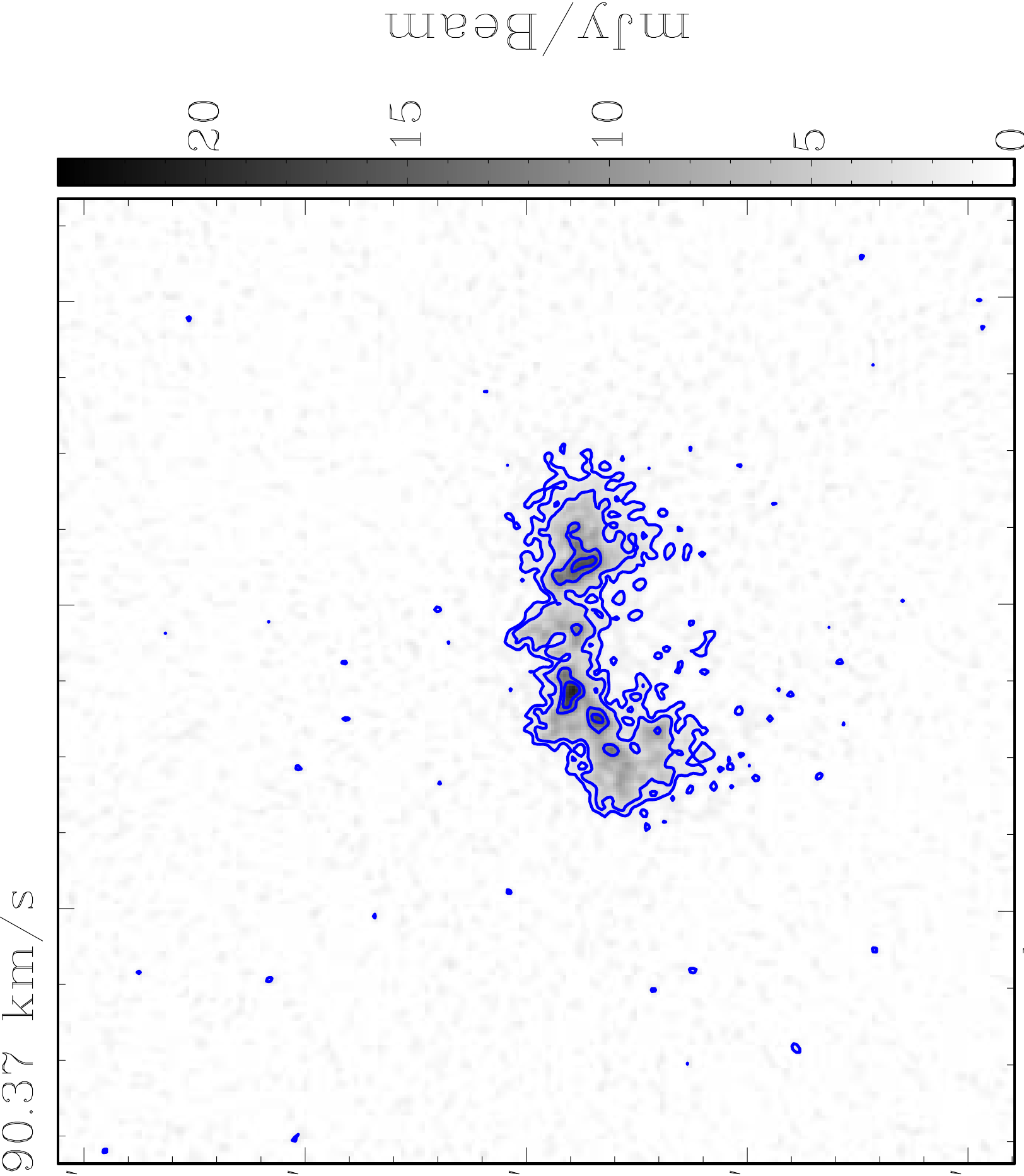}
\includegraphics[width=0.24\textwidth,angle=270,bb=0 0 612 792,keepaspectratio=true,clip=true,trim=33pt 92pt 112pt 132pt]{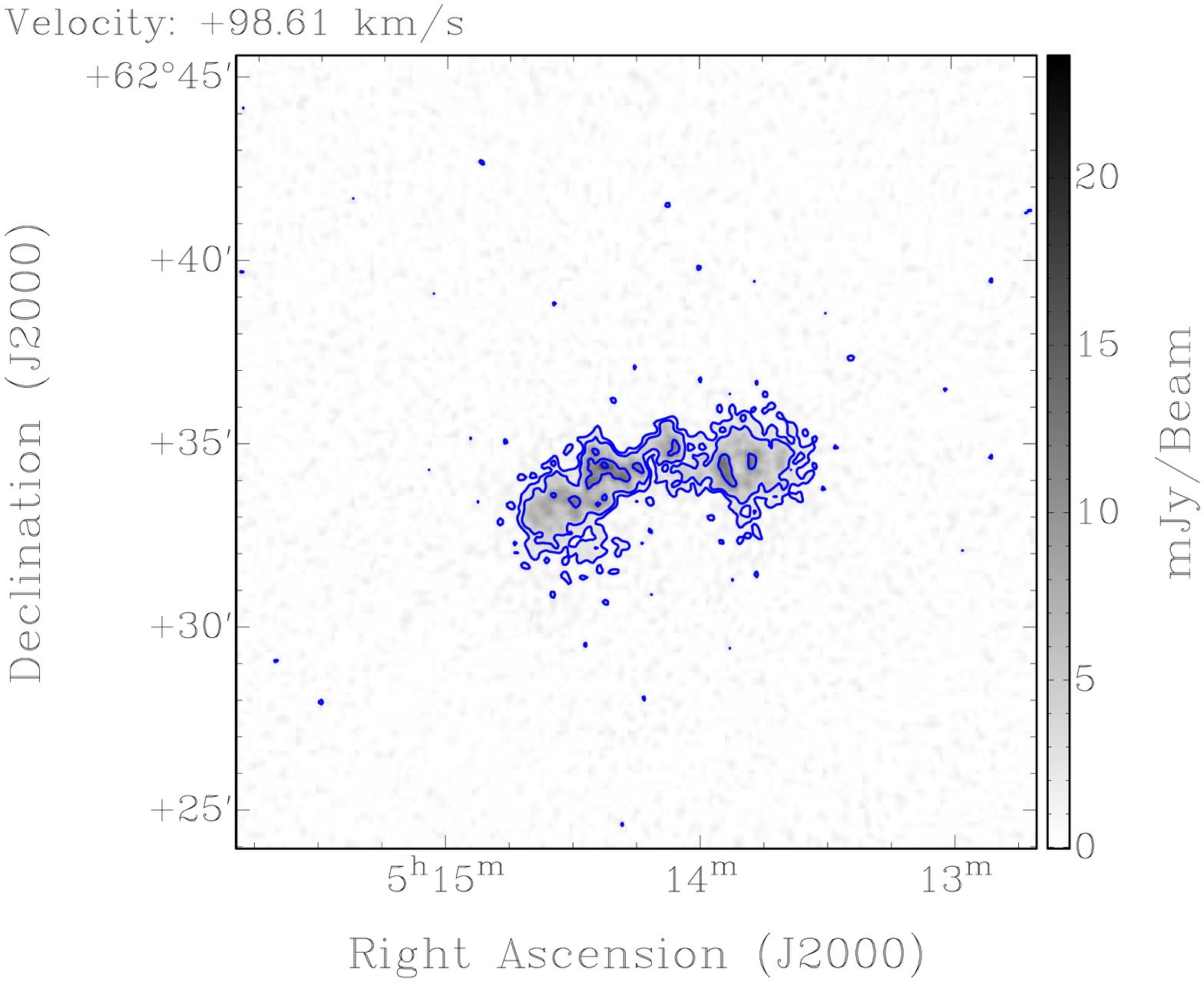}\includegraphics[width=0.24\textwidth,angle=270,bb=0 0 612 792,keepaspectratio=true,clip=true,trim=33pt 214pt 112pt 132pt]{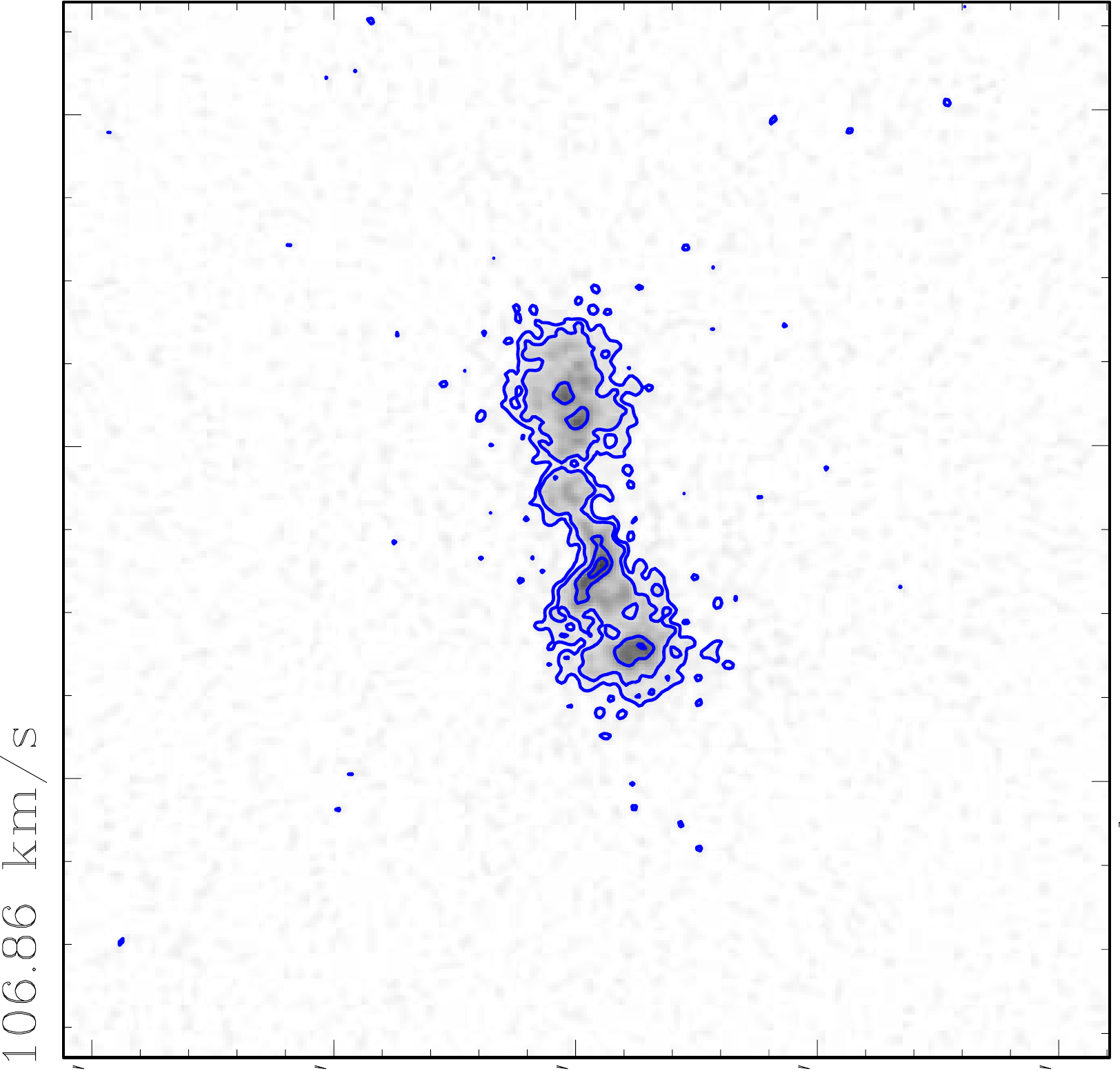}\includegraphics[width=0.24\textwidth,angle=270,bb=0 0 612 792,keepaspectratio=true,clip=true,trim=33pt 214pt 112pt 132pt]{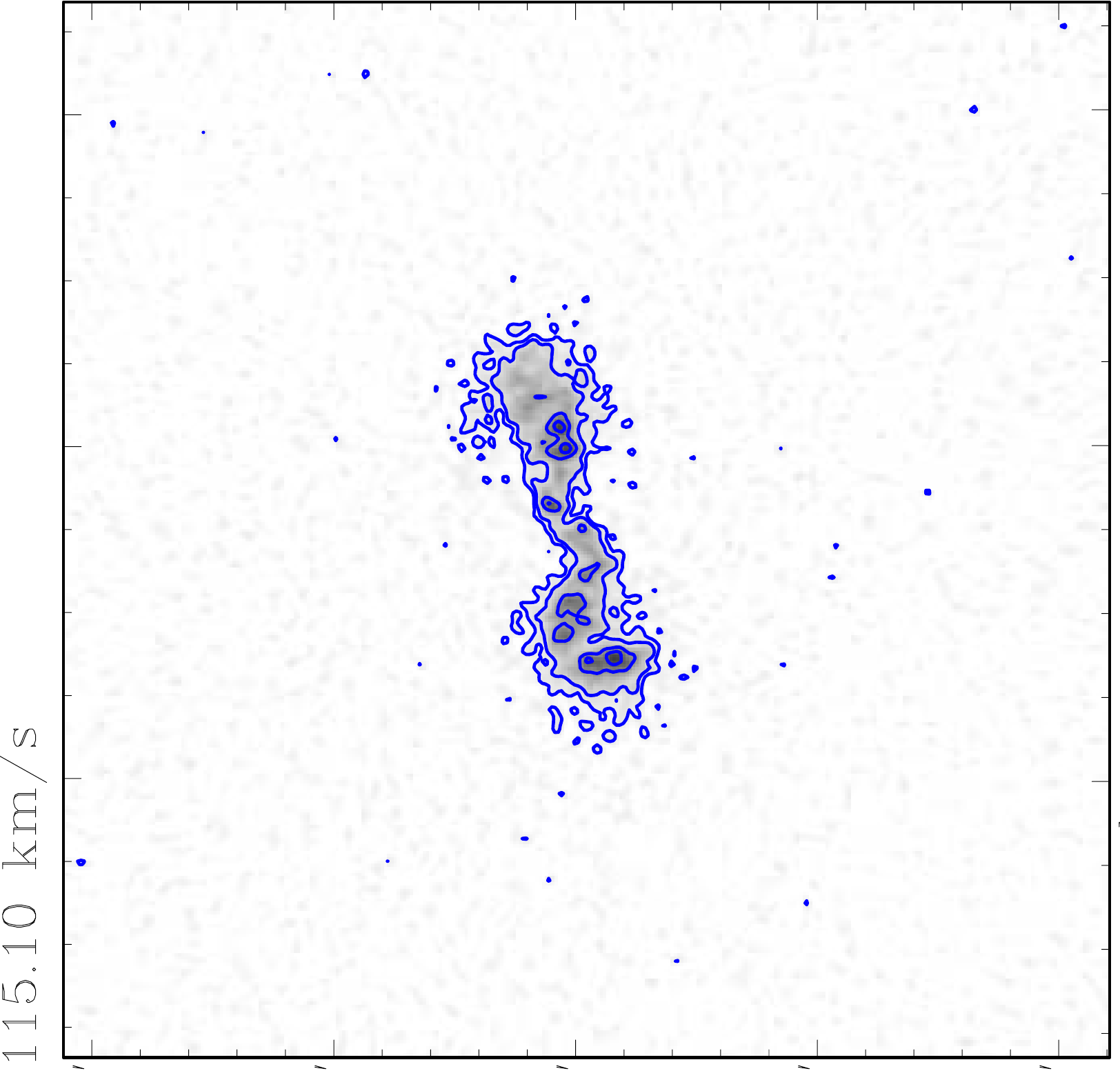}\includegraphics[width=0.24\textwidth,angle=270,bb=0 0 612 792,keepaspectratio=true,clip=true,trim=33pt 214pt 112pt 42pt]{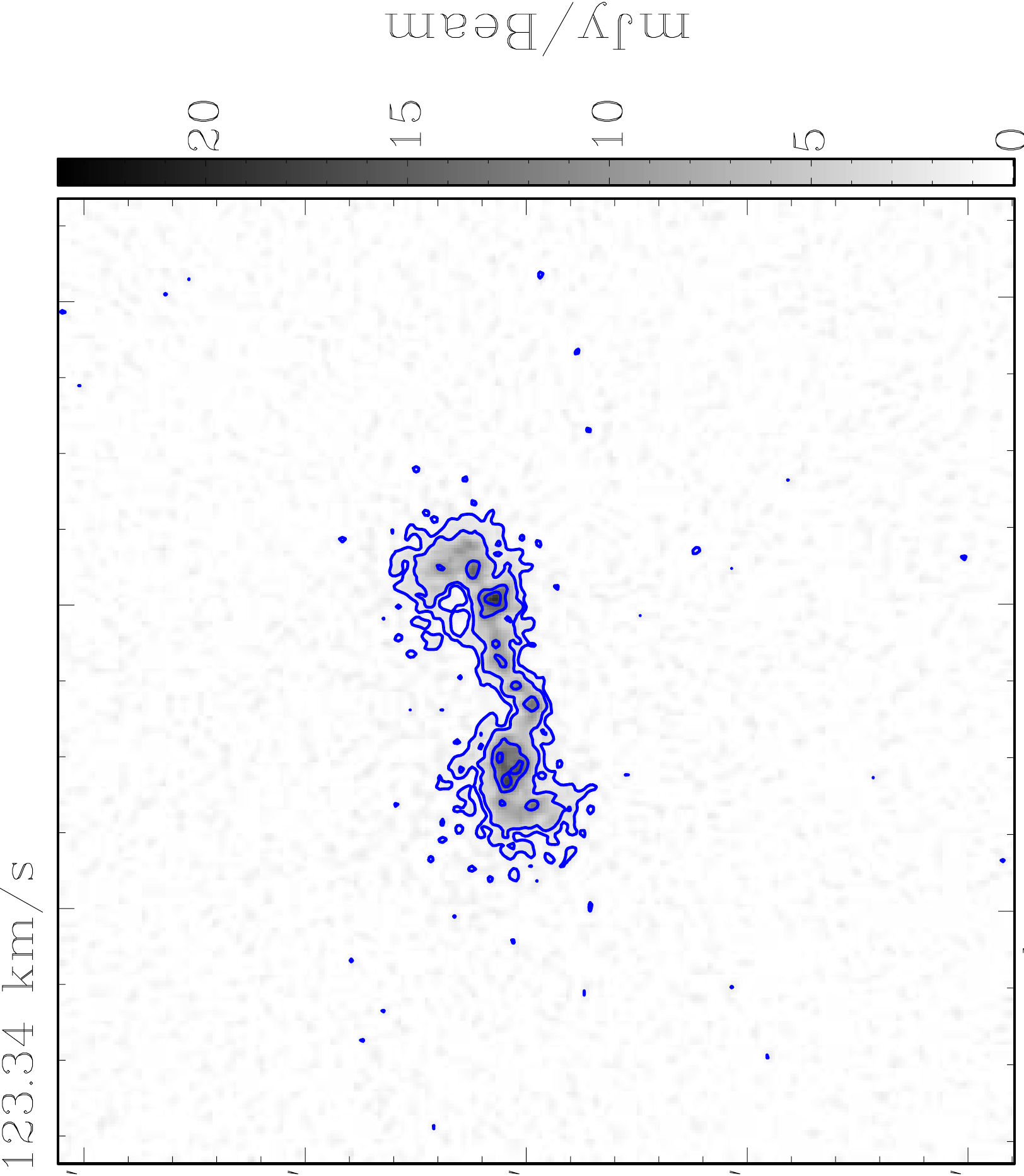}
\includegraphics[width=0.24\textwidth,angle=270,bb=0 0 612 792,keepaspectratio=true,clip=true,trim=33pt 92pt 112pt 132pt]{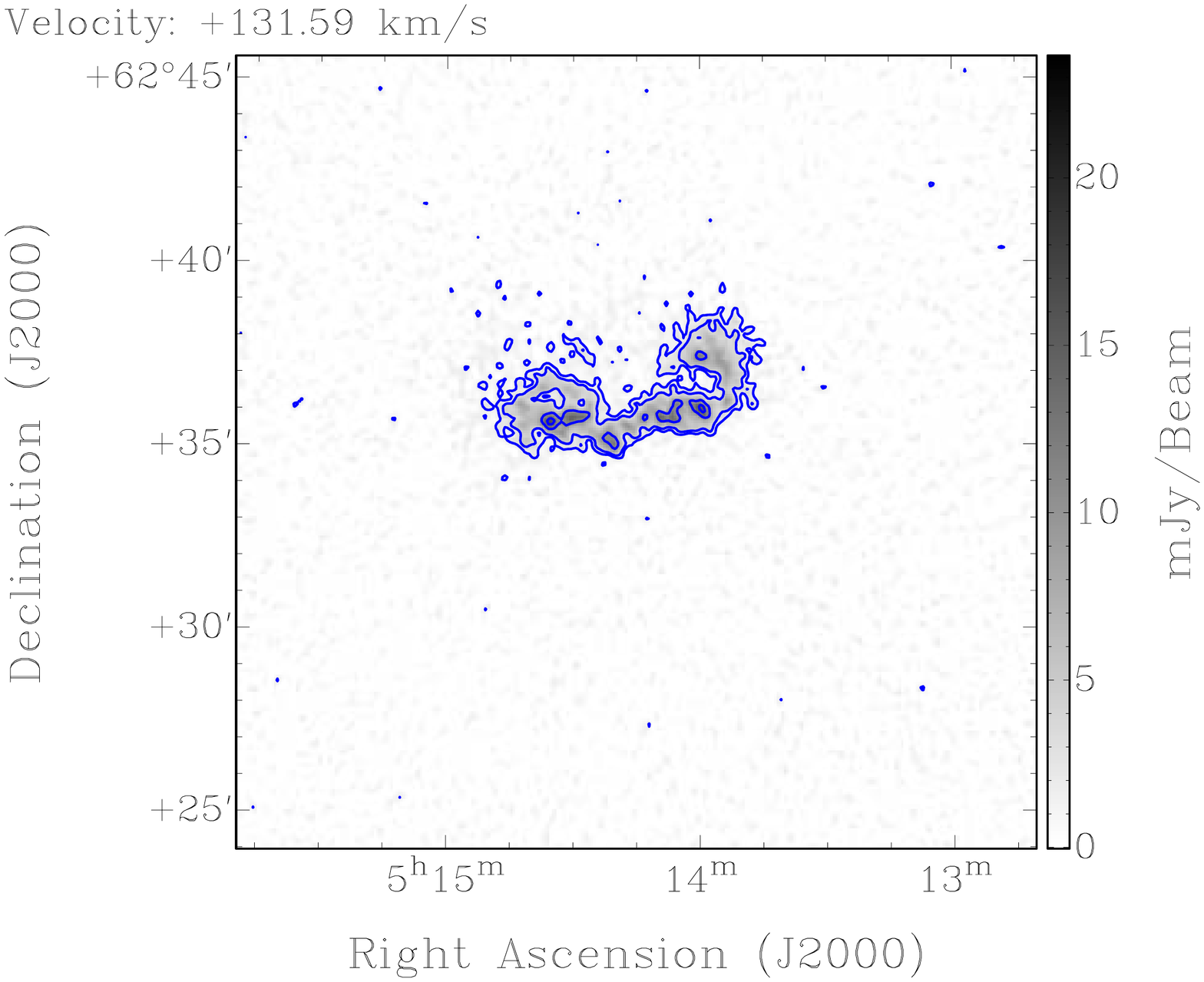}\includegraphics[width=0.24\textwidth,angle=270,bb=0 0 612 792,keepaspectratio=true,clip=true,trim=33pt 214pt 112pt 132pt]{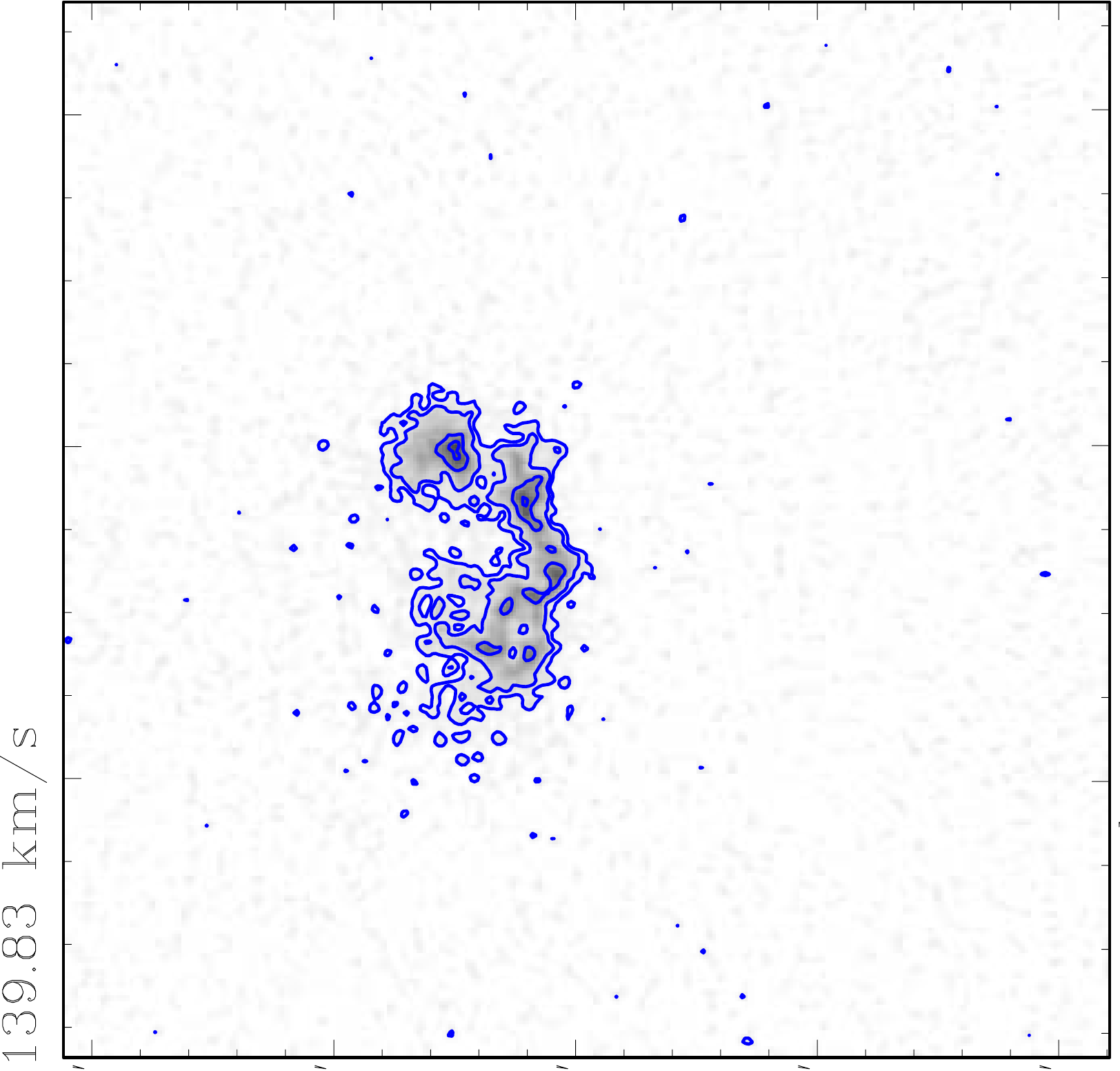}\includegraphics[width=0.24\textwidth,angle=270,bb=0 0 612 792,keepaspectratio=true,clip=true,trim=33pt 214pt 112pt 132pt]{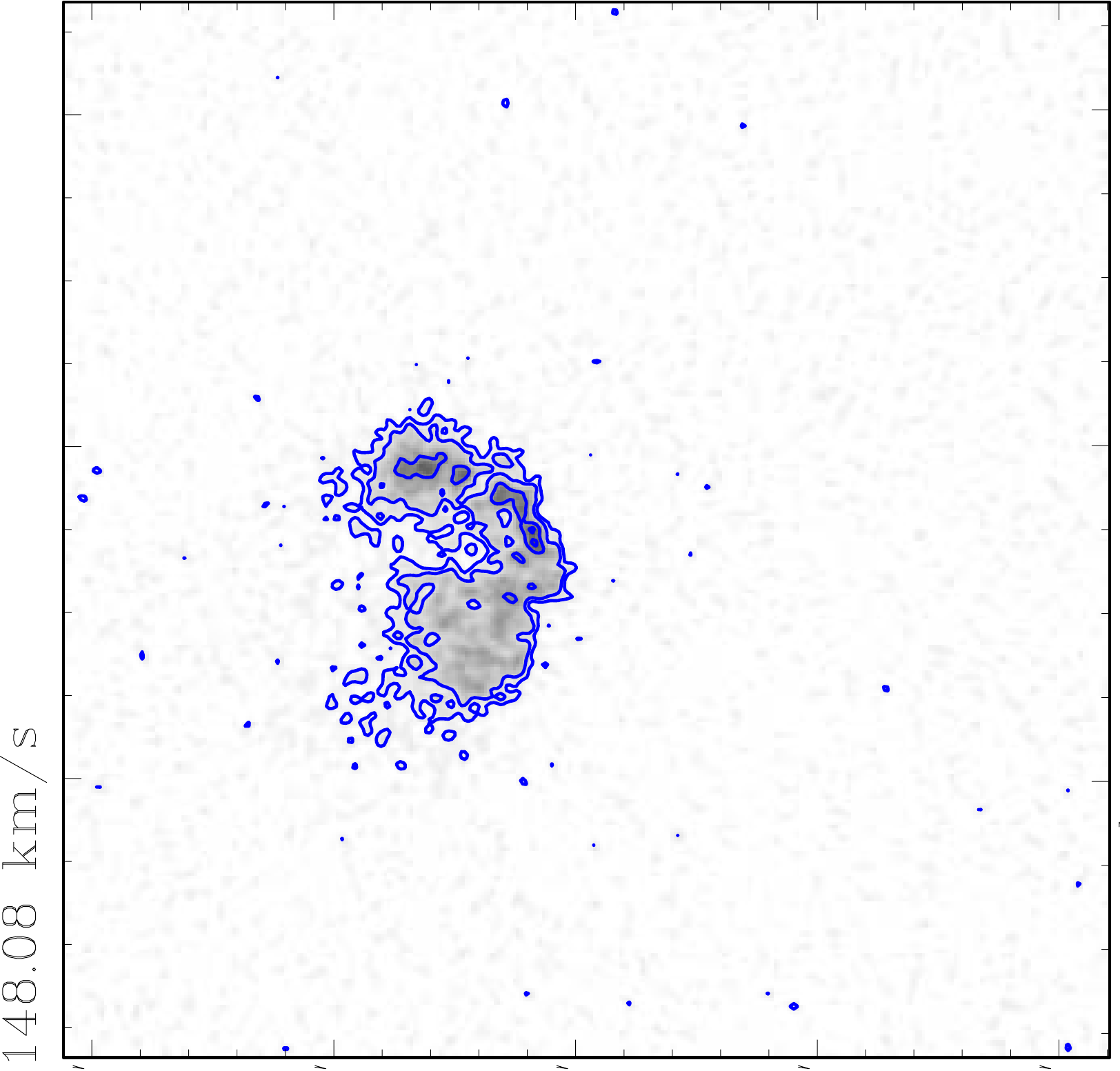}\includegraphics[width=0.24\textwidth,angle=270,bb=0 0 612 792,keepaspectratio=true,clip=true,trim=33pt 214pt 112pt 42pt]{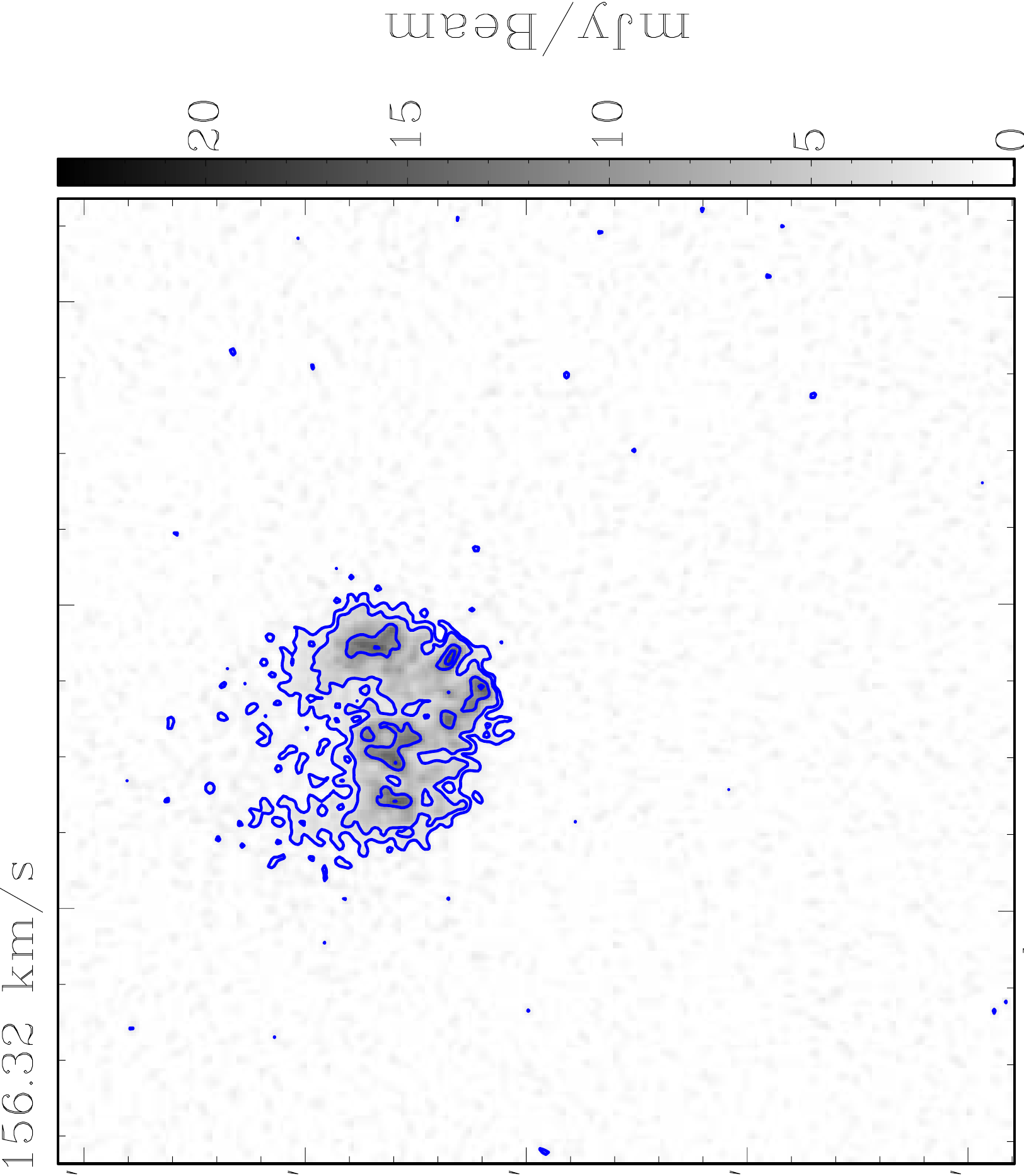}
\includegraphics[width=0.29728\textwidth,angle=270,bb=0 0 612 792,keepaspectratio=true,clip=true,trim=33pt 92pt 0pt 132pt]{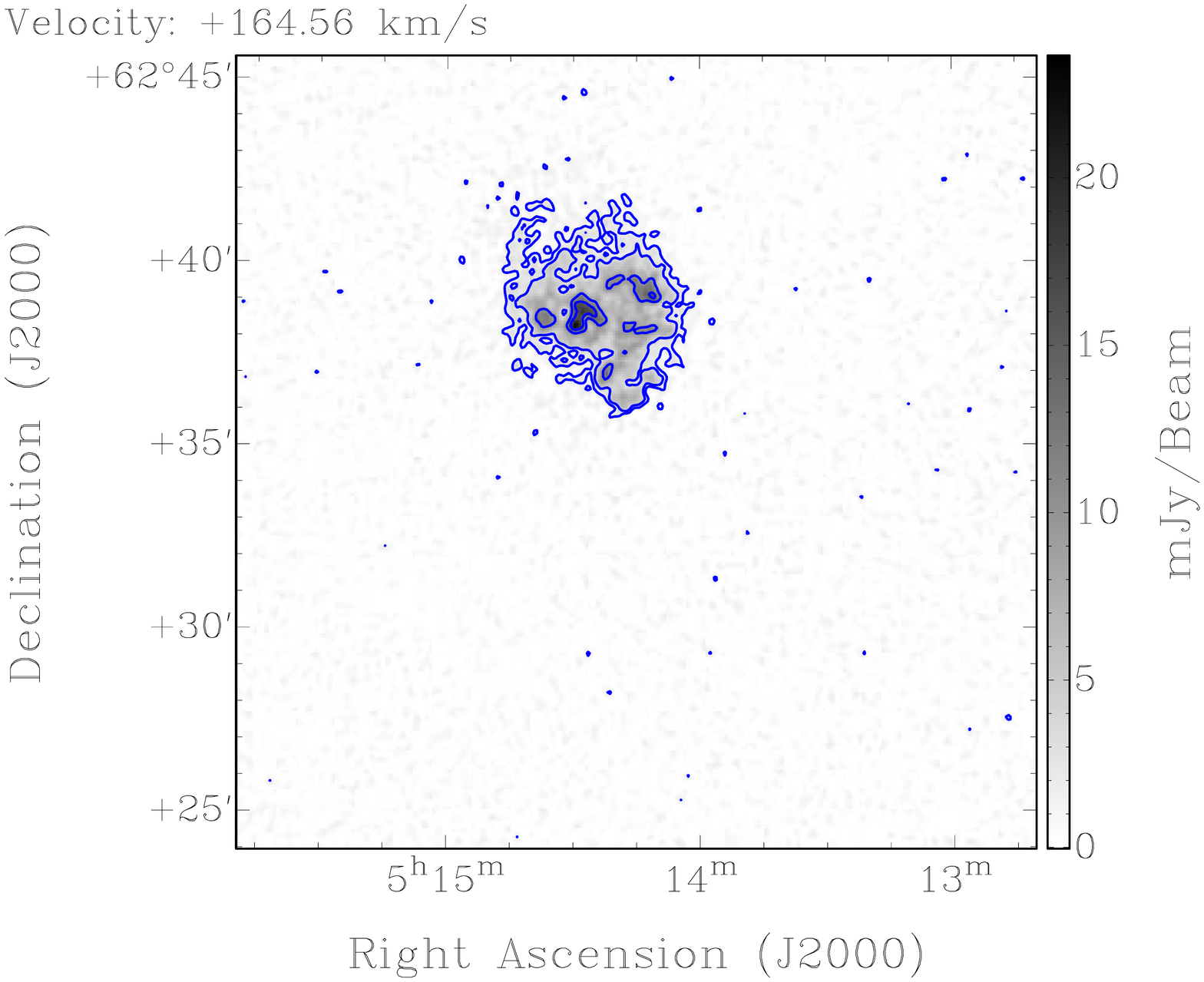}\includegraphics[width=0.29728\textwidth,angle=270,bb=0 0 612 792,keepaspectratio=true,clip=true,trim=33pt 214pt 0pt 132pt]{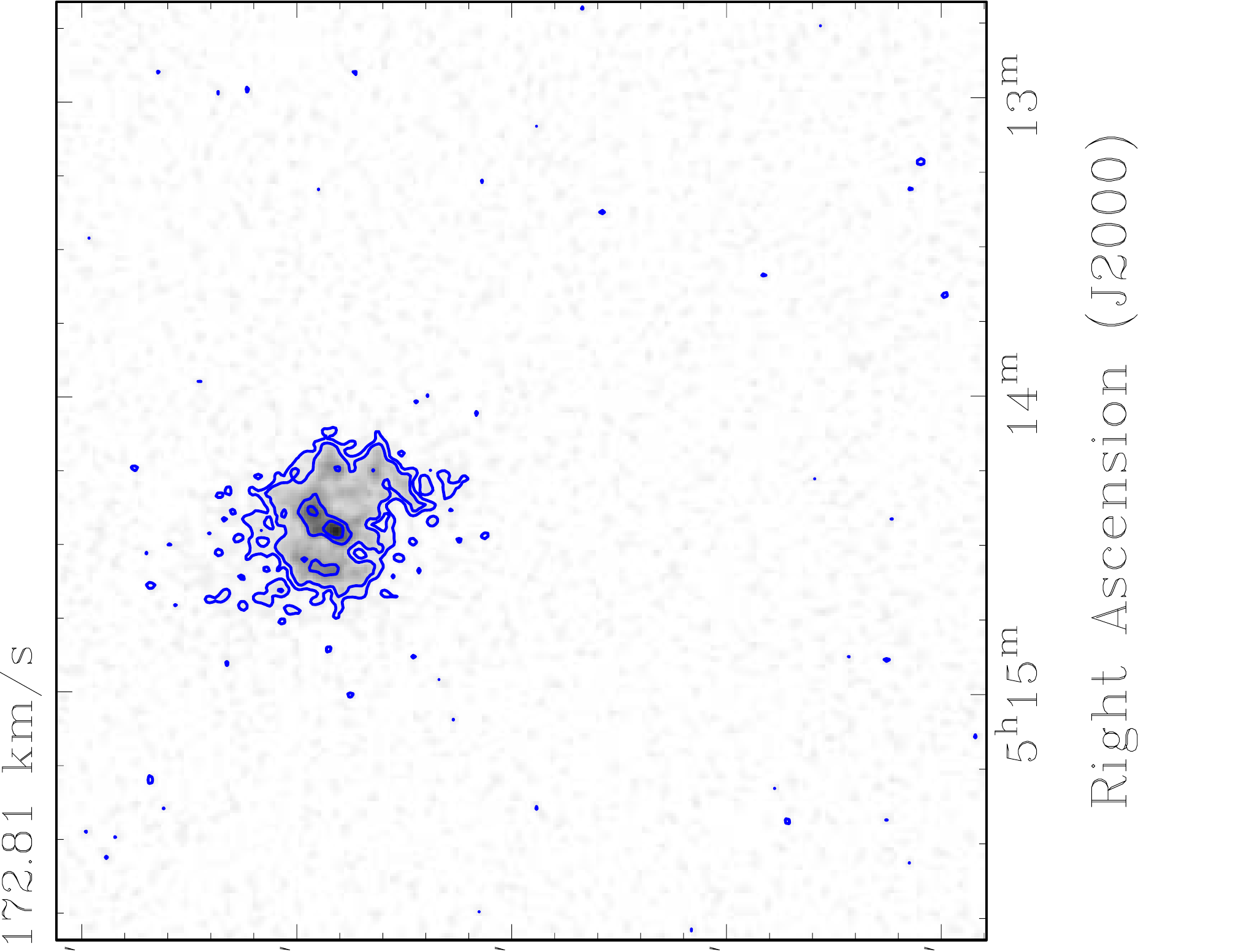}\includegraphics[width=0.29728\textwidth,angle=270,bb=0 0 612 792,keepaspectratio=true,clip=true,trim=33pt 214pt 0pt 132pt]{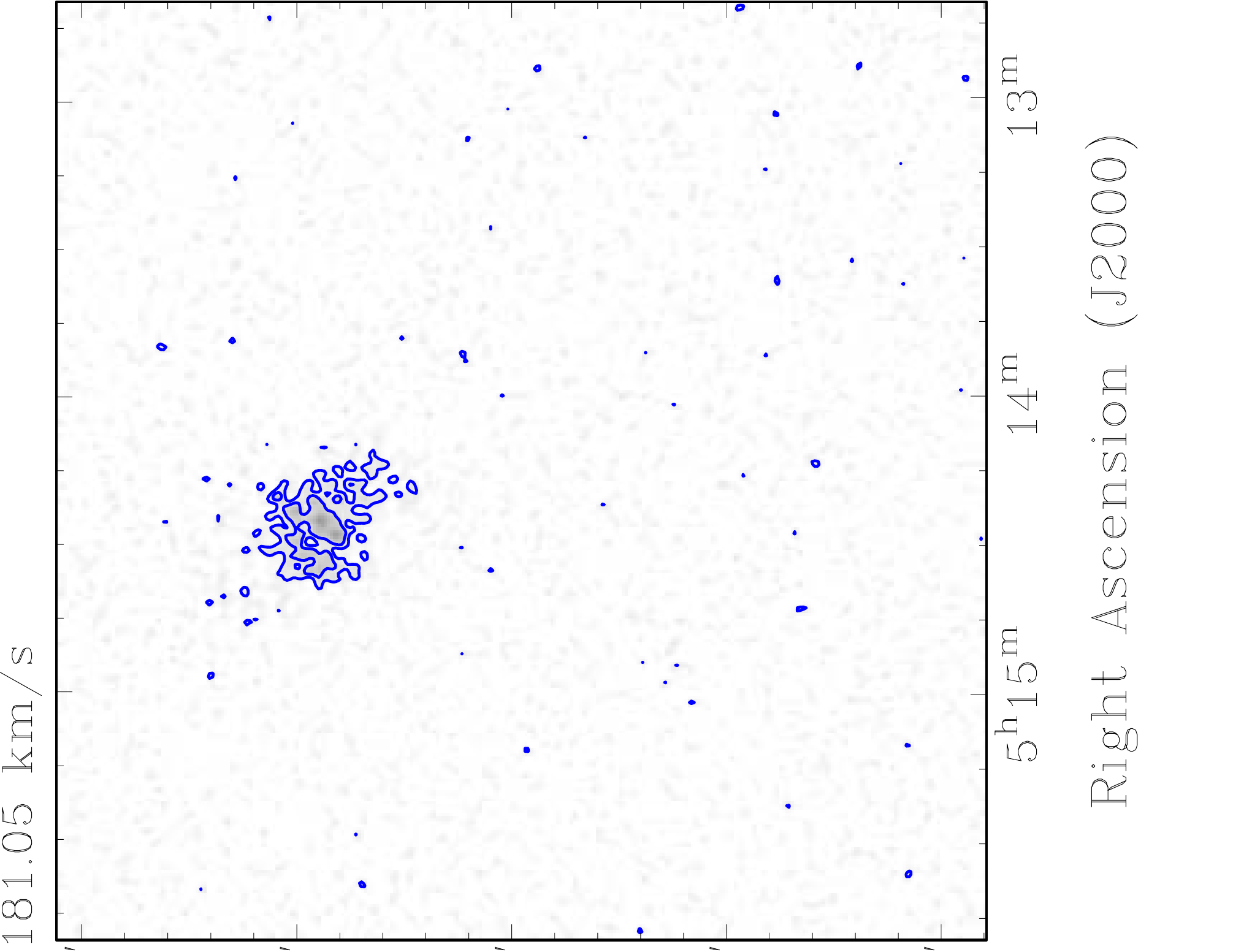}\includegraphics[width=0.29728\textwidth,angle=270,bb=0 0 612 792,keepaspectratio=true,clip=true,trim=33pt 214pt 0pt 42pt]{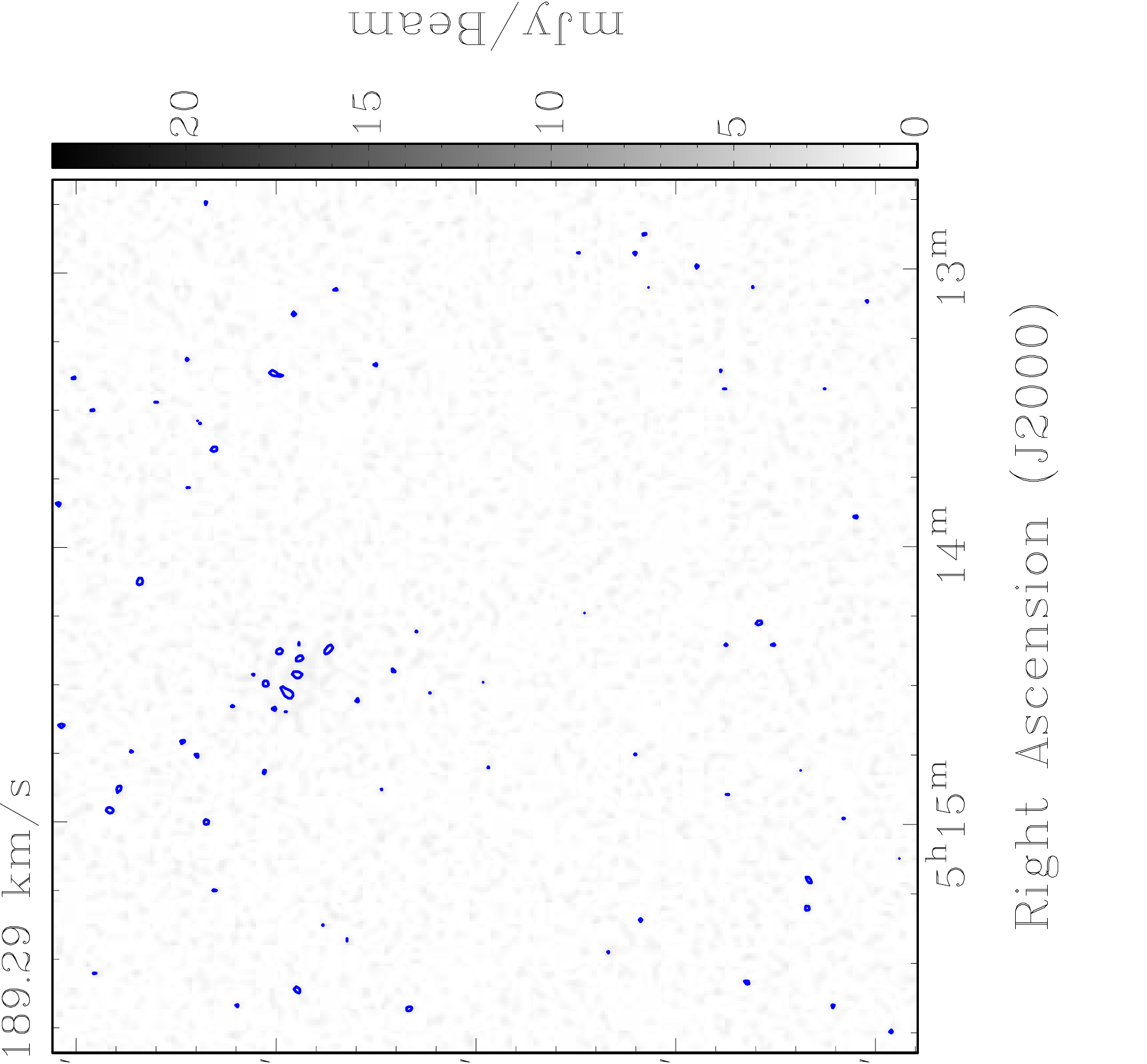}
 \caption{Channel maps from the 14\farcs84 $\times$ 13\farcs14 resolution data cube of UGCA\,105, observed with the WSRT. Contour levels are at 1.8 (3 $\sigma$), 3.5, 10, and 15\,mJy\,beam$^{-1}$. The heliocentric velocity is noted at the upper left corner of each map.}
\label{fig:02}
\end{figure*}

\begin{figure*}
 \centering
 \includegraphics[scale=0.45,angle=270,bb=0 0 612 792,clip=true,trim=0pt 117pt 45pt 75pt]{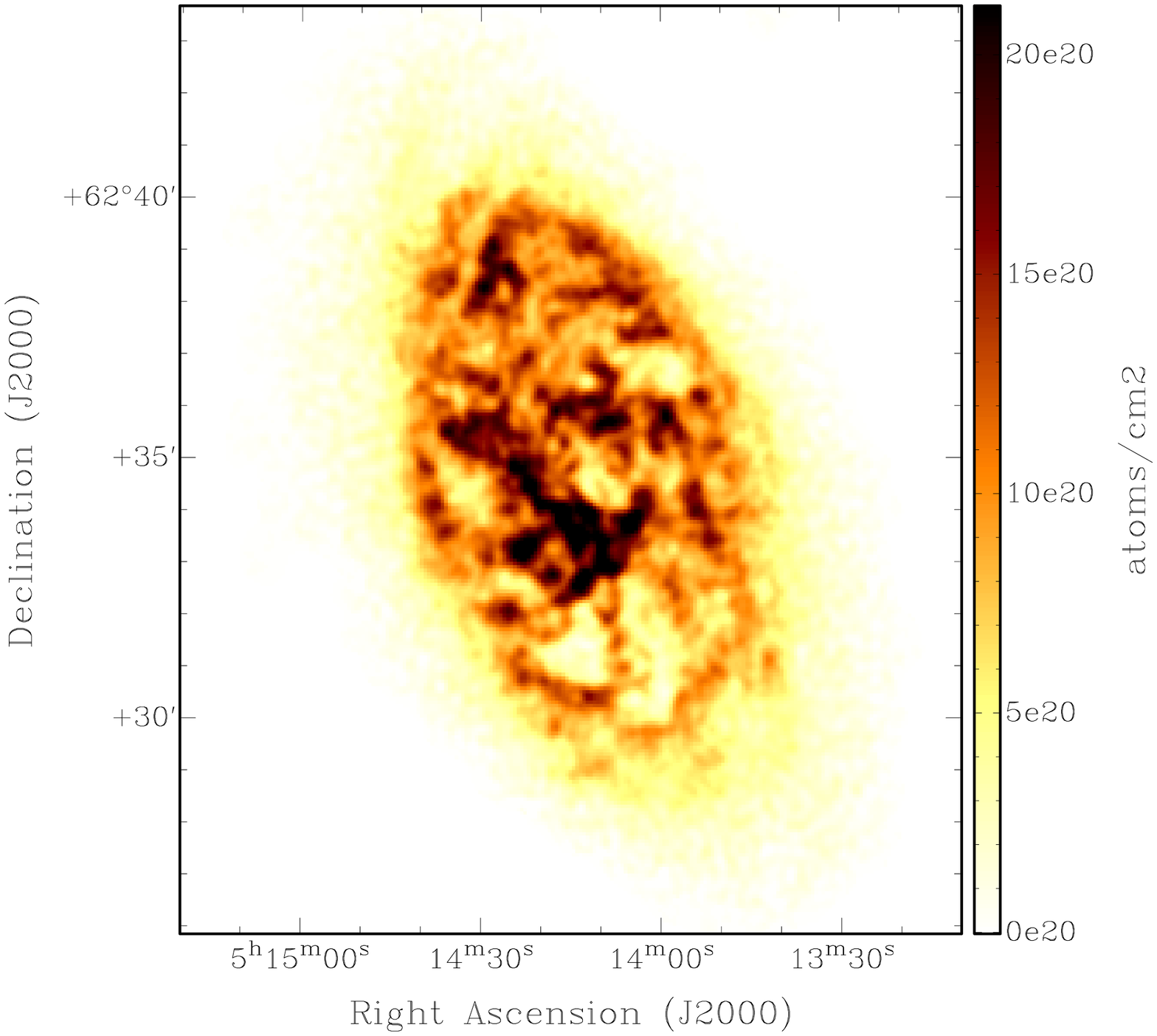}\includegraphics[scale=0.45,angle=270,bb=0 50 612 792,clip=true,trim=0pt 150pt 45pt 75pt]{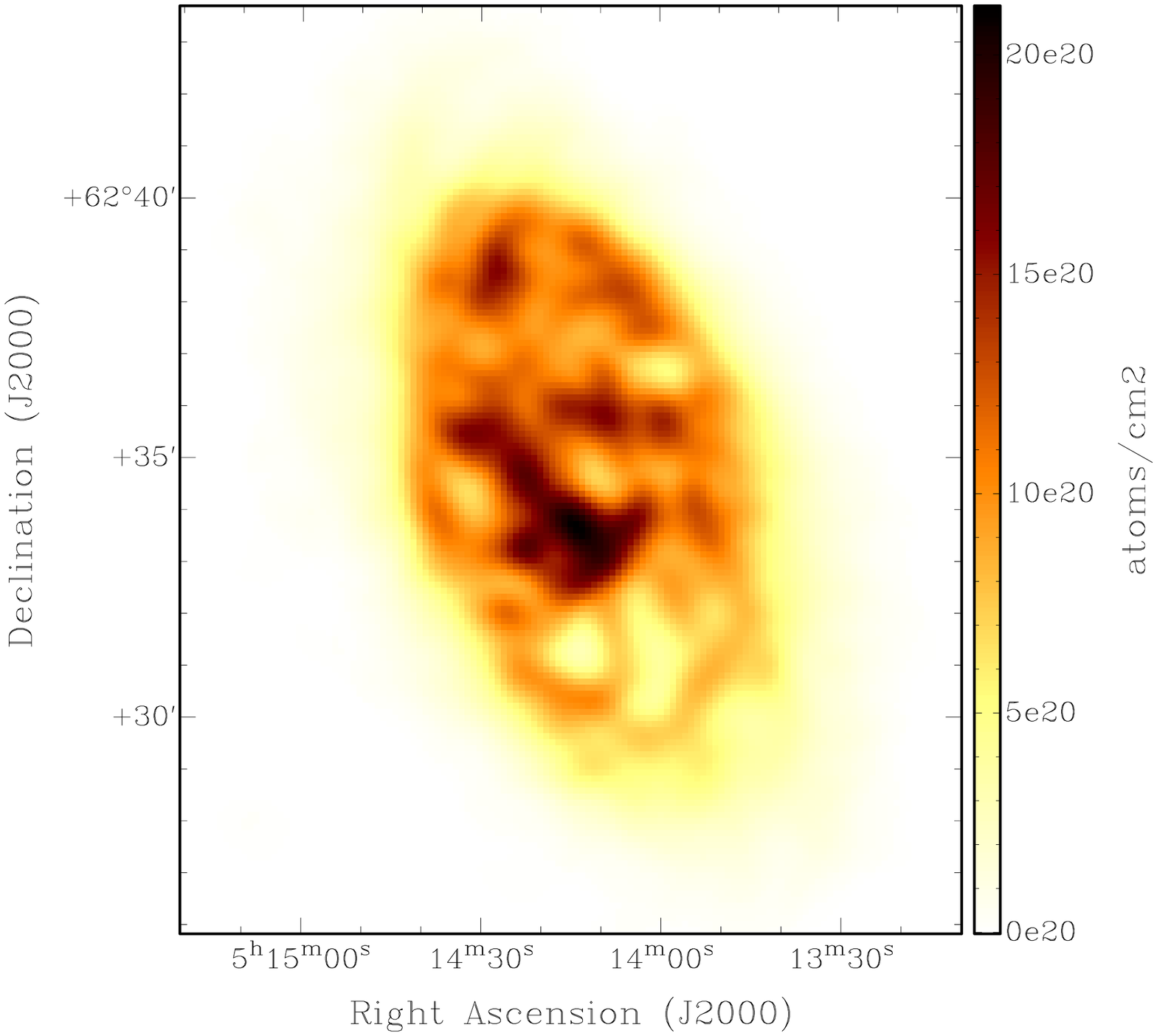}
 \caption{$\ion{H}{i}$ column density distribution of UGCA\,105 at 14\farcs84 $\times$ 13\farcs14 (left) and $39\farcs86 \times 38\farcs36$ (right) resolution. An angular distance of 1\arcmin corresponds to 1.01 kpc.}
\label{fig:03}
\end{figure*}

\begin{figure*}
 \centering
 \includegraphics[scale=0.45,angle=270,bb=0 0 612 792,clip=true,trim=0pt 0pt 41pt 260pt]{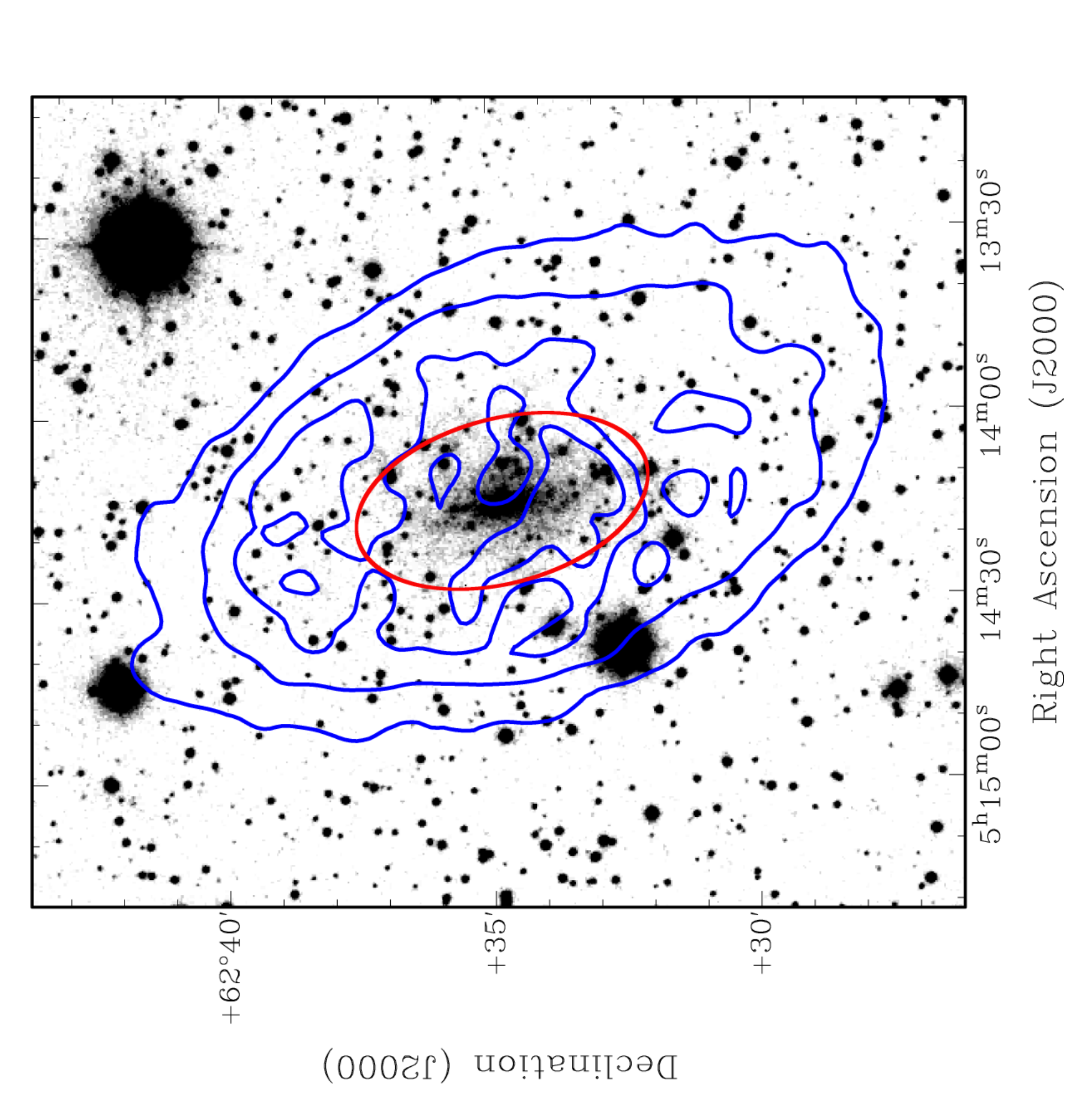}\includegraphics[scale=0.61,angle=270,bb=0 50 612 792,clip=true,trim=107pt 245pt 70pt 100pt]{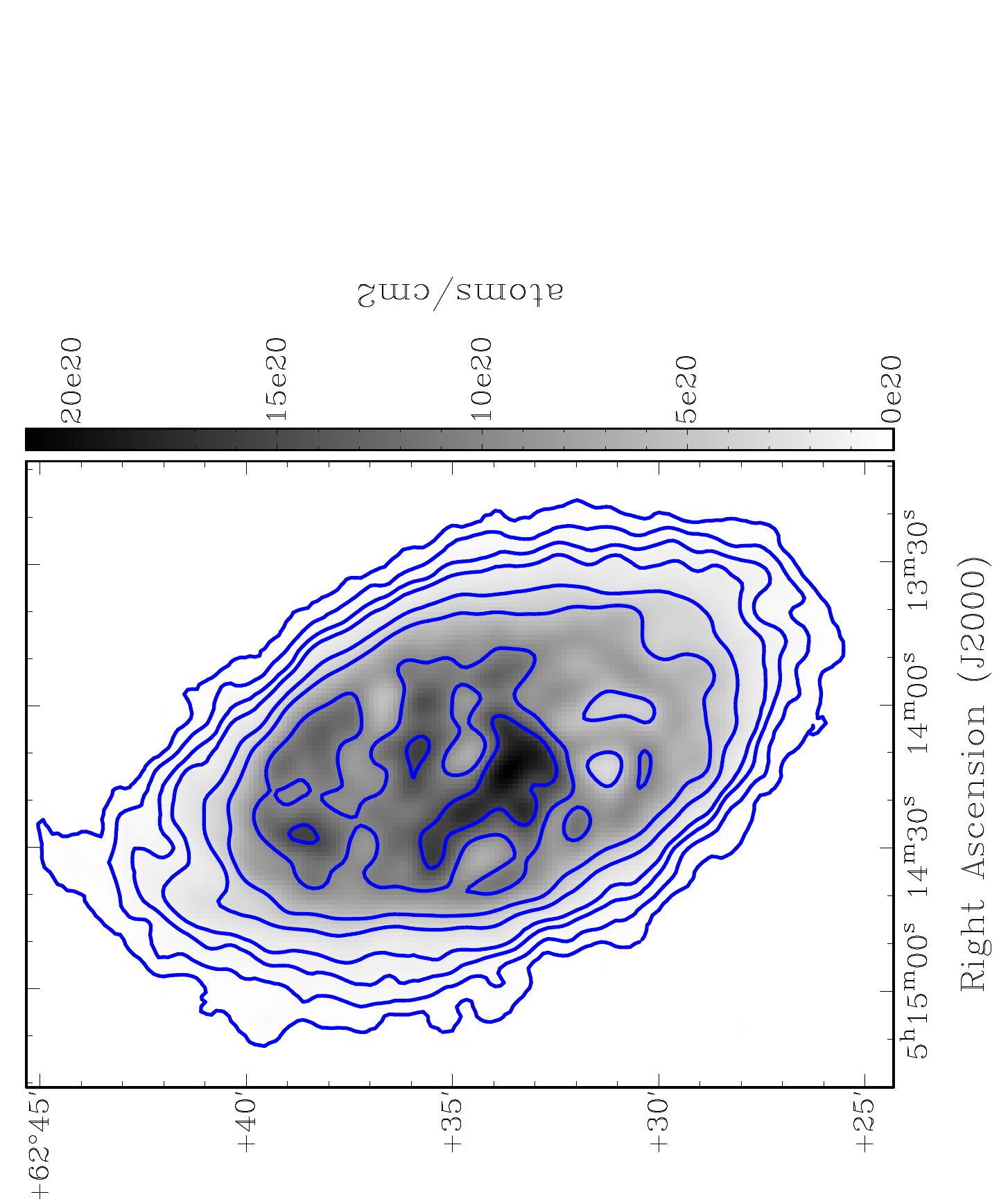}
 \caption{\bf Left: \normalfont DSS $I$-band image of UGCA\,105, overlaid with contours of the 39\farcs86 $\times$ 38\farcs36 resolution column density map at 1.5, 5, 10, and $15\times10^{20}$ atoms\,cm$^{-2}$. The red ellipse represents the extent of the optical disk ($r_{25}$). 
\bf Right: \normalfont $\ion{H}{i}$ column density map at the same resolution. Contour levels are 0.15, 0.5, 1, 1.5, 3, 5, 10, and $15\times10^{20}$ atoms\,cm$^{-2}$.}
\label{fig:04}
\end{figure*}

\begin{figure*}
 \centering
 \includegraphics[scale=0.58,angle=270,clip=true,trim=50pt 200pt 95pt 168pt]{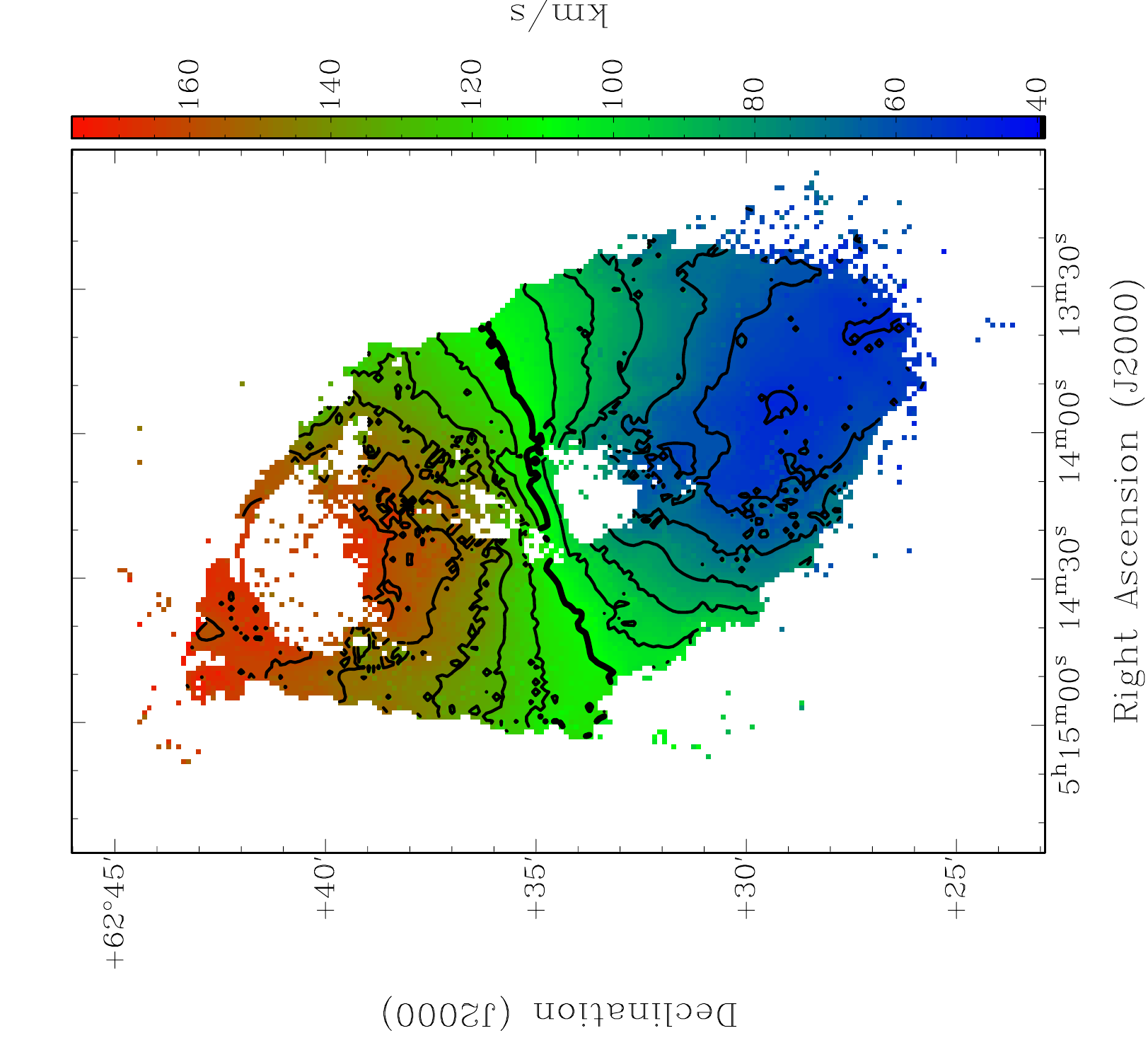}\includegraphics[scale=0.66,angle=270,clip=true,trim=54pt 230pt 105pt 100pt]{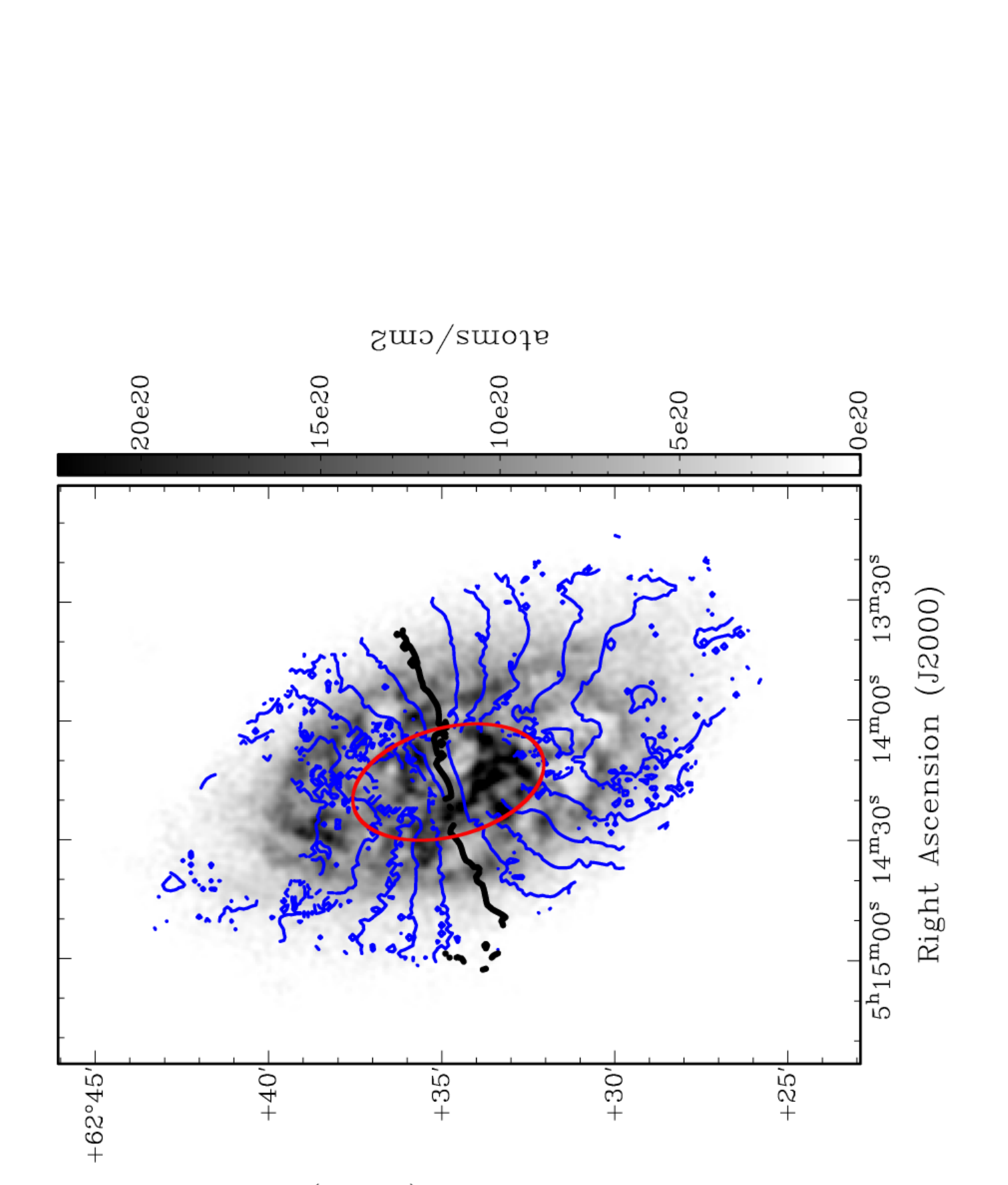}
 \caption{\bf Left: \normalfont Bulk-motion velocity field of UGCA\,105, derived from the low-resolution data cube, with contour levels ranging from 50 to 170\,$\rm{km\,s}^{-1}$, in steps of 10\,$\rm{km\,s}^{-1}$. The thick line represents the systemic velocity of 111\,$\rm{km\,s}^{-1}$. \bf Right: \normalfont Full-resolution $\ion{H}{i}$ column density map, overlaid with velocity contours as shown in the left-hand panel. The red ellipse shows the extent of the optical disk ($r_{25}$).}
\label{fig:05}
\end{figure*}

\begin{figure*}
 \centering
 \includegraphics[angle=0,scale=0.33,clip=true,trim=0pt 0pt 0pt 0pt]{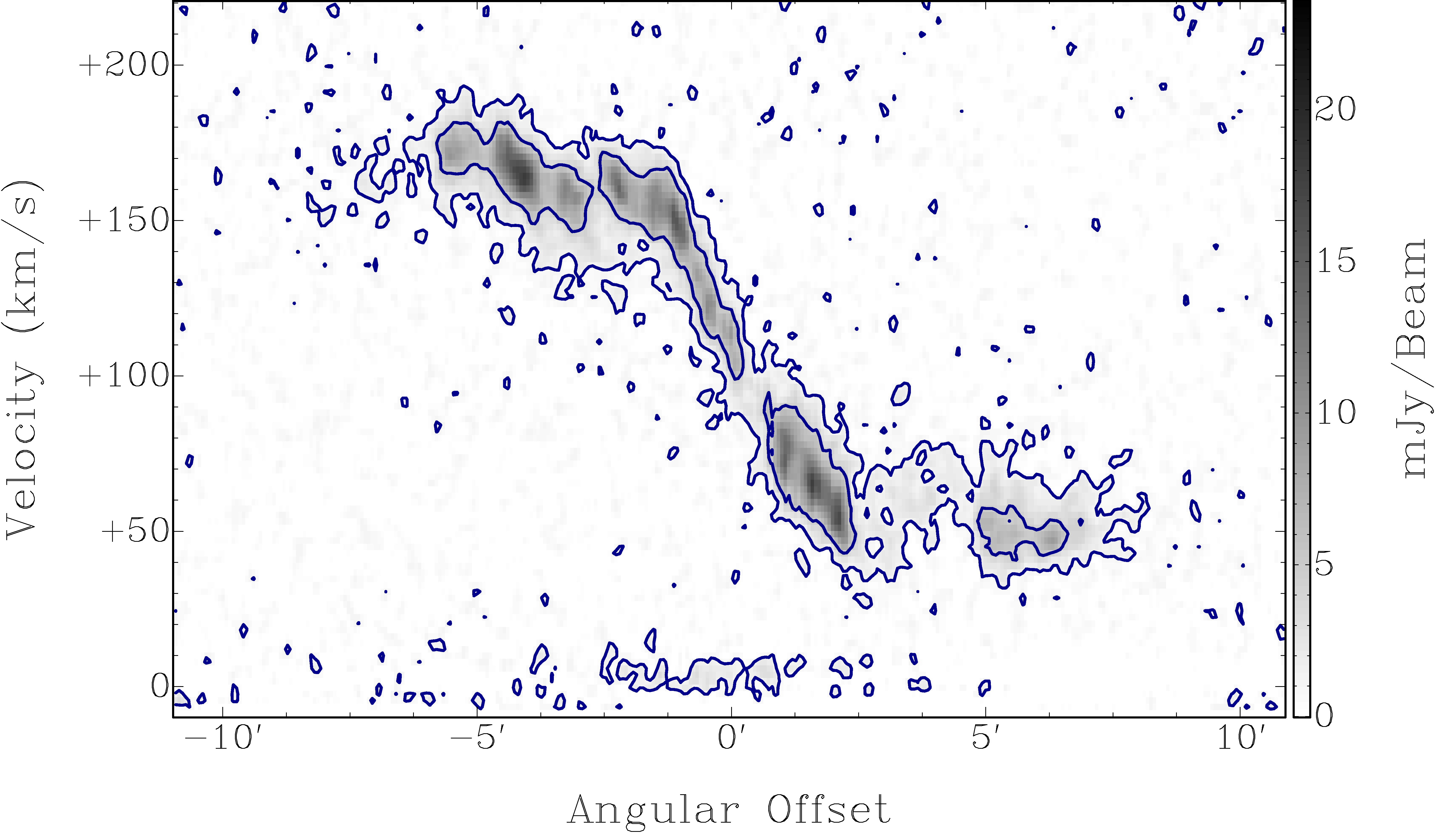}\includegraphics[angle=0,scale=0.33,clip=true,trim=24pt 0pt 0pt 0pt]{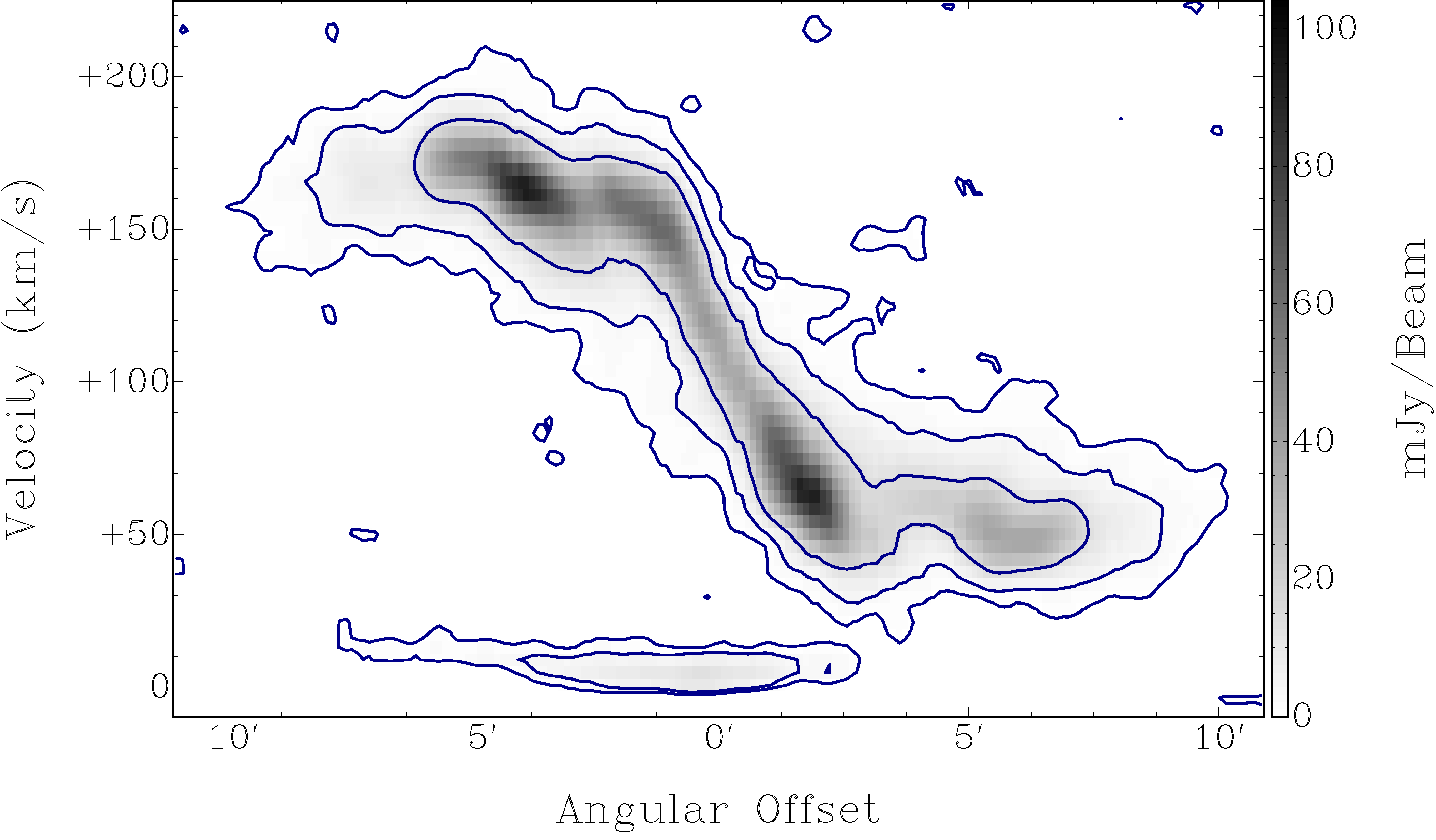}
 \caption{\bf Left: \normalfont Position-velocity (PV) diagram along the kinematical major axis of the 14\farcs84 $\times$ 13\farcs14 resolution cube of UGCA\,105, with contour levels of 1.2 (2 $\sigma$) and 4.8\,mJy\,beam$^{-1}$. \bf Right: \normalfont Position-velocity (PV) diagram along the kinematical major axis of the 39\farcs86 $\times$ 38\farcs36 resolution cube of UGCA\,105, with contour levels of 0.9 (2 $\sigma$), 3.6, and 14.4\,mJy\,beam$^{-1}$. The low-resolution data shows a ``beard'' of material at anomalous velocities, indicating the presence of a lagging extraplanar gaseous component. The feature visible at velocities near zero in both panels is Galactic $\ion{H}{i}$ emission.}
\label{fig:07}
\end{figure*}

\section{Results of $\ion{H}{i}$ observations}

\label{sect:U105cube}
Individual velocity channel maps of the high-resolution $\ion{H}{i}$ data cube of UGCA\,105 are presented in Fig.~\ref{fig:02}. Every fourth channel within the frequency range containing $\ion{H}{i}$ emission is shown. A quite regular and well-ordered rotation pattern of the $\ion{H}{i}$ disk is evident in the channel maps, but a multitude of hole-like structures is also revealed.

The $\ion{H}{i}$ total-intensity maps computed from both the high- (Robust-0) and low-resolution (Robust-0.4) cubes are shown in Fig.~\ref{fig:03}. The $\ion{H}{i}$ distribution shows an intricate pattern of low-density holes and high-density clumps and filaments, while the overall shape of the disk looks fairly regular and symmetric. A close look at the outermost diffuse northern and southern edges of the disk suggests a faint spiral structure, corresponding to counter-clockwise rotation.
As expected, the column densities tend to be higher in the inner parts of the galaxy; the radial gas distribution is, however, far from symmetric. There is a large, yet highly fragmented region of maximum column density located just southeast of the centre, with a peak value of $N_{\rm{\ion{H}{i}}}=2.98\times10^{21}$ atoms\,cm$^{-2}$, and an extension in the northeast. Another, much less extended maximum is located in the far northern part. Directly south of the maximum-intensity region there is an equally large area of exceptionally low column density, comprising one or several (overlapping) kiloparsec-sized holes. The largest holes are particularly prominent in the low-resolution $\ion{H}{i}$ map; one of them is located right in the galactic centre, surrounded by some of the most notable high column density regions. The northern part of the disk shows much lower density contrasts between holes and the gas in their proximity, thus displaying more of a filamentary structure of higher- and lower-density 
material rather than clearly defined holes or shells.

In the left-hand panel of Fig.~\ref{fig:04}, an I-band image of UGCA\,105 from the Digitized Sky Survey (DSS) is overlaid with contours of the low-resolution $\ion{H}{i}$ column density map. With a diameter of 16\,kpc (determined from the $1.5\times10^{20}$ \,cm$^{-2}$ contour in the low-resolution total intensity map), the $\ion{H}{i}$ disk is almost three times more extended than its stellar counterpart ($R_{25}$). The overlay also shows that the region of maximum $\ion{H}{i}$ emission is offset from the optical centre, and that the major axis orientation (i.e. the position angle) of the $\ion{H}{i}$ disk differs slightly from that of the optical one.

To derive a velocity field, we used the so-called bulk velocity method, filtering secondary velocity components in a multiple-component structure, which resulted in a cleaner velocity field \citep{oh08}. The velocity field extracted from the low-resolution data cube of UGCA\,105 is shown in Fig.~\ref{fig:05}. We found the large-scale velocity distribution of the gas to be much more regular and symmetric than typically observed in dwarf irregulars. On smaller scales, however, the velocity field is characterised by irregularities and disturbances, such as small wiggles in the isovelocity contours. Closer inspection reveals that the major axis of the velocity field is slightly twisted (suggesting a warped $\ion{H}{i}$ disk).

Remarkably enough, the $\ion{H}{i}$ disk of UGCA\,105 is extended enough to show differential rotation over a wide range of galacto-centric distances -- which is rarely seen in dwarf galaxies, as they typically exhibit solid-body rotation up to the outer edges of their detectable $\ion{H}{i}$ disks. The transition from solid-body- to differential rotation in UGCA\,105 takes place at a rather small distance ($\sim1.5$\,kpc) from the centre, comparable to the optical radius.

{The impression of a spiral pattern is mainly induced by the projected shearing of the density contrasts in the same, clockwise, direction towards the outskirts of the galaxy, in the transition region from solid-body to differential rotation. We presume that the spiral pattern is mainly due to the coherent (clockwise) elongation of holes and shells through differential rotation. In the following we hence assumed counterclockwise rotation, and, with that, the west side of the galaxy to point towards the observer.}

A more detailed inspection of the velocity field shows a slight twist of the minor with respect to the major kinematical axis at small radii (near the stellar body). This indicates the presence of non-circular motions in the galaxy, the most simple one being a radial component.

Figure~\ref{fig:07} shows position-velocity (PV) diagrams along the major axis of both data cubes, with intensity contours down at the 2-$\sigma$ level. The slice through the low-resolution cube shows faint wings of emission visible on the approaching and receding side, extending coherently towards the systemic velocity (and beyond, indicating non-circular motions), that is, there is a ``beard'' of anomalous $\ion{H}{i}$ moving at lower velocities than the overall galaxy rotation. The anomalous emission appears to be very faint and rather diffuse, so that it is hardly visible in the Robust-0 cube. As confirmed by our tilted-ring analysis (Sect.~\ref{sect:TRM}), the presence of the beard is a sign of a lagging extraplanar gas component.

\section{Tilted-ring modelling}
\label{sect:TRM}


We modelled UGCA~105 by means of a tilted-ring model \citep[][]{rogstad74}. In its simplest form, this method is based on the assumption of circular orbits of the gas, such that the kinematics of a galactic disk at any radius can be described by the following quantities: the circular velocity, the systemic velocity, the central position of the orbit, the inclination, and the position angle, the latter two defining the local orientation of the disk. Tilted-ring models of galaxies therefore do not only provide rotation curves, but {they} also describe the radial change in orientation of {the galactic} disks (i.e. symmetric warping). 
We use the least-squares fitting routine \texttt{TiRiFiC} (\textit{Ti}lted \textit{Ri}ng \textit{Fi}tting \textit{C}ode, \citealt{Jozsa07a}), which allows for a number of extensions to the simple tilted-ring model. We mention those we tested in the following description where necessary. \texttt{TiRiFiC} achieves a best-fit solution by optimising the match of a model data cube and the data, which makes it possible to investigate the 3D structure of our target galaxy.

In our model we need to take into account all constituents of the \ion{H}{i} disk. Since significant parts of the \ion{H}{i} at anomalous velocities are apparently smoothly distributed, it is detected best in the low-resolution data cube. We hence based our model on an analysis of the low-resolution data cube. We used ring widths of at least 30$\arcsec$ for all fit parameters. To avoid ambiguities and to keep the number of free parameters as low as possible, we modelled an inner disk of 120$\arcsec$ radius with a constant orientation and scale height in all cases. Furthermore, we treated the centre position (RA, DEC) and systemic velocity as free fit-parameters in \texttt{TiRiFiC}, but since the $\ion{H}{i}$ distribution and velocity field are sufficiently symmetric, these parameters were treated as being independent of radius.

The model can reproduce the major features of the data cubes sufficiently well, retaining the simplicity of the model as much as possible. Because a kinematical model is not constrained by physics, only by the data, a virtually infinite number of models may reflect the real \ion{H}{i} structure of the galaxy. We present a selection of different models (carrying different physical interpretations) to argue that among the represented model families we can find and quantify a best solution. However, it cannot be excluded that even our best-fit solution is ambiguous with respect to the used parametrisation scheme. We describe eight different (best-fit) model solutions for the $\ion{H}{i}$ cube of UGCA\,105, and assess their quality based on a comparison of position-velocity diagrams along the kinematical major and minor axes of the data and model cubes, respectively (see Fig.~\ref{fig:slicemodels}). The critical features of 
the data that we aim to reproduce in the modelling process are not only the central beard emission seen in the major-axis PV slice, but also the skewness of the contours in the minor-axis PV diagram.

\subsection{Warped model}
\label{sect:warp}

{Initially, we} applied a basic tilted-ring model. The obtained best-fit solution (as shown in Fig.~\ref{fig:13}) is characterised by a 26$^{\circ}$ inclination warp of the $\ion{H}{i}$ disk, as well as a 10$^{\circ}$ warp in position angle, and we therefore refer to it as the warped model throughout this paper (a warp is present in our subsequent models as well, albeit with a distinctly lower amplitude). The derived surface density distribution shows the highest values for the region between 1 and 2.5\,kpc, with a rather sharp drop towards the innermost region (corresponding to the central $\ion{H}{i}$ hole), and a roughly exponential decline towards the outer edge, except for a secondary maximum between 4 and 5\,kpc. The fitted scale height, the vertical density profile assumed throughout the paper to be determined by a $sech^2$ law, 
\begin{equation}
\rho_{\rm HI}(r,z) = \sigma_{\rm HI}(r)\cdot h(z) = \sigma_{\rm HI}(r)\cdot (2\,z_0(r))^{-1}\cdot sech^2(z\cdot z_0^{-1}(r))\;, 
\end{equation}
with $\rho_{\rm HI}$ being the $\ion{H}{i}$ density, $z_0$ the scale height, and $\sigma_{\rm HI}(r)$ the surface-density, shows a strong flare of the disk, with values exceeding 2\,kpc for the outermost rings. The rotation curve indicates solid-body rotation in the inner part of the galaxy, with a transition to differential rotation at a radius of $\sim$1.5\,kpc, and shows an {over}-Keplerian decline at the largest radii. {As we show below, this unphysical decline vanishes for the models where an additional, lagging component above the disk is taken into account.}

\subsection{Central bar streaming}
\label{sect:bar}

From Fig.~\ref{fig:slicemodels} (panel \textit{a)}) it is evident that while our warped model provides a reasonable approximation to the overall kinematics including the \ion{H}{i} with anomalous velocities at larger radii, it clearly fails to reproduce the anomalous gas kinematics in the centre and the shape of the fainter emission in the minor-axis PV diagram. Before proceeding to introduce vertical velocity gradients, we tried to improve the warped model by considering elliptical streaming due to a central bar (as observations of the stellar disk suggest, see Sect.~\ref{sect:sample}), since the skewness of the minor-axis slice is indicative of non-circular motions. This is realised by including second-order harmonic terms in tangential and radial velocity (see e.g. \citealt{Franx94}, \citealt{Schoenmakers97}, \citealt{Spekkens07}). Testing various combinations of bar amplitudes and phases, we found the best-fitting model at a second-order rotational amplitude of $20\,\rm{km\,s}^{-1}$ within $2.0\,\rm{
kpc}$ from the centre and a bar orientation of $60^{\circ}$ clockwise from the major axis. Panel \textit{b)} of Fig.~\ref{fig:slicemodels} indicates only slight improvement with respect to the major-axis diagram, with the inner beard still largely uncaptured. In the minor-axis slice, the model gives a better approximation of the upper-central bump in the contours (see red arrow in panel \textit{a)}), but the upper-right and lower-left bumps (red circles in panel \textit{a)}) in the outer contour still cannot not be reproduced by this {model}.

\subsection{Introducing a rotational lag}
\label{sect:DVRO}

Since the models described above do not yet provide satisfactory fits to the anomalous gas, we introduced a lagging extraplanar component by fitting a vertical gradient of the rotation velocity $DVRO$. Panel \textit{c)} of Fig.~\ref{fig:slicemodels} shows a model with $DVRO\approx-60\,\rm{km\,s}^{-1}\,\rm{kpc}^{-1}$ for the inner disk, increasing to $DVRO\approx-30\,\rm{km\,s}^{-1}\,\rm{kpc}^{-1}$, which is reached at $r=4.0\,\rm{kpc}$, and without a strong warp and bar streaming. While $DVRO$ has a similar effect on the low-column-density gas at large radii as the warp of our initial model, the fit of the minor-axis slice is poorer than in the previous cases because it now lacks in skewness, which the bar (and to a lesser extent also the warp) provided. 

Adding the previously used bar motion parameters (panel \textit{d)}) leads to a minor improvement in terms of the beard, while for the minor-axis diagrams a satisfactory match is still not achieved.

\subsection{Radial motion}
\label{sect:VRAD}

To obtain a better approximation for the outer contours in the minor-axis PV slice, we constructed a model with an inward radial velocity component $\varv_{\rm{rad}}$, whose amplitude decreases linearly from $\varv_{\rm{rad}}=-30\,\rm{km\,s}^{-1}$ at the centre to $\varv_{\rm{rad}}=0$ at $r=5.6\,\rm{kpc}$. The radial velocity component is constant with height above the plane. We also kept the $DVRO$ from the previous two models (note, however, that neither a bar nor a strong warp is included in this or any of the following models). The corresponding PV-slices are shown in panel \textit{e)}. We find that while the beard in the major-axis slice can in principle be reproduced by choosing a strong enough radial velocity component, the resulting skewness of the minor-axis slice is too large at high column densities, even though the upper-right and lower-left bumps of the outer contour are reasonably well approximated. From this, we {deduce} that the (inward) radial component to the 
velocity increases with height above the plane. We only consider an extraplanar radial velocity component in the following steps.

\subsection{Adding a gradient in radial velocity}
\label{sect:DVRA}

The model shown in panel \textit{f)} of Fig.~\ref{fig:slicemodels} features a vertical gradient in (inward) radial velocity $DVRA$, decreasing linearly from $-85\,\rm{km\,s}^{-1}\,\rm{kpc}^{-1}$ at the centre to $-60\,\rm{km\,s}^{-1}\rm{kpc}^{-1}$ at $r=2.0\,\rm{kpc}$, and decreasing further to $DVRA=0$ for $r\geq4.0\,\rm{kpc}$. Neither $DVRO$ nor $\varv_{\rm{rad}}$ is included. While $DVRA$ on its own does not provide any major-axis beard emission at all, we see that it is crucial to reproducing the skewness of the minor-axis PV slice also at low column densities.

Our next model includes both $DVRO$ and $DVRA$, each having the values specified above. As apparent from panel \textit{g)}, this provides an excellent fit to the beard of anomalous $\ion{H}{i}$ emission, and {an additional}, albeit marginal, improvement of the minor-axis PV contours.

As a final step, onset heights of the gradients $ZDRO$ and $ZDRA$ were introduced (panel \textit{h)}), since gradients near the mid-plane result in slightly too low velocities at low column densities, as a comparison of the major-axis PV diagrams in panel \textit{g)} illustrates. This is also evident in individual channels of the respective data cubes (Fig.~\ref{fig:modchannels}). We found the best fit to the data using $ZDRO=ZDRA=0.2$\,kpc. Notice that while this onset height above the plane is below the resolution of the analysed data cube, a non-zero choice nevertheless leaves its signature on the modelled data, since the complete vertical profile is affected.

\begin{figure*}
 \centering
 \includegraphics[angle=0,scale=0.21,bb=0 0 612 792,clip=true,trim=0pt 290pt 75pt 0pt]{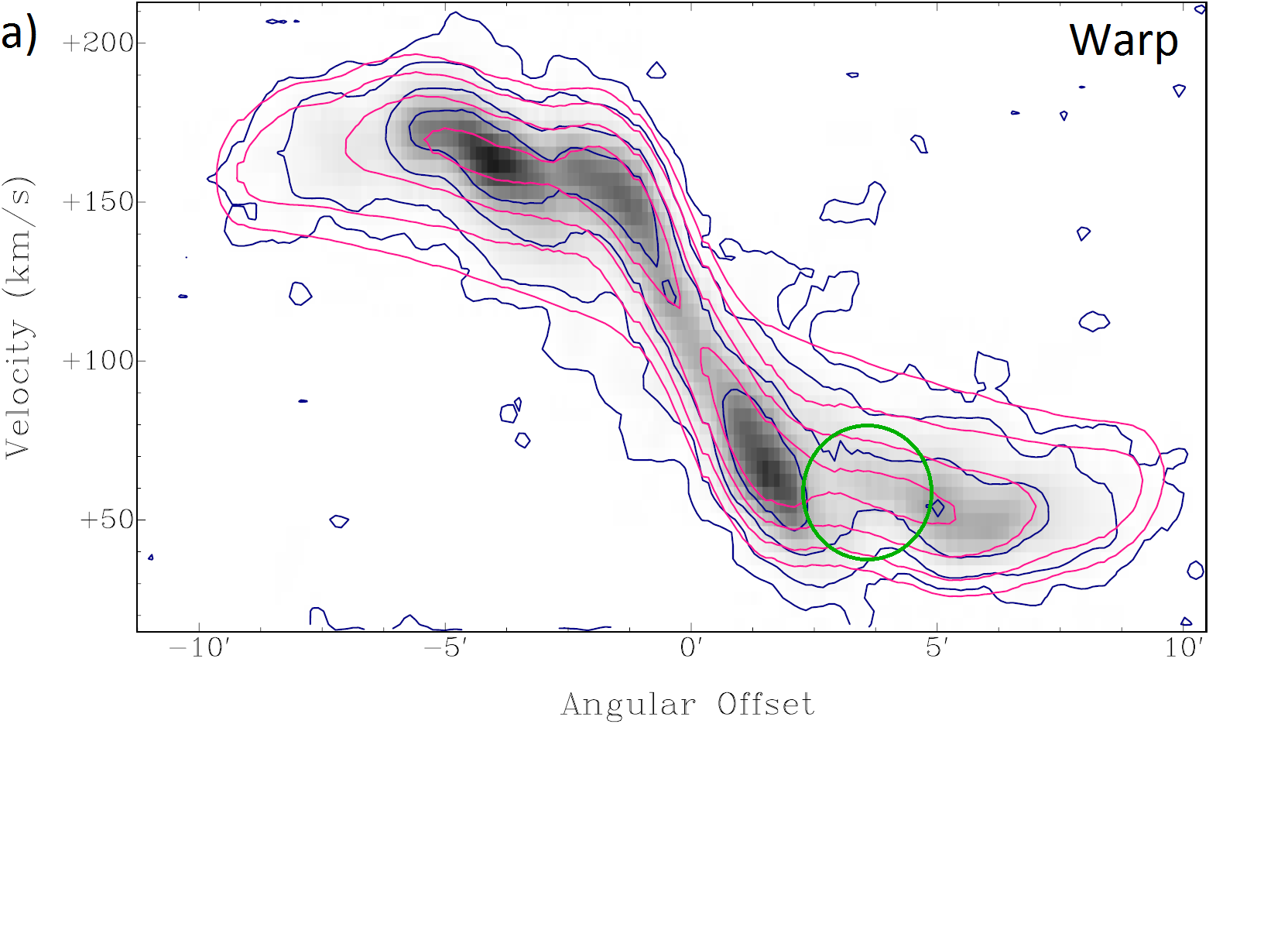}\includegraphics[angle=0,scale=0.21,bb=0 0 612 792,clip=true,trim=133pt 290pt 0pt 0pt]{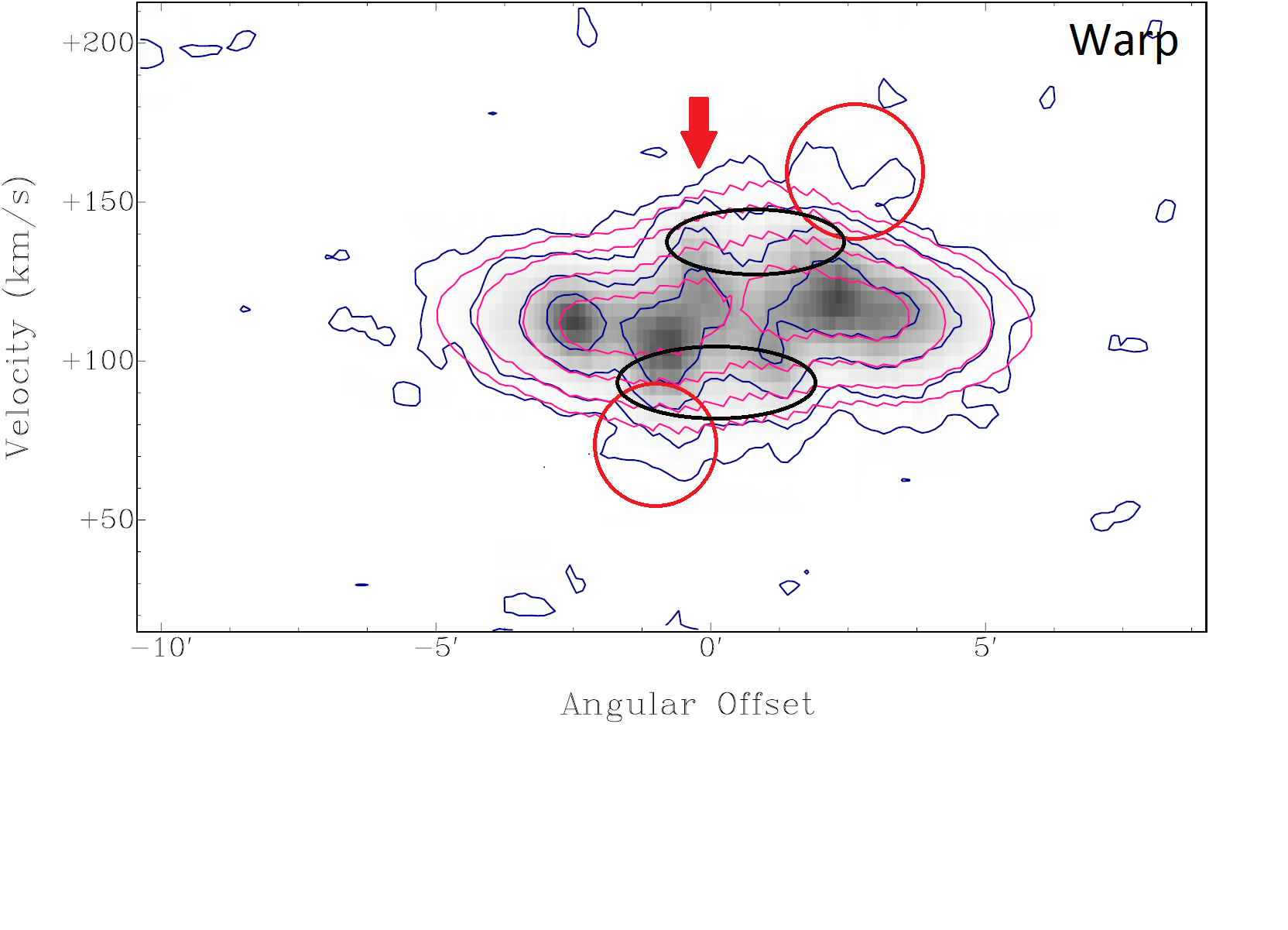}
\includegraphics[angle=0,scale=0.21,bb=0 0 612 792,clip=true,trim=0pt 290pt 75pt 0pt]{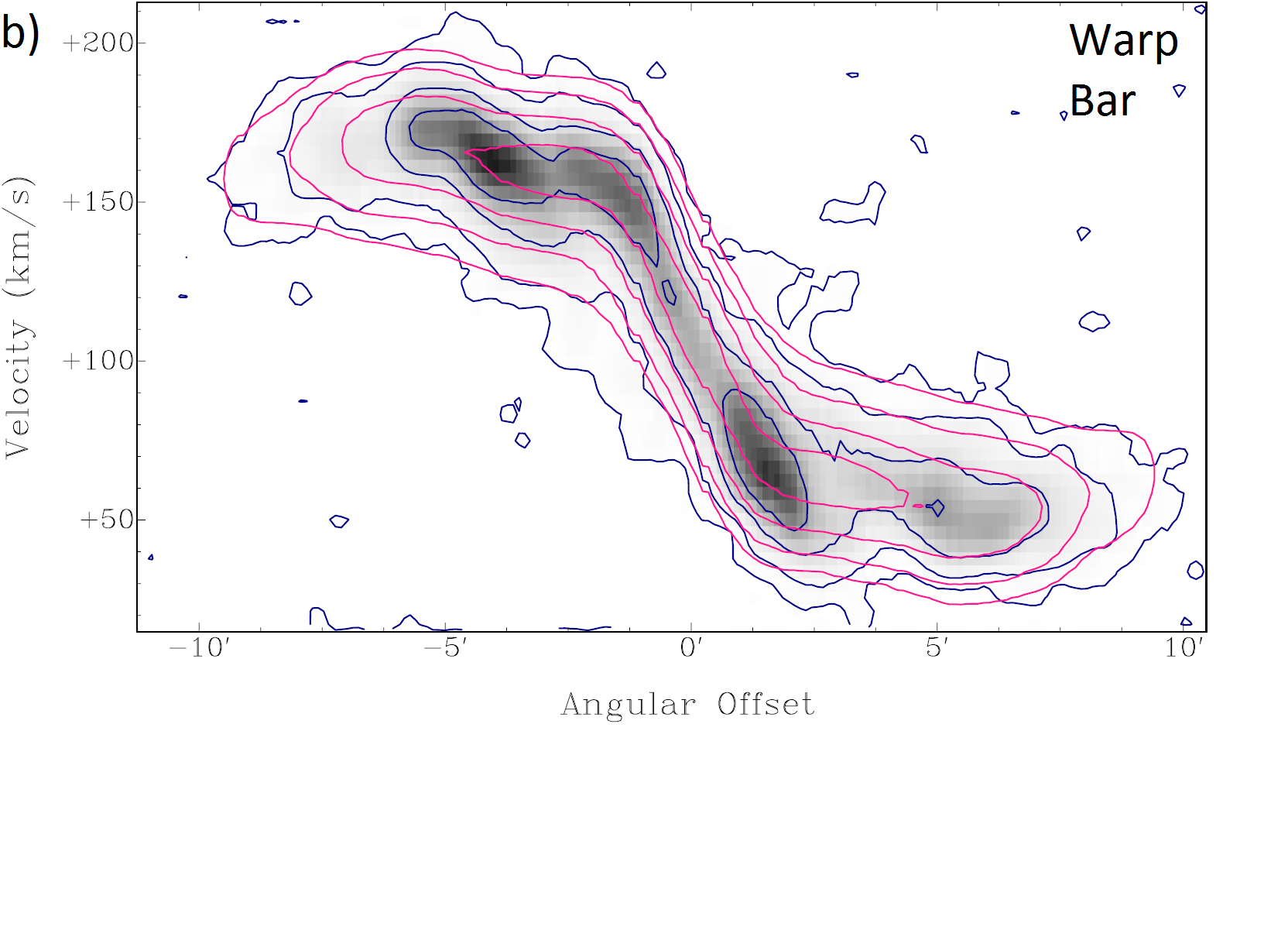}\includegraphics[angle=0,scale=0.21,bb=0 0 612 792,clip=true,trim=133pt 290pt 0pt 0pt]{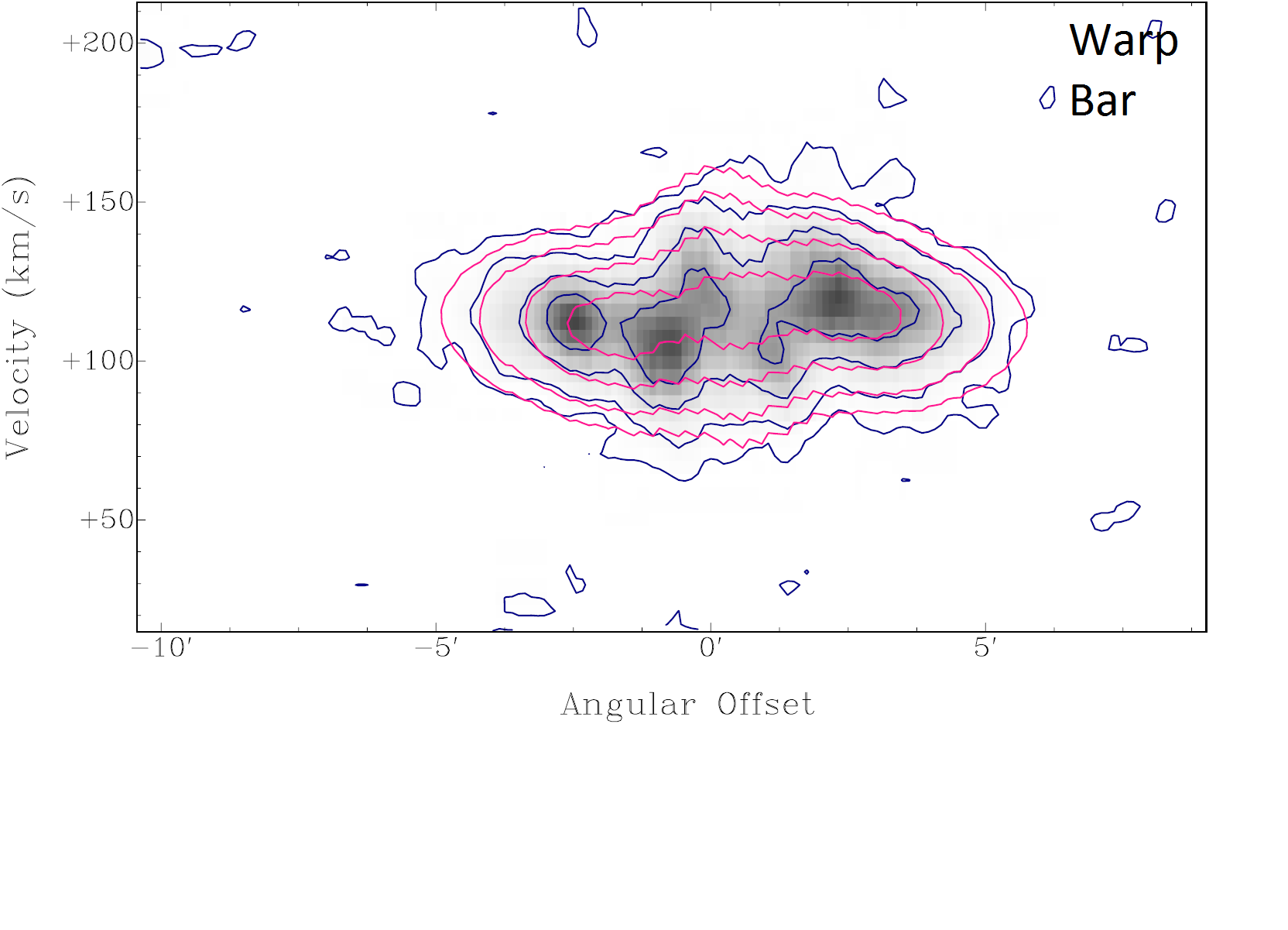}
\includegraphics[angle=0,scale=0.21,bb=0 0 612 792,clip=true,trim=0pt 290pt 75pt 0pt]{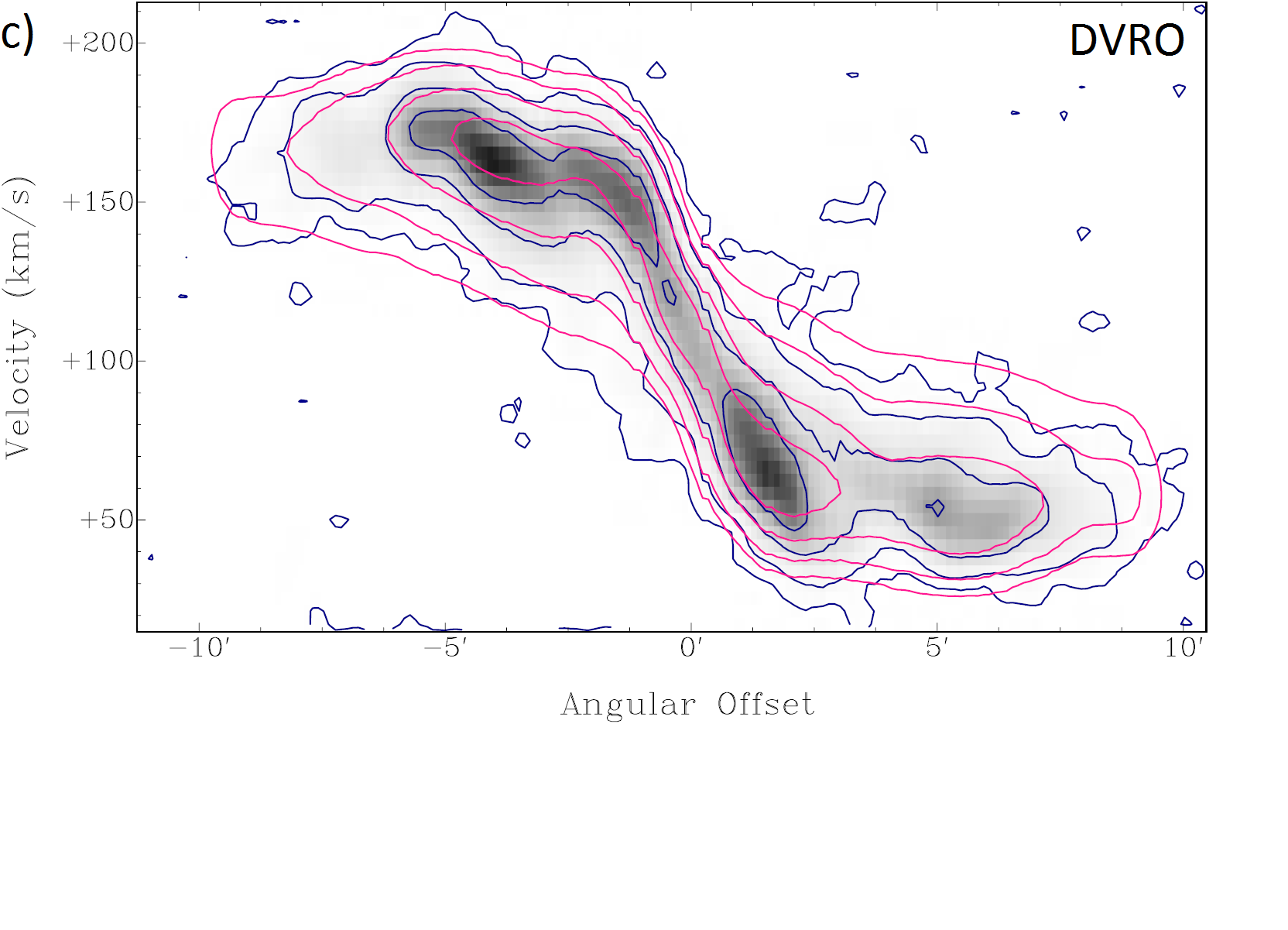}\includegraphics[angle=0,scale=0.21,bb=0 0 612 792,clip=true,trim=133pt 290pt 0pt 0pt]{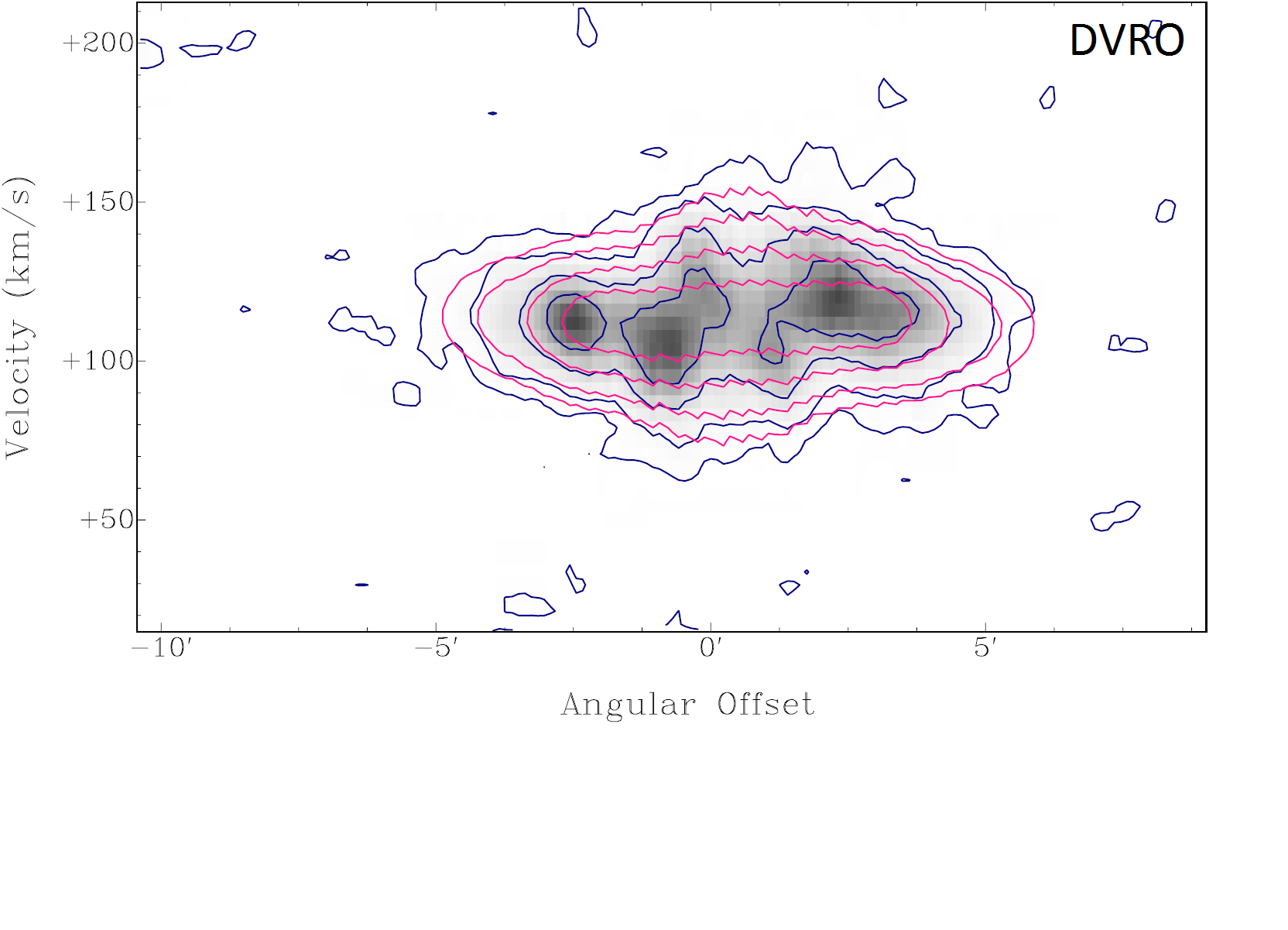}
\includegraphics[angle=0,scale=0.21,bb=0 0 612 792,clip=true,trim=0pt 150pt 75pt 0pt]{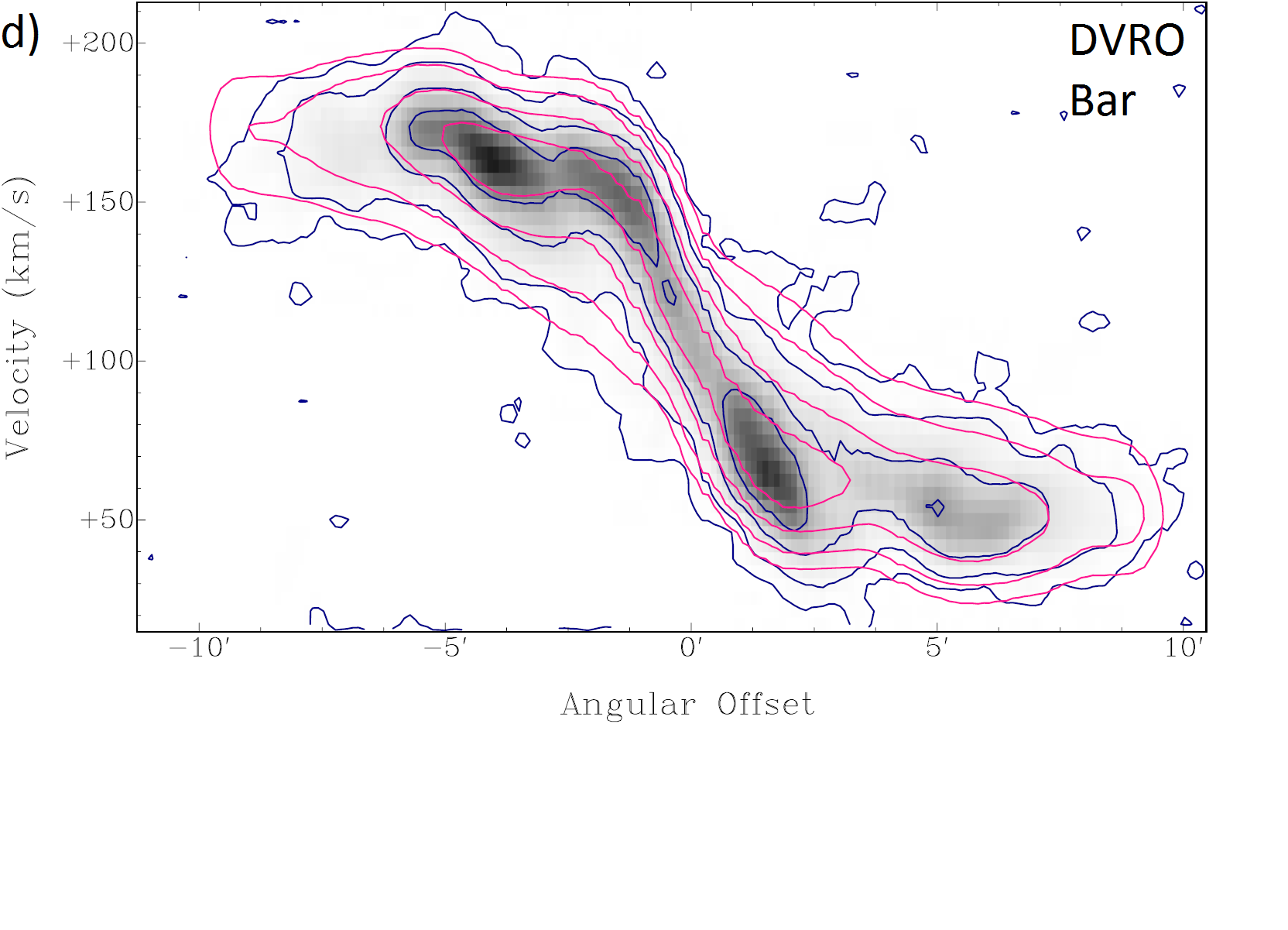}\includegraphics[angle=0,scale=0.21,bb=0 0 612 792,clip=true,trim=133pt 150pt 0pt 0pt]{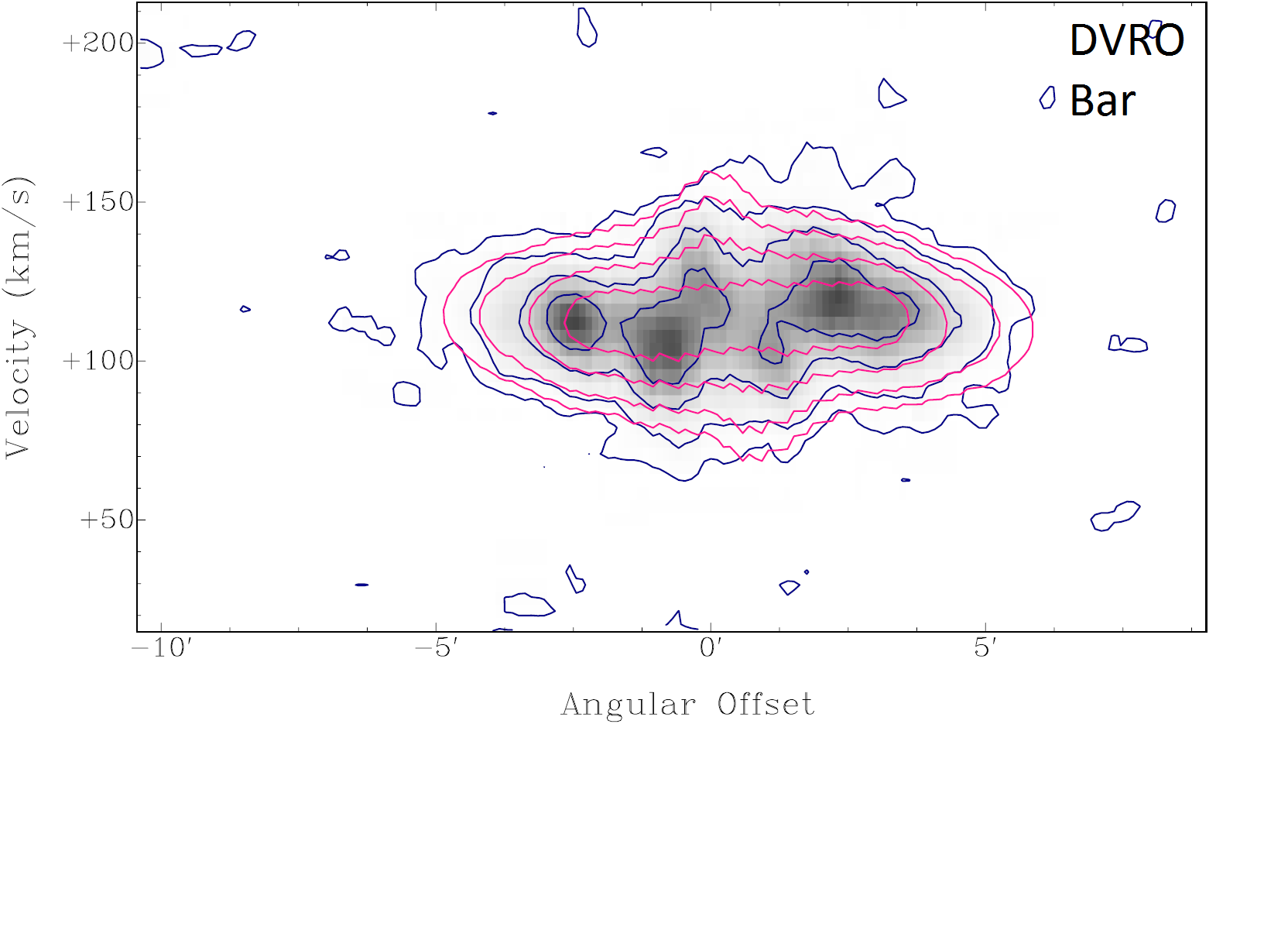}
 \caption{\bf (continued on next page) \normalfont PV slices along the kinematical major (left column) and minor axis (right column) of the 39\farcs86 $\times$ 38\farcs36 resolution data cube of UGCA\,105 (greyscale and blue contours) in comparison with the corresponding PV slices through different model cubes (pink contours). Contour levels are 0.9 (2\,$\sigma$), 3.6, 14.4, and 40\,mJy\,beam$^{-1}$. The characteristic features of each model are given in the top right corner of the respective panels (DVRO: vertical gradient of rotation velocity, DVRA: vertical gradient of radial velocity, Vrad: radial velocity, ZDRO: onset height for vertical gradient of rotation velocity, ZDRA: onset height for vertical gradient of radial velocity; see Sect.~\ref{sect:warp}-~\ref{sect:DVRO} for details). The arrow, circles, and ellipses in panel \textit{a)} denote unusual features in the data, some of which are successfully reproduced by certain models, while others are not (see text for details).
}
 \label{fig:slicemodels}
\end{figure*}

\begin{figure*}
 \centering
 \includegraphics[angle=0,scale=0.21,bb=0 0 612 792,clip=true,trim=0pt 290pt 75pt 0pt]{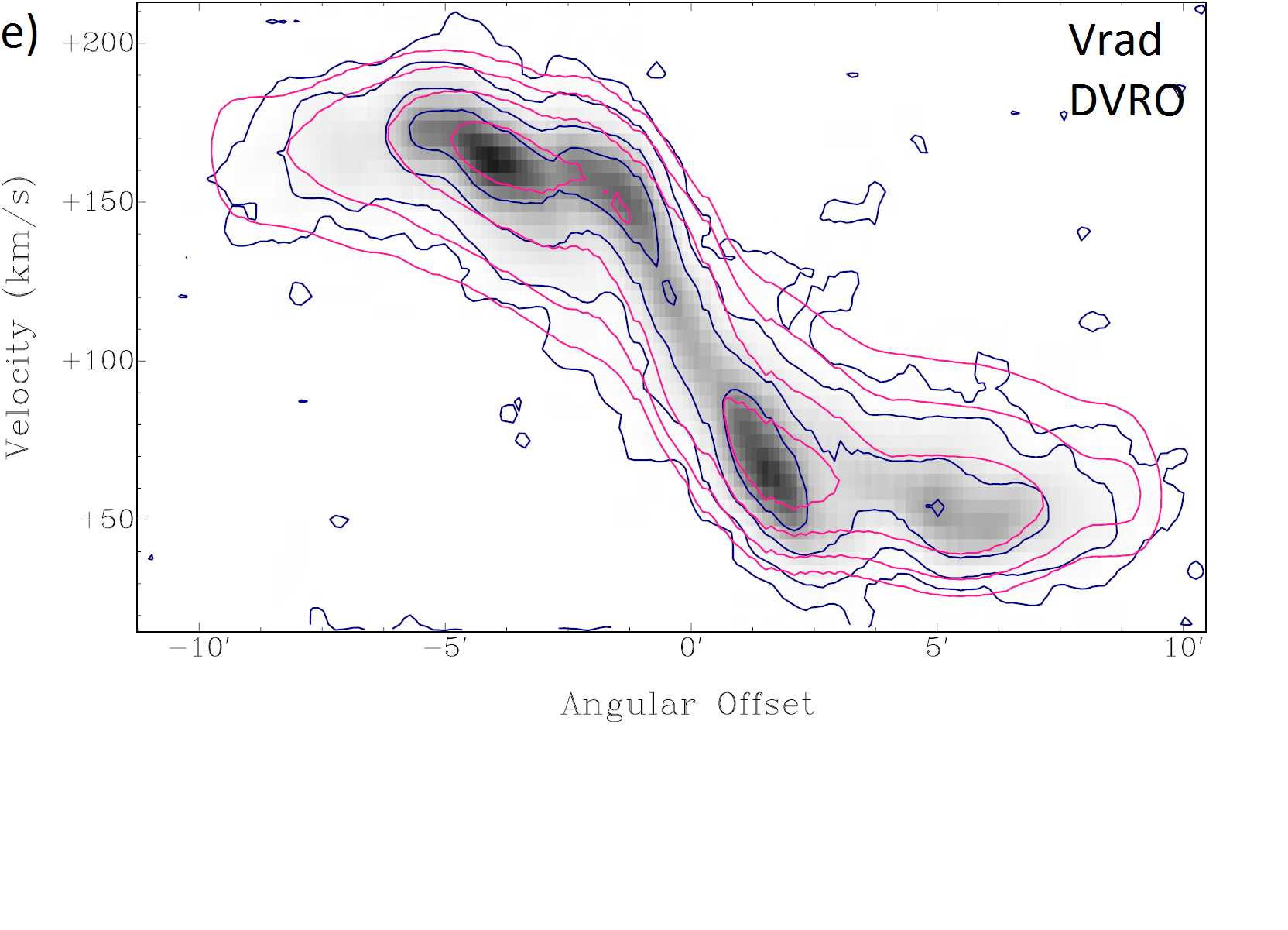}\includegraphics[angle=0,scale=0.21,bb=0 0 612 792,clip=true,trim=133pt 290pt 0pt 0pt]{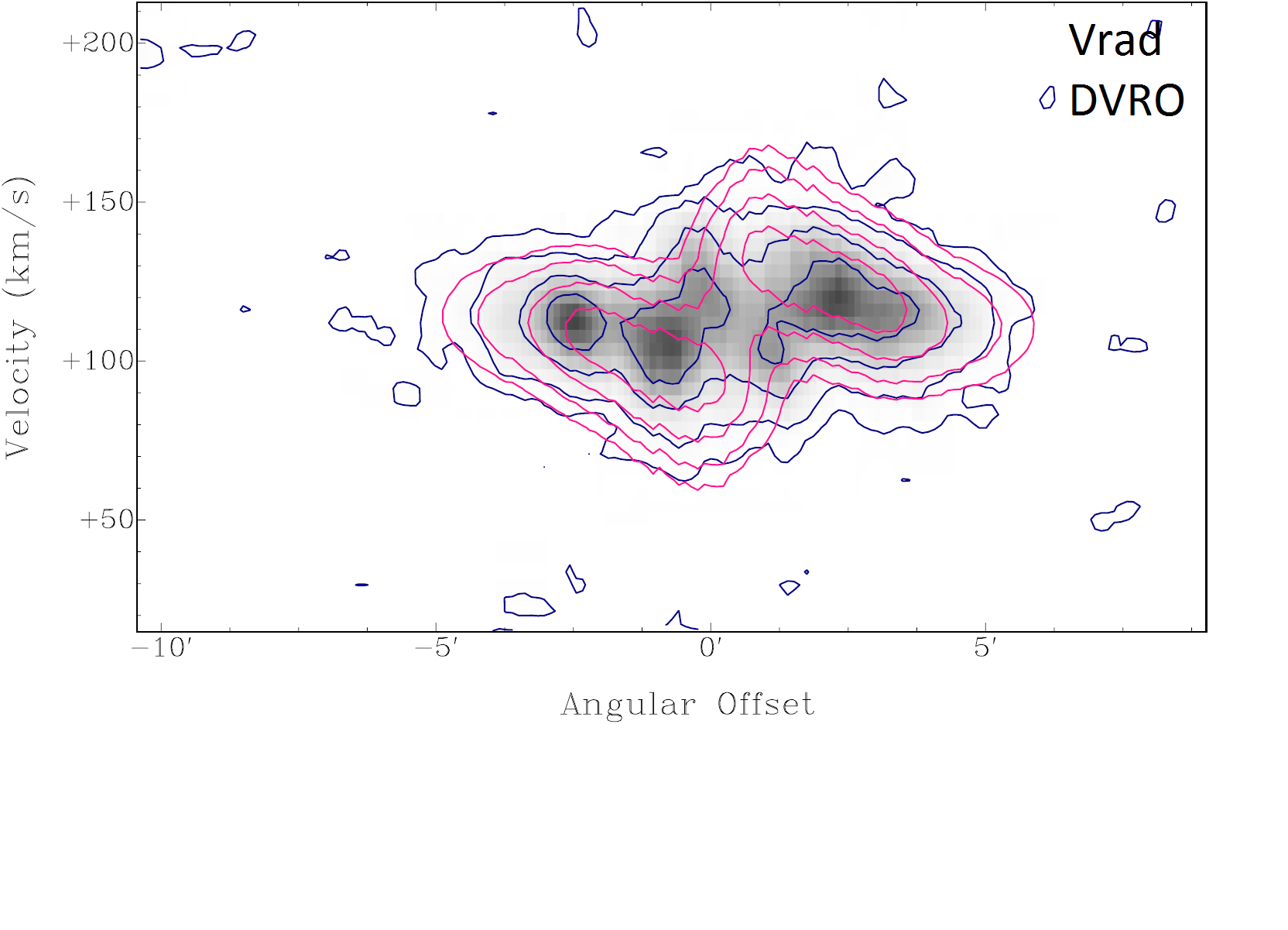}
\includegraphics[angle=0,scale=0.21,bb=0 0 612 792,clip=true,trim=0pt 290pt 75pt 0pt]{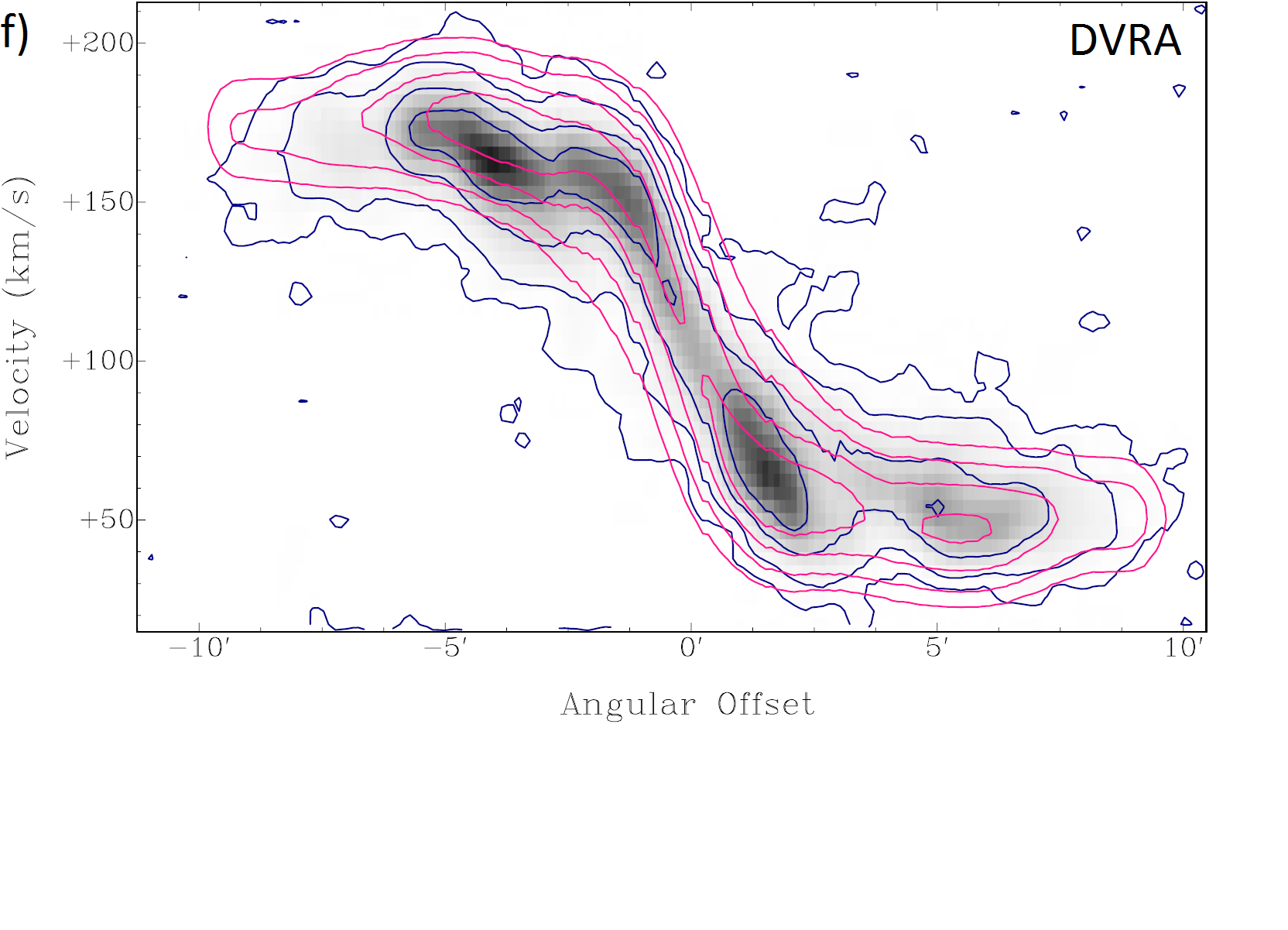}\includegraphics[angle=0,scale=0.21,bb=0 0 612 792,clip=true,trim=133pt 290pt 0pt 0pt]{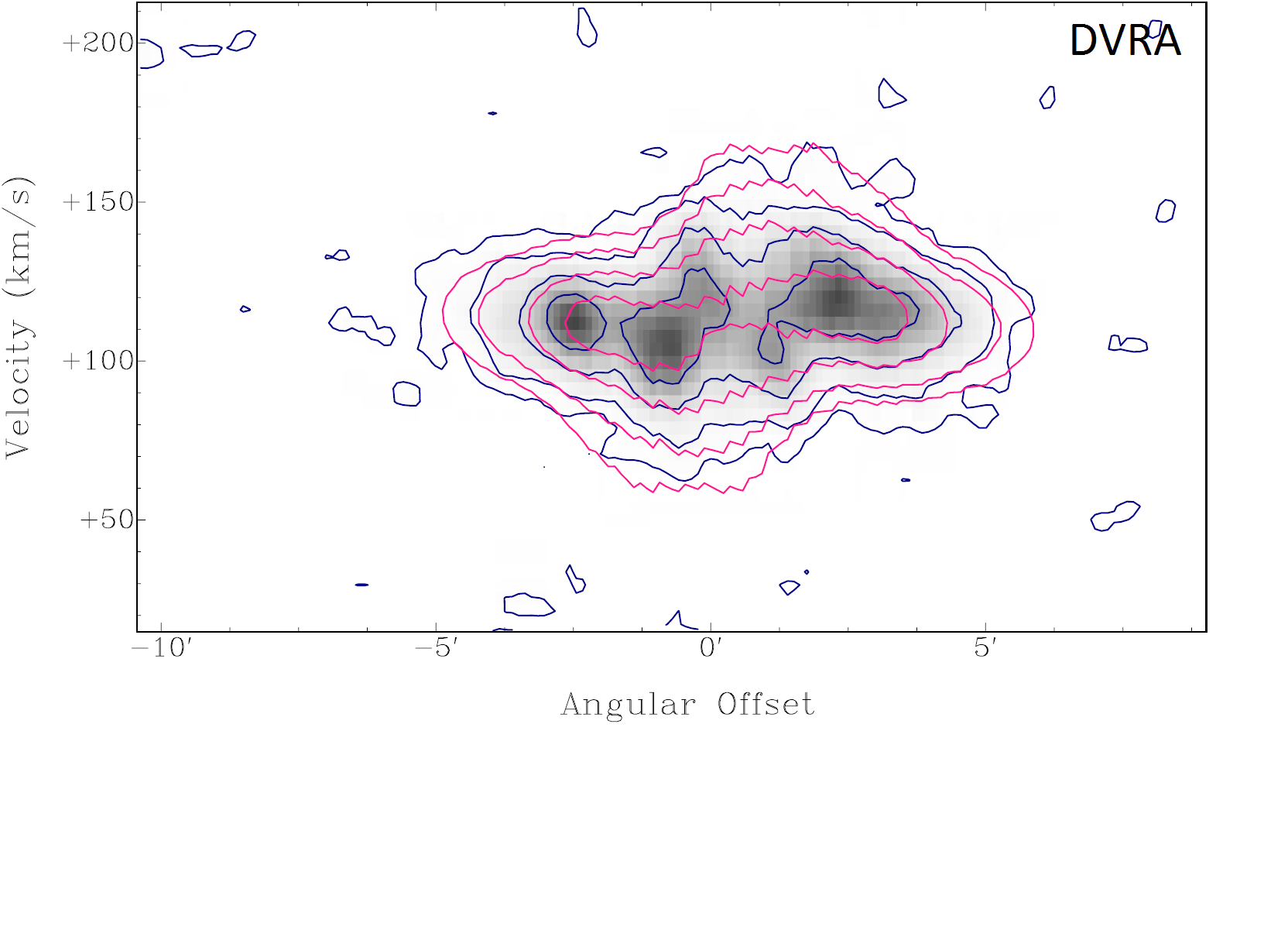}
\includegraphics[angle=0,scale=0.21,bb=0 0 612 792,clip=true,trim=0pt 290pt 75pt 0pt]{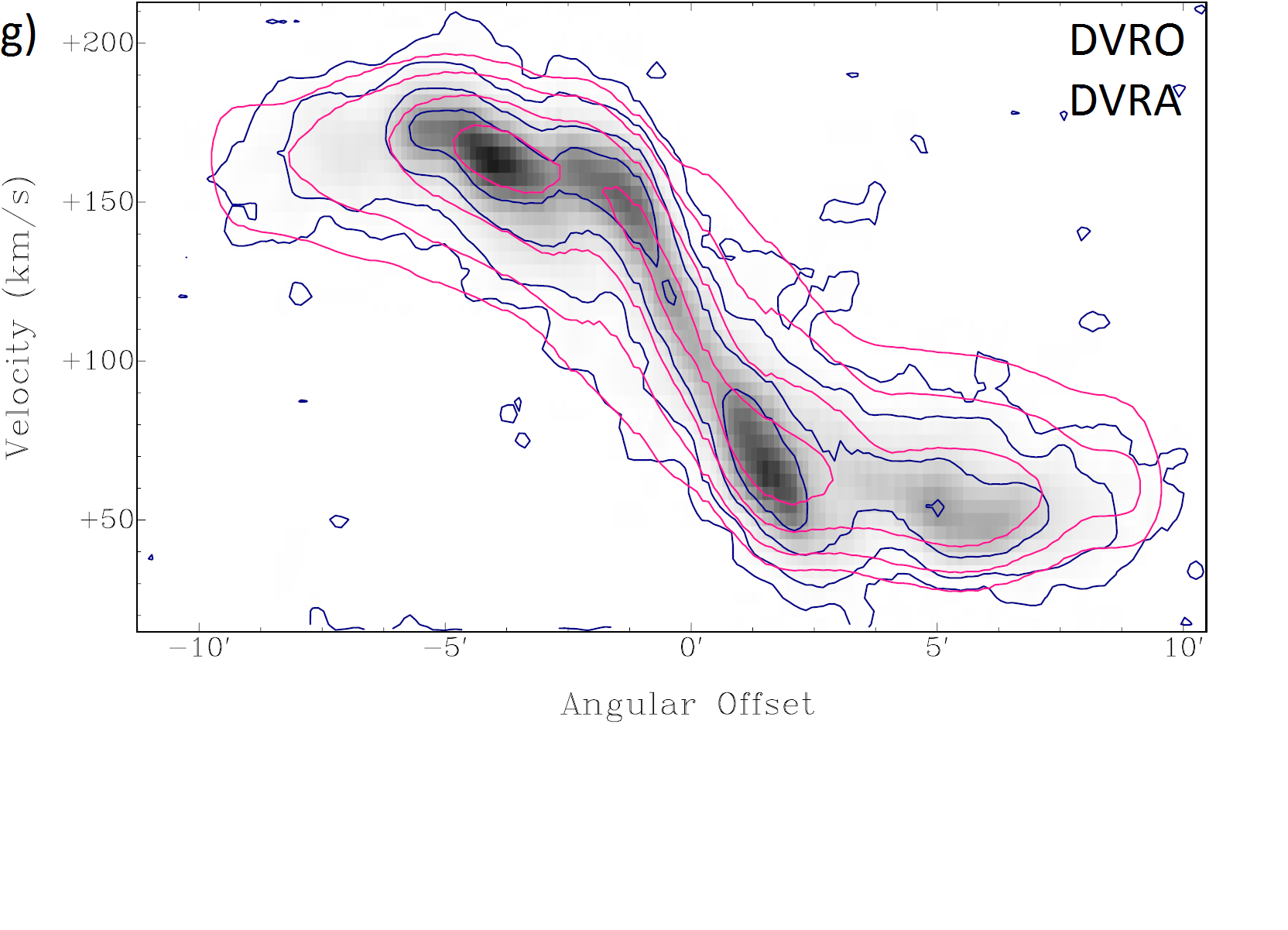}\includegraphics[angle=0,scale=0.21,bb=0 0 612 792,clip=true,trim=133pt 290pt 0pt 0pt]{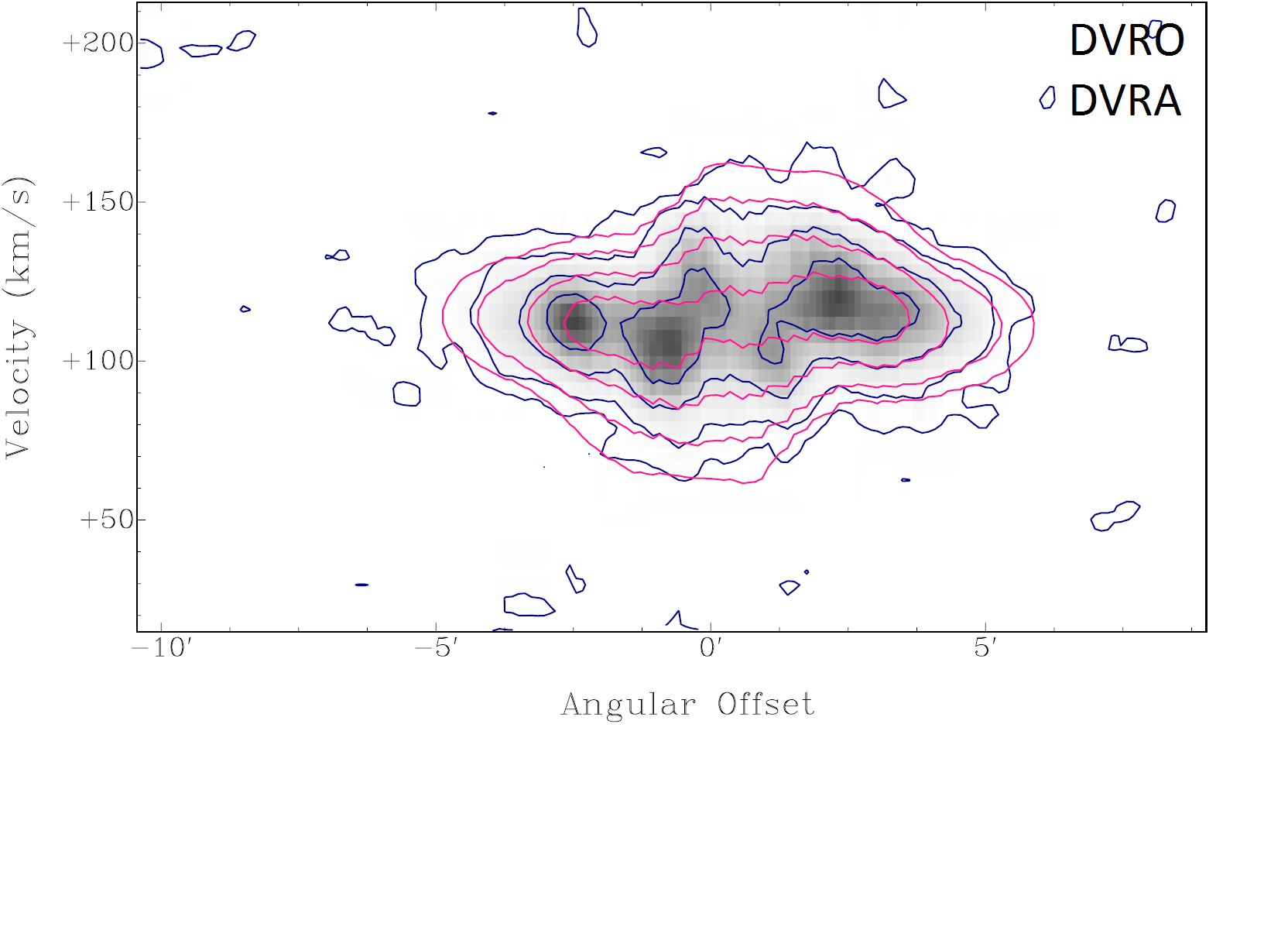}
\includegraphics[angle=0,scale=0.21,bb=0 0 612 792,clip=true,trim=0pt 150pt 75pt 0pt]{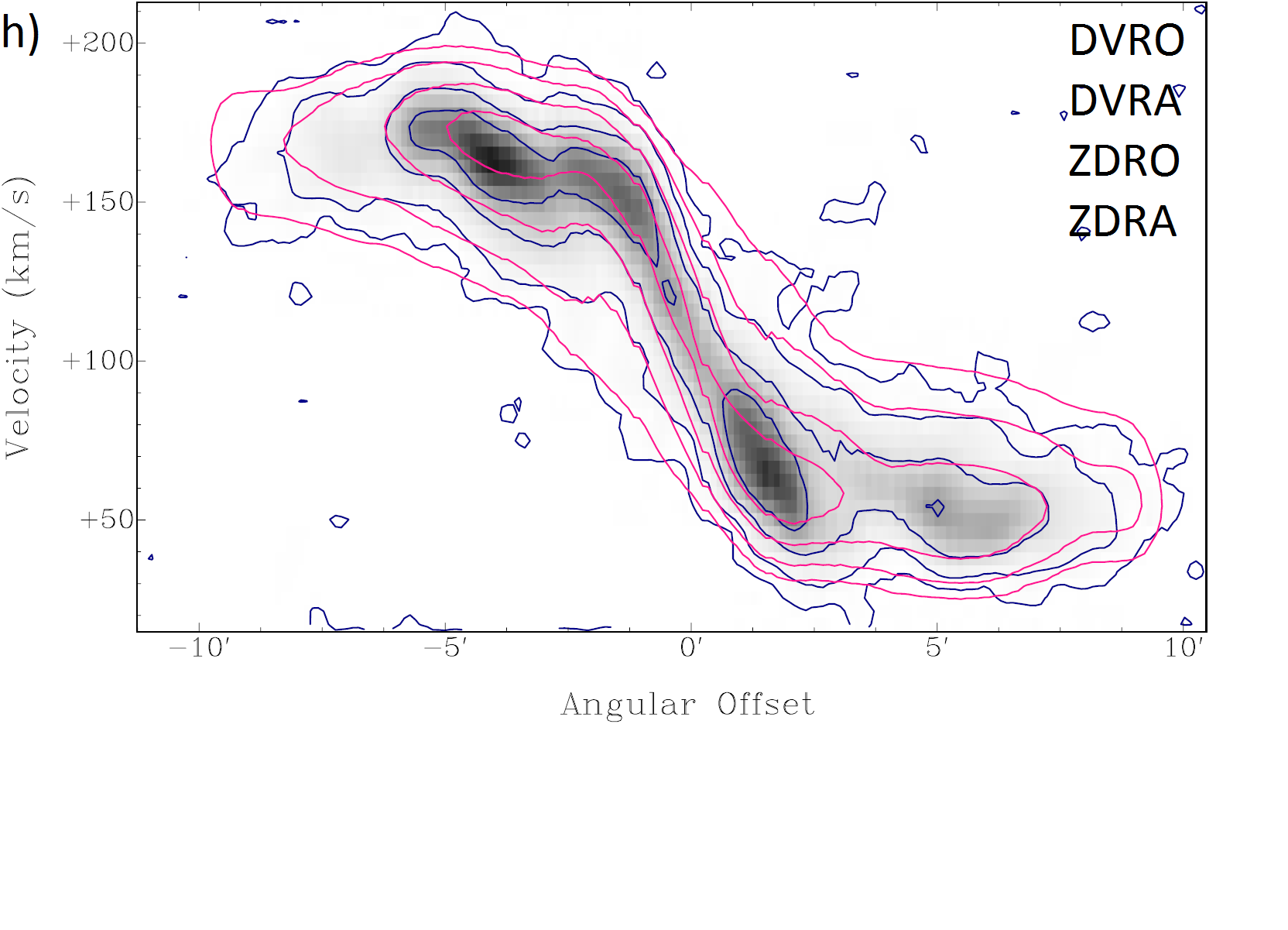}\includegraphics[angle=0,scale=0.21,bb=0 0 612 792,clip=true,trim=133pt 150pt 0pt 0pt]{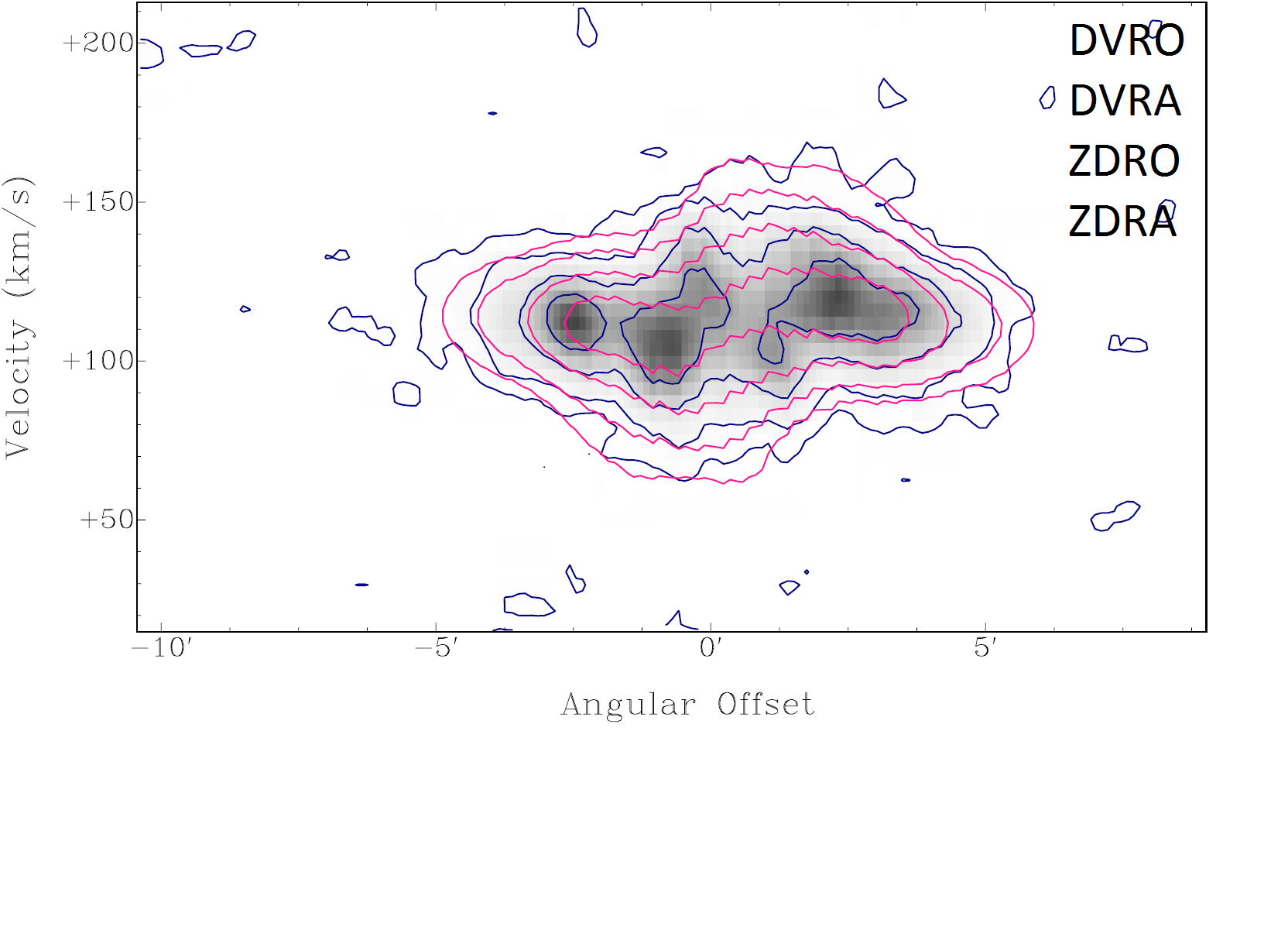} 
 \begin{flushleft}
  \bf Fig.~\ref{fig:slicemodels}  \normalfont (continued). See Sect.~\ref{sect:VRAD}-~\ref{sect:finalmod} for details. Clearly, the kinematics of UGCA\,105 are best described by the model featuring vertical gradients in both rotation and radial velocity as well as the respective onset heights (cf. panel \textit{h)}).
 \end{flushleft}
\end{figure*}

\subsection{Final model}
\label{sect:finalmod}

\begin{figure*}
 \centering
 \includegraphics[scale=0.33,clip=true,trim=28pt 195pt 90pt 122pt]{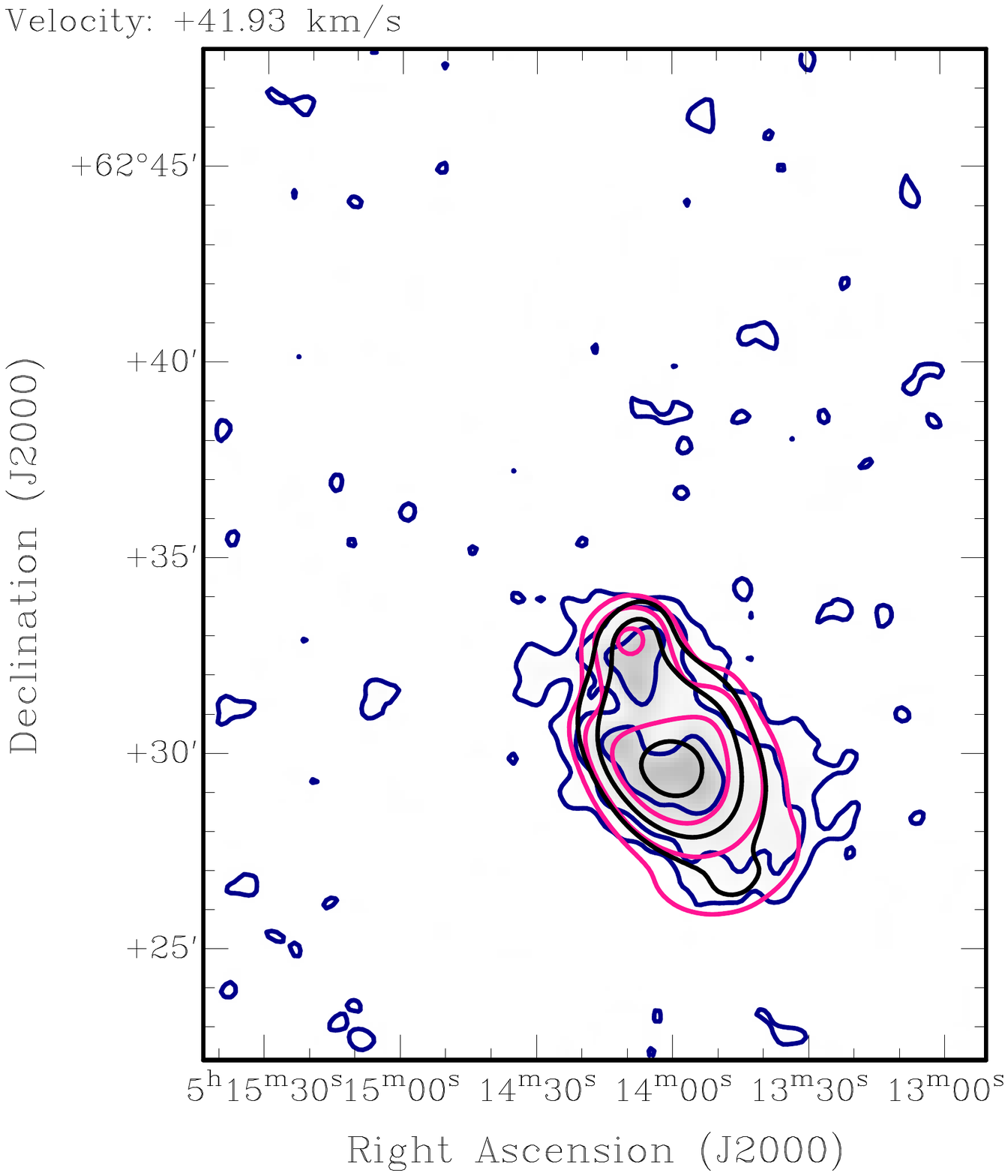}\includegraphics[scale=0.33,clip=true,trim=122pt 195pt 90pt 122pt]{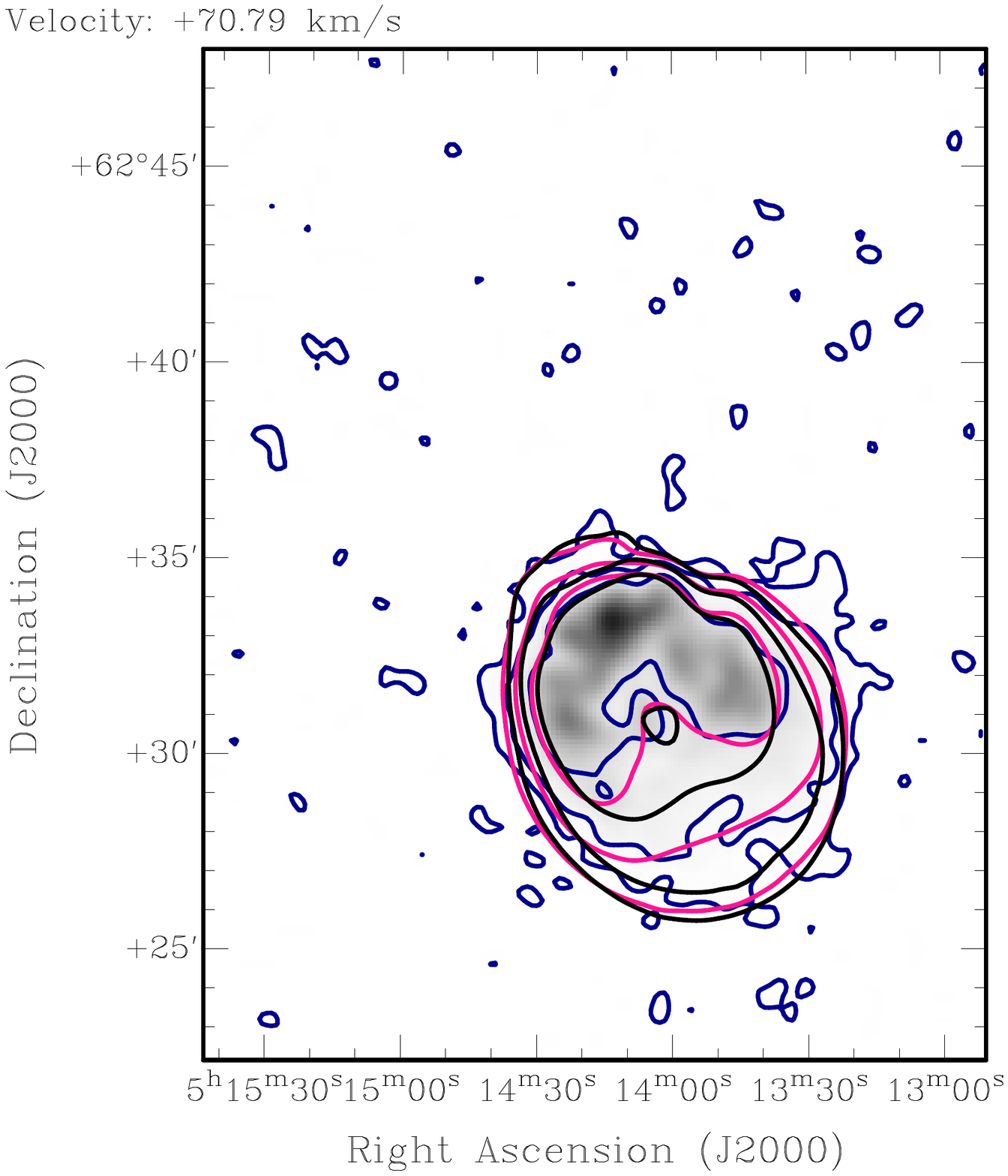}\includegraphics[scale=0.33,clip=true,trim=122pt 195pt 90pt 122pt]{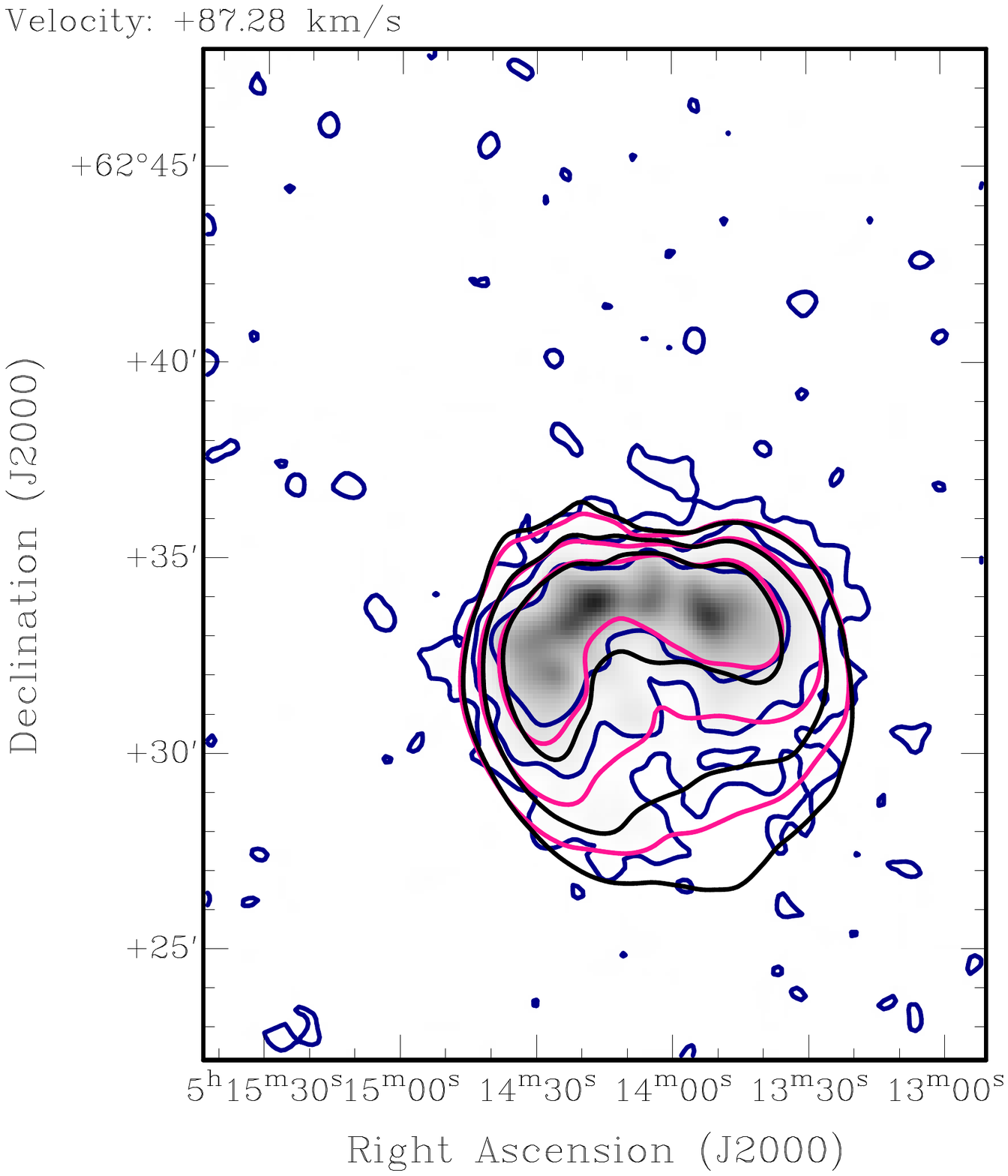}\includegraphics[scale=0.33,clip=true,trim=122pt 195pt 90pt 122pt]{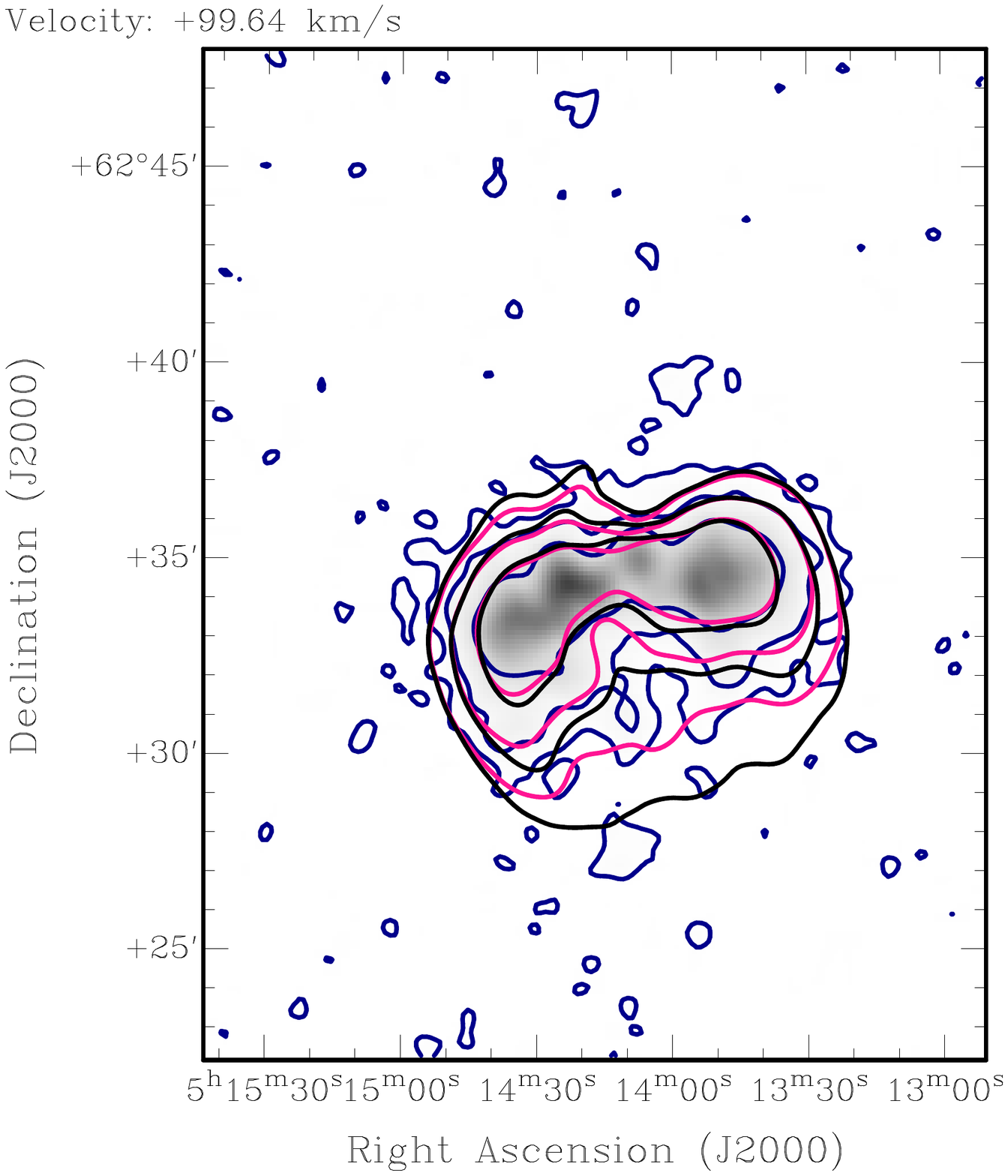}
 \includegraphics[scale=0.33,clip=true,trim=28pt 140pt 90pt 122pt]{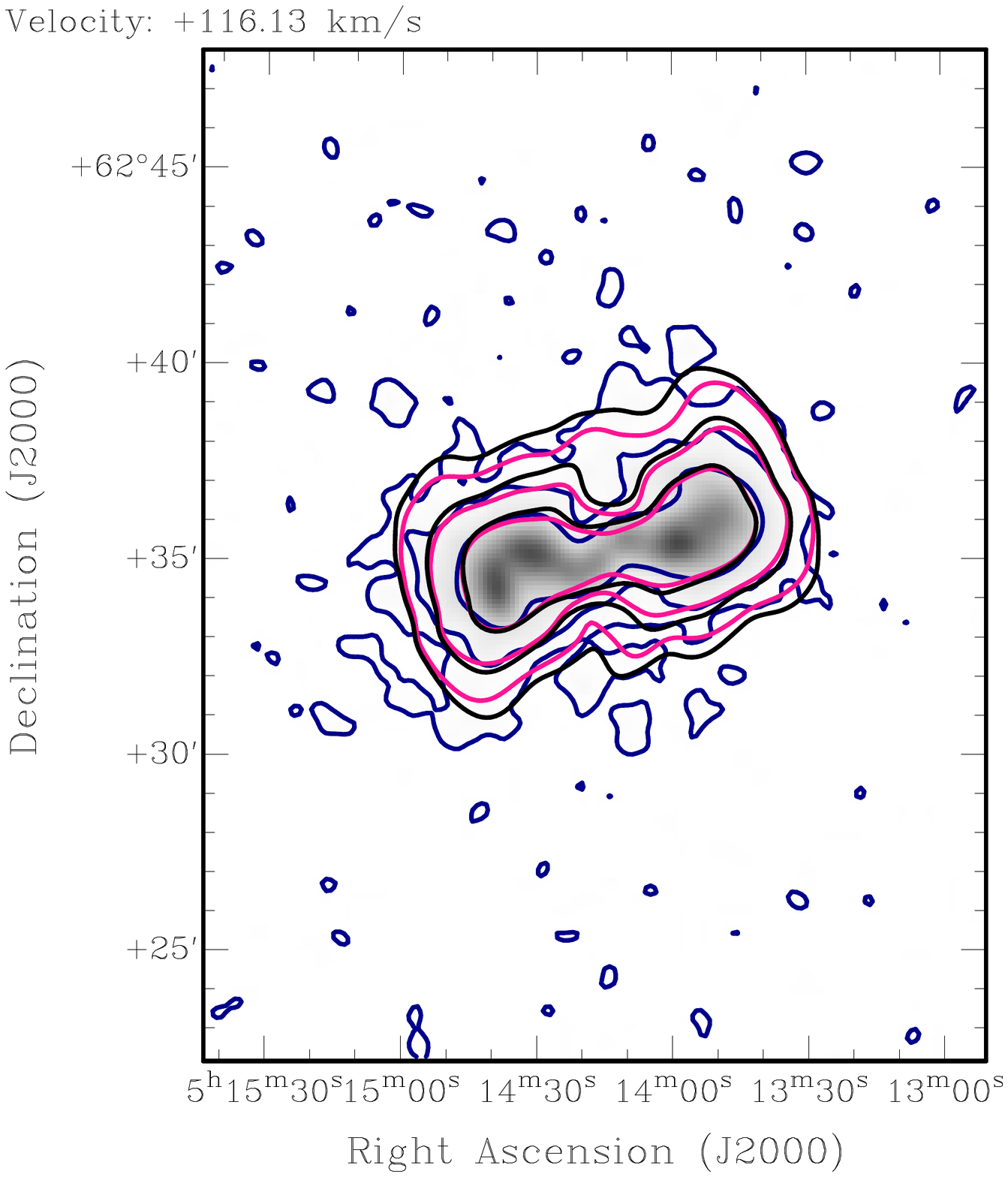}\includegraphics[scale=0.33,clip=true,trim=122pt 140pt 90pt 122pt]{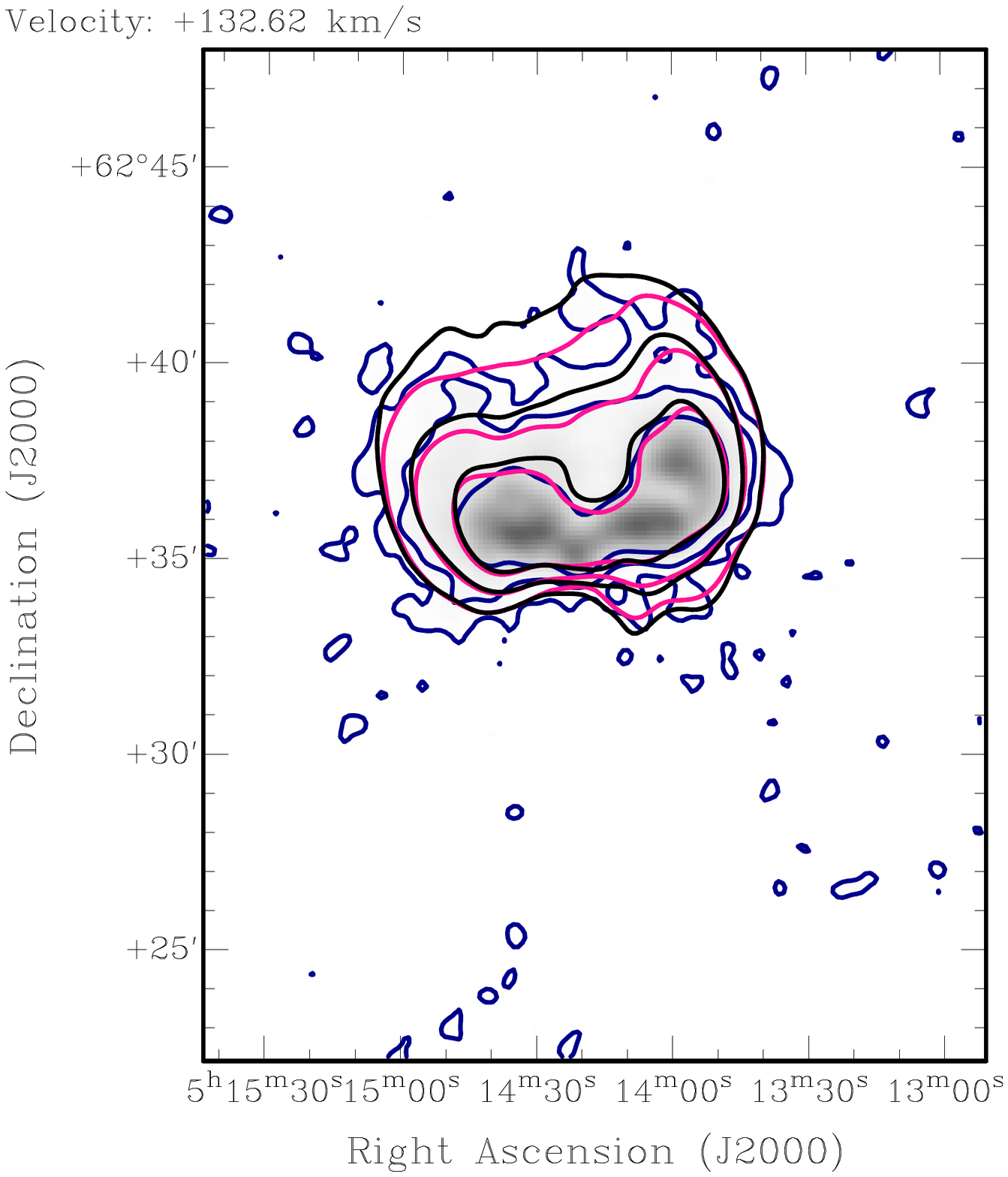}\includegraphics[scale=0.33,clip=true,trim=122pt 140pt 90pt 122pt]{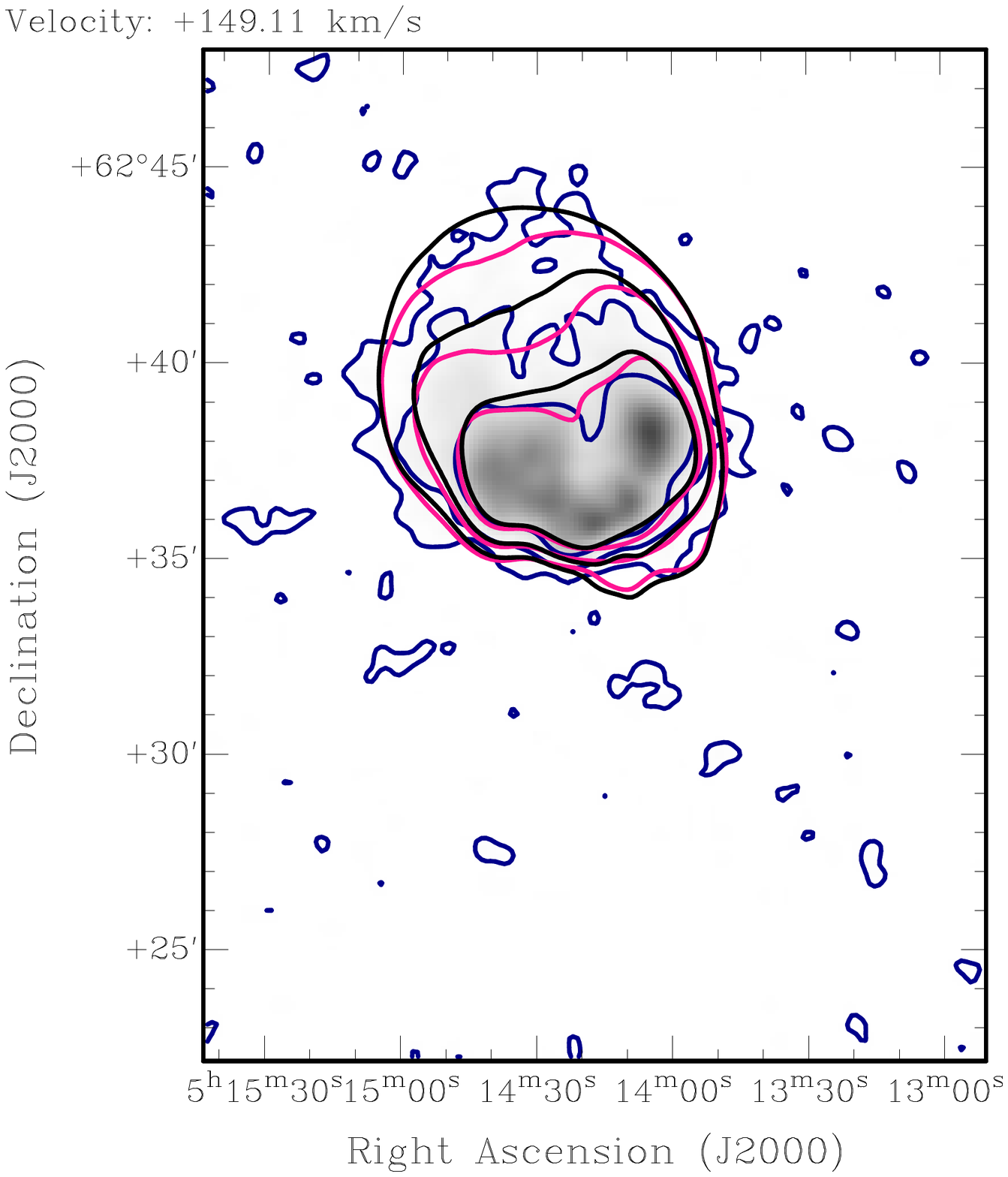}\includegraphics[scale=0.33,clip=true,trim=122pt 140pt 90pt 122pt]{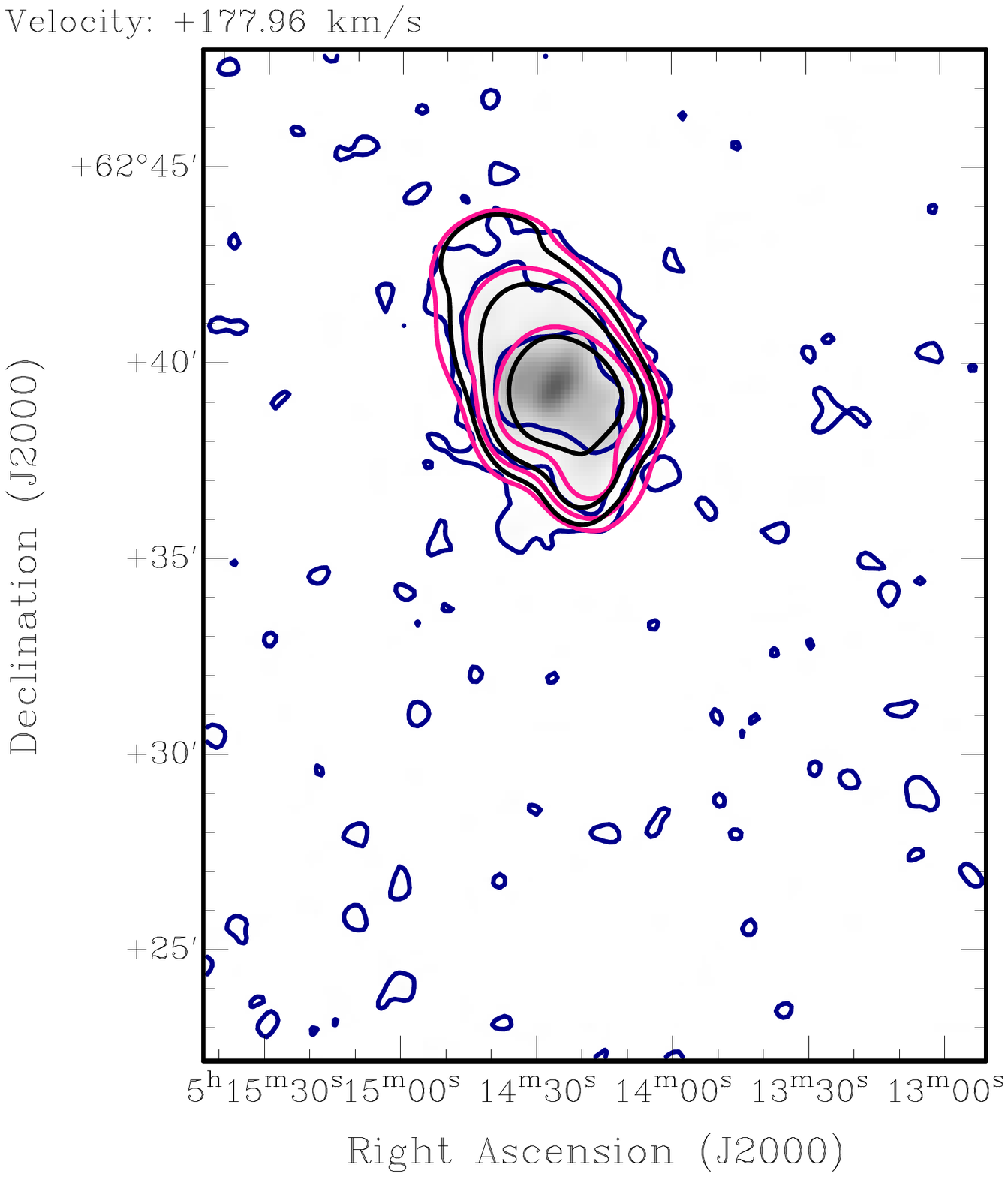}
 \caption{Selected channel maps from the 39\farcs86 $\times$ 38\farcs36 resolution data cube of UGCA\,105 (greyscale and blue contours) in comparison with the corresponding channel maps of the \texttt{TiRiFiC} model including both vertical velocity gradients DVRO, DVRA but no onset heights ZDRO, ZDRA (as shown in panel \textit{g)} of Fig.~\ref{fig:slicemodels}; black contours) and of our best-fit \texttt{TiRiFiC} model (as shown in Fig.~\ref{fig:14} and panel \textit{h)} of Fig.~\ref{fig:slicemodels}; pink contours). Contour levels are 0.9 (2\,$\sigma$), 3.6, and 14.4\,mJy\,beam$^{-1}$. The heliocentric velocity is noted at the upper left corner of each map.}
 \label{fig:modchannels}
\end{figure*}

\begin{figure}[h!]
 \centering
 \includegraphics[angle=270,scale=0.44,clip=true,trim=0pt 125pt 29pt 130pt]{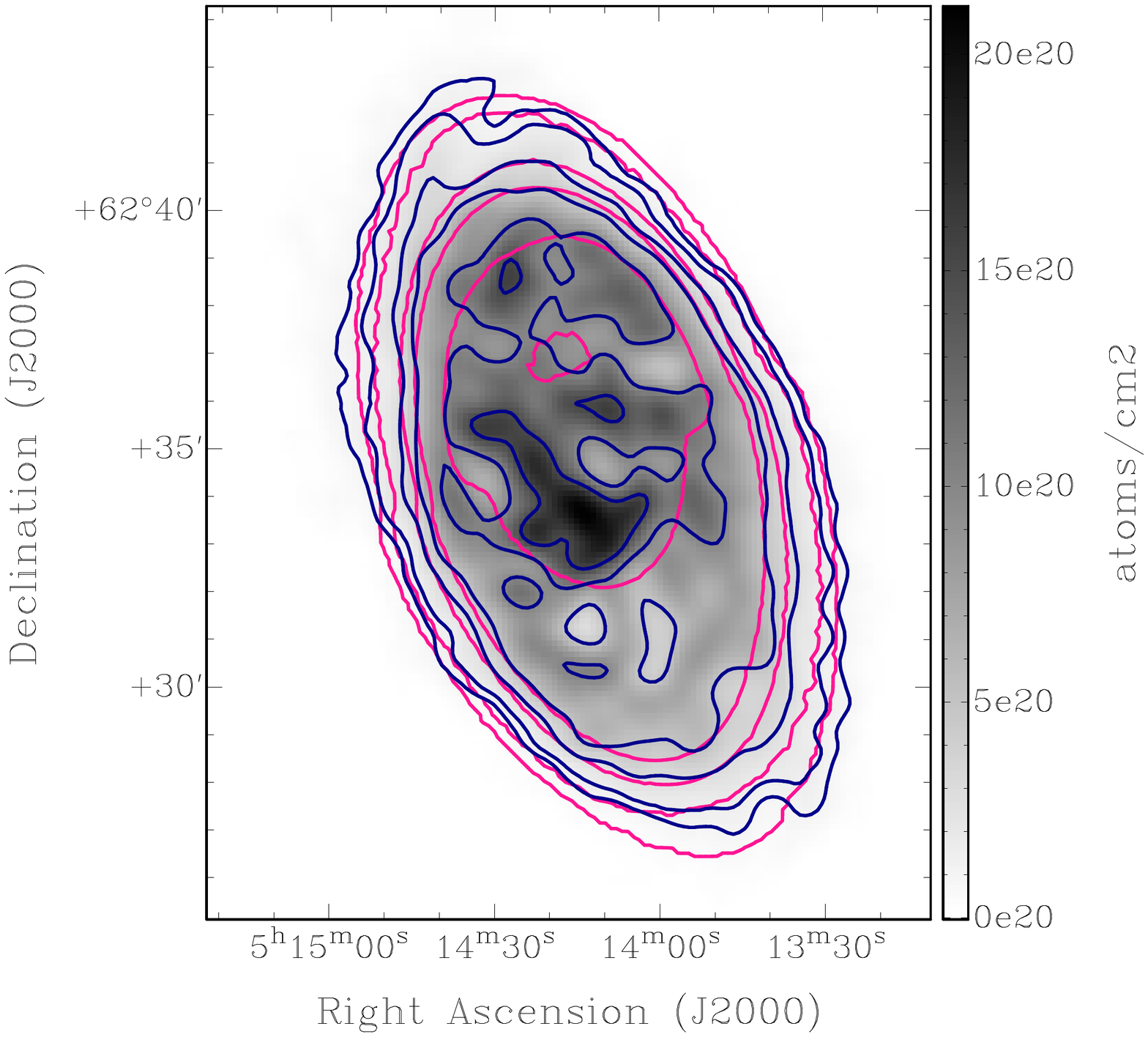}
 \caption{$\ion{H}{i}$ column density map of UGCA\,105 at 39\farcs86 $\times$ 38\farcs36 resolution (greyscale and blue contours) in comparison with the column density map of our best-fit \texttt{TiRiFiC} model (as shown in Fig.~\ref{fig:14} and panel \textit{h)} of Fig.~\ref{fig:slicemodels}; pink contours). Contour levels are 1, 1.5, 3, 5, 10, and $15\times10^{20}$ atoms\,cm$^{-2}$.}
 \label{fig:mom0compare}
\end{figure}

We found the model shown in panel \textit{h)} of Fig.~\ref{fig:slicemodels} to be our best approximation to the UGCA\,105  $\ion{H}{i}$ data cube. Figures~\ref{fig:modchannels} and \ref{fig:mom0compare} demonstrate the overall consistency of this model with the data, based on eight representative spectral channel maps and on the column density distribution, respectively. The model parameters are plotted against radius in Fig.~\ref{fig:14}.

To obtain stable fit results while keeping our model as simple as possible, all parameters were sampled with ring separations significantly larger than the beam size. Since $\chi^{2}$ minimisation did not lead to satisfactory results for all data points, a number of model values had to be set manually to achieve a good approximation to the data cube. Note that in case of the onset heights, even though the program found the minimum $\chi^{2}$ for $ZDRO=ZDRA\approx0.1$\,kpc, we found a slightly better agreement with the data from visual inspection after doubling these parameter values, and estimate the error in this parameter as the difference between the manually fixed- and the best-fit value.
 
Since the distribution of holes and filaments in the $\ion{H}{i}$ disk is largely asymmetric, we fitted separate $\ion{H}{i}$ surface density profiles to the approaching and the receding side of the disk, while such a treatment was not necessary for the other model parameters. As a notable difference to the warped model, we found that the best-fitting rotation curve now shows a somewhat higher overall amplitude (up to 83\,$\rm{km\,s}^{-1}$) and a far less pronounced decrease at the outermost radii. In addition, the new model does not require a flare as strong as in the warped model -- we obtain a maximum scale height of $\sim$0.7\,kpc for the outer disk. Note that the rotation curve, surface density profiles, orientation, and scale height are roughly the same for all of the above models that include vertical velocity gradients (i.e. panels \textit{c)} to \textit{h)} in Fig.~\ref{fig:slicemodels}).

In \texttt{TiRiFiC}, negative gradients correspond to decreasing velocities with increasing height above the disk plane. Assuming that the spiral structure in the outer disk is trailing, we were able to assign the direction of the radial motion. This means that we observe gas above the inner part of the disk with a lag in circular velocity relative to the bulk motion as well as an inward velocity component increasing with $z$. The data are consistent with a negligible vertical velocity component or its gradient ($DVVE$), which, for completeness, we fitted to the inner disk, and (using the same onset height above the plane) derived a gradient of $DVVE\approx-7\,\rm{km\,s}^{-1}\,\rm{kpc}^{-1}$ for $r<2.0\,\rm kpc$.

A feature that is not well reproduced by our models is the wiggle in the high-column-density contours on both sides of the minor-axis PV slice (marked by the black ellipses in panel \textit{a)} of Fig.~\ref{fig:slicemodels}). Since we managed to at least marginally fit this structure in the two barred models, we do not rule out the possibility that it is caused by bar streaming motions; however, the feature might as well be due to local inhomogeneities of the $\ion{H}{i}$ distribution (e.g. the central hole and its surroundings). Moreover, if there is indeed a bar flow that could affect the determination of the inner rotation curve, we do not expect the necessary correction to be larger than the statistical uncertainties of these data points.

Similarly, models and data do not perfectly agree in terms of the inner contours of the major-axis PV diagram, again primarily because of local inhomogeneities, such as the remarkable $\ion{H}{i}$ depression in the approaching half (green circle in panel \textit{a)}). We also would like to point out again that our \texttt{TiRiFiC} models are symmetric (except for the radial surface density profile), and hence it is expected that they do not match the approaching and receding side of the data cube equally well (as is also evident from Fig.~\ref{fig:modchannels}). The differences between the two sides are taken into account by accordingly assigning errors to the model parameters (see Sect.~\ref{sect:errors}).{We point out again that we can take into account most of an apparent global lopsidedness by assuming different surface brightness for the receding and approaching half of the galaxy, such that on a global scale, we believe to be able to explain an apparent lopsidedness by variations in surface 
densities, not the kinematics, and that hence our major conclusions are not much affected by these effects.}

As emphasised above, we cannot claim to have found a unique kinematic model solution. In particular, having found a satisfactory parametrisation using a single, coherent disk model, we {did not} {investigate} the {possibility of} a two-disk structure to the \ion{H}{i}, which is a frequently used approach in the literature to quantify the structure of anomalous gas (e.g. \citealt{Swaters97,Gentile13}). This might offer a possibility to reproduce our data equally well, but we do not expect that such a solution would change the basic picture. In summary, our kinematic \ion{H}{i} modelling indicates that (see also Fig.~\ref{fig:14})
\begin{itemize}
\item{While UGCA~105 possesses a warp, it is not sufficient to explain the appearance of \ion{H}{i} at anomalous velocities in projection against the inner disk of the galaxy.}
\item{Bar streaming provides a better fit to the data, but is not sufficient to reproduce the kinematic structure of the extraplanar \ion{H}{i}.}
\item{An extended disk with gas that is rotating slower above the mid-plane than the rotation speed in the central plane is required to reproduce the data. The effect is strongest in the range of the optical disk.}
\item{Approximately inside the optical radius $r_{25}$, inwards motion is required to reproduce the data.}
\item{The mid-plane does not exhibit inwards motion, while it becomes significant at some height above the mid-plane.}
\item{A flare is present. The scale height of the disk increases beyond the optical radius of the galaxy.}
\item{The rotation curve of UGCA~105 is flat {within the errors}.}
\end{itemize}

\begin{figure}
 \centering
 \includegraphics[clip=true,scale=0.46,page=2,trim=10pt 140pt 20pt 86pt]{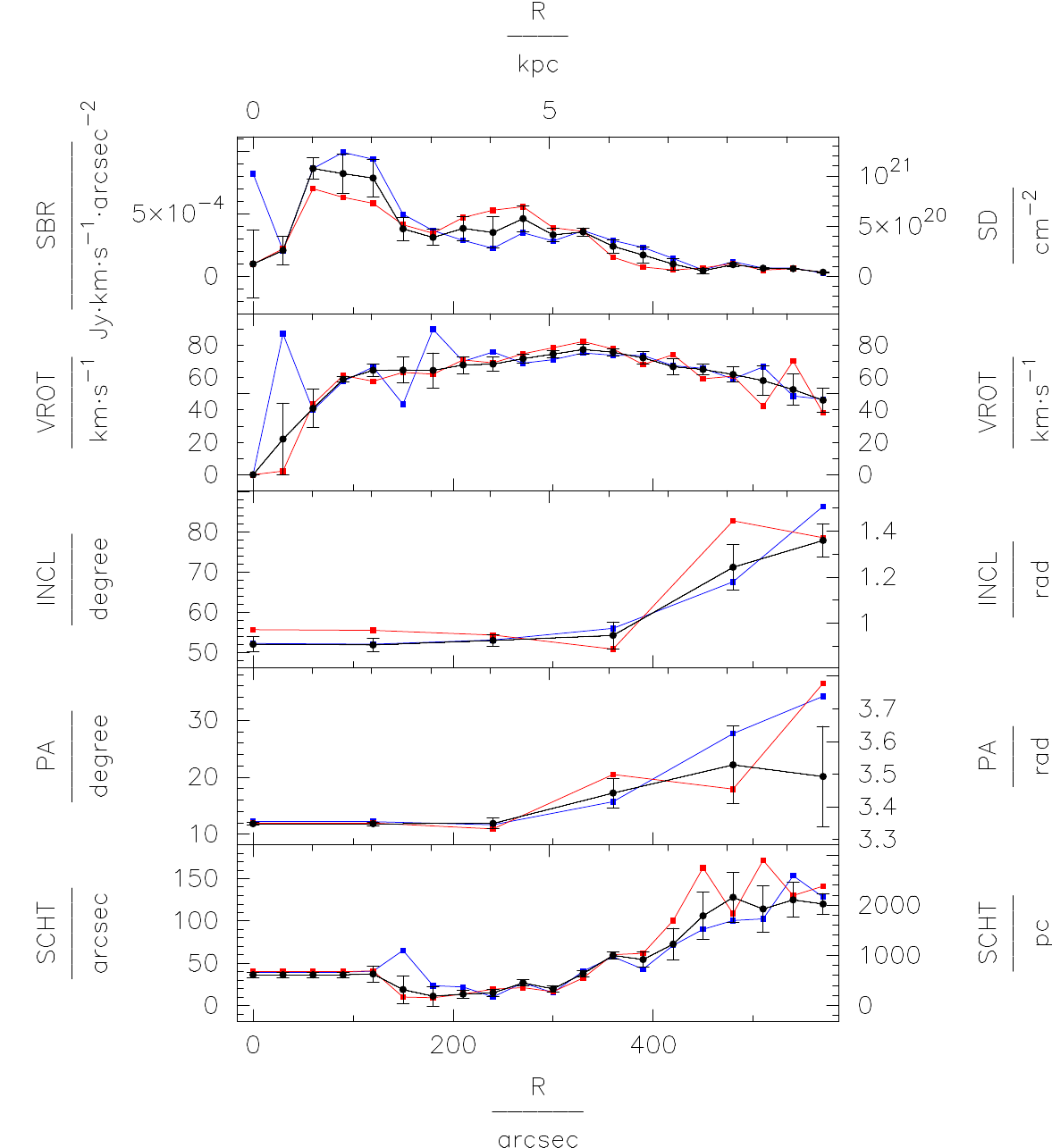}
 \caption{Best-fit solution for UGCA\,105 obtained with \texttt{TiRiFiC} when no bar and no velocity gradients are considered, i.e. the warped model described in Sect.~\ref{sect:warp} and shown in panel \textit{a)} of Fig.~\ref{fig:slicemodels}. Fit parameters are the surface brightness or -density (SBR/SD), rotation velocity (VROT), inclination (INCL), position angle (PA), and scale height (SCHT). Black dots and lines represent the fit to the entire velocity field or data cube, while the blue (red) squares and lines denote the solution for the case in which only the approaching (receding) side was fitted.}
\label{fig:13}
\end{figure}

\subsection{Error determination}
\label{sect:errors}
\begin{figure}
 \centering
 \includegraphics[clip=true,scale=0.46,page=2,trim=10pt 205pt 20pt 86pt]{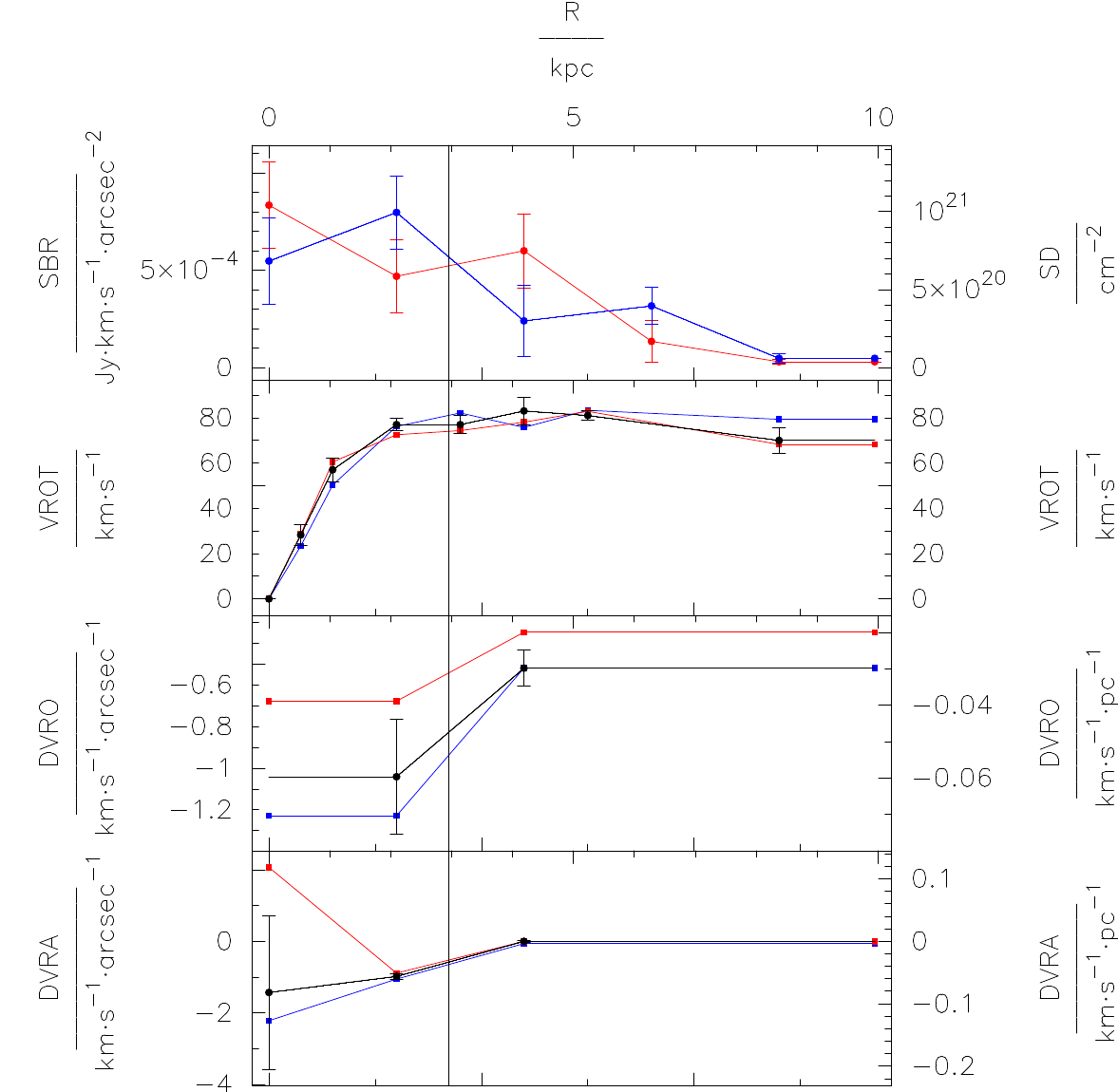}
 \includegraphics[clip=true,scale=0.46,page=2,trim=10pt 140pt 20pt 160pt]{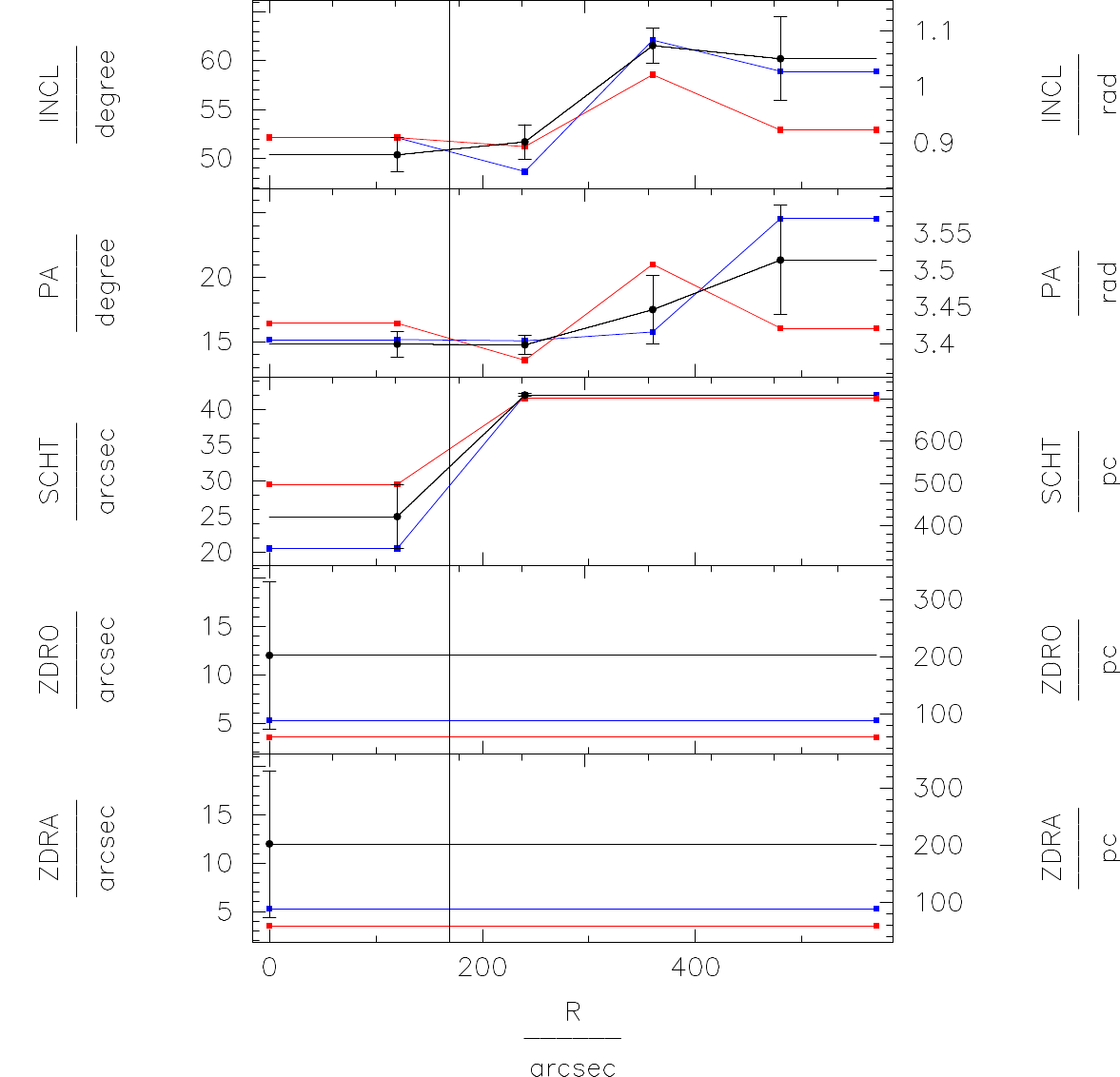}
 \caption{Parametrisation of our best-fit \texttt{TiRiFiC} model for UGCA\,105 (cf. panel \textit{h)} of Fig.~\ref{fig:slicemodels}): surface brightness or -density (SBR/SD), rotation velocity (VROT), vertical gradient of rotation velocity (DVRO), vertical gradient of radial velocity (DVRA), inclination (INCL), position angle (PA), scale height (SCHT), onset height for vertical gradient of rotation velocity (ZDRO), onset height for vertical gradient of radial velocity (ZDRA). The vertical line in each panel is drawn at the optical radius $R_{25}=2.9$\,kpc. Colours have the same meaning as in Fig.~\ref{fig:13}.}
\label{fig:14}
\end{figure}

Since a reliable technique for error determination in \texttt{TiRiFiC} is yet to be developed, {we estimated the uncertainties} by fitting the approaching and receding part of the galaxy separately. As the routine calculates azimuthal averages for each ring, it is reasonable to assume error margins to be represented by deviations from radial symmetry. The error bars we adopted are given for each data point by the average of the deviations (i.e. from the approaching and receding side, respectively). To prevent the error bars from becoming unrealistically small or large, Hanning smoothing was applied to the errors for every fit parameter, that is, we weighted each error by a factor of one half, and added one quarter of the errors of the two adjacent data points. In Figs.~\ref{fig:13} and \ref{fig:14}, the modelling results for UGCA\,105 are plotted along with the respective solutions for the approaching and receding side for comparison.



\section{Extraplanar gas}
\label{sect:extra} 
We consider the implications of our specific kinematical model. While depending on the implementation of a tilted-ring model (e.g. as a one- or two-disk model), the derived numbers might change by some substantial factor, we expect the global finding{s, as highlighted at the end of Sect.~\ref{sect:finalmod},} to {qualitatively remain the same. However, despite an extensive search in parameter space, we cannot completely exclude that an alternative model with different physical implications can fit the data equally well or better, while a variety of alternative model families have been shown to fail in satisfactorily fitting the data.}

As shown in Sect.~\ref{sect:TRM} (see also Fig.~\ref{fig:slicemodels}), a tilted-ring model that can reproduce the major features of the data at least requires a substantial vertical gradient in rotation velocity ($-60\,\rm km\,s^{-1}\,kpc^{-1}$) and an inwards velocity component within the radius of the optical disk, increasing with height above the plane (with a gradient of $-70\,\rm km\,s^{-1}\,kpc^{-1}$). The onset of both the decrease in rotation velocity and the increase in radial velocity starts at some height above the plane ($ZDRA = ZDRO = 12^{\prime\prime}\,\widehat{=}\, 0.2\,{\rm kpc}$). Below that height, the gas is co-rotating with the mid-plane and shows no radial velocity component.

\begin{figure}
 \centering
 \includegraphics[clip=true,scale=0.48,page=1]{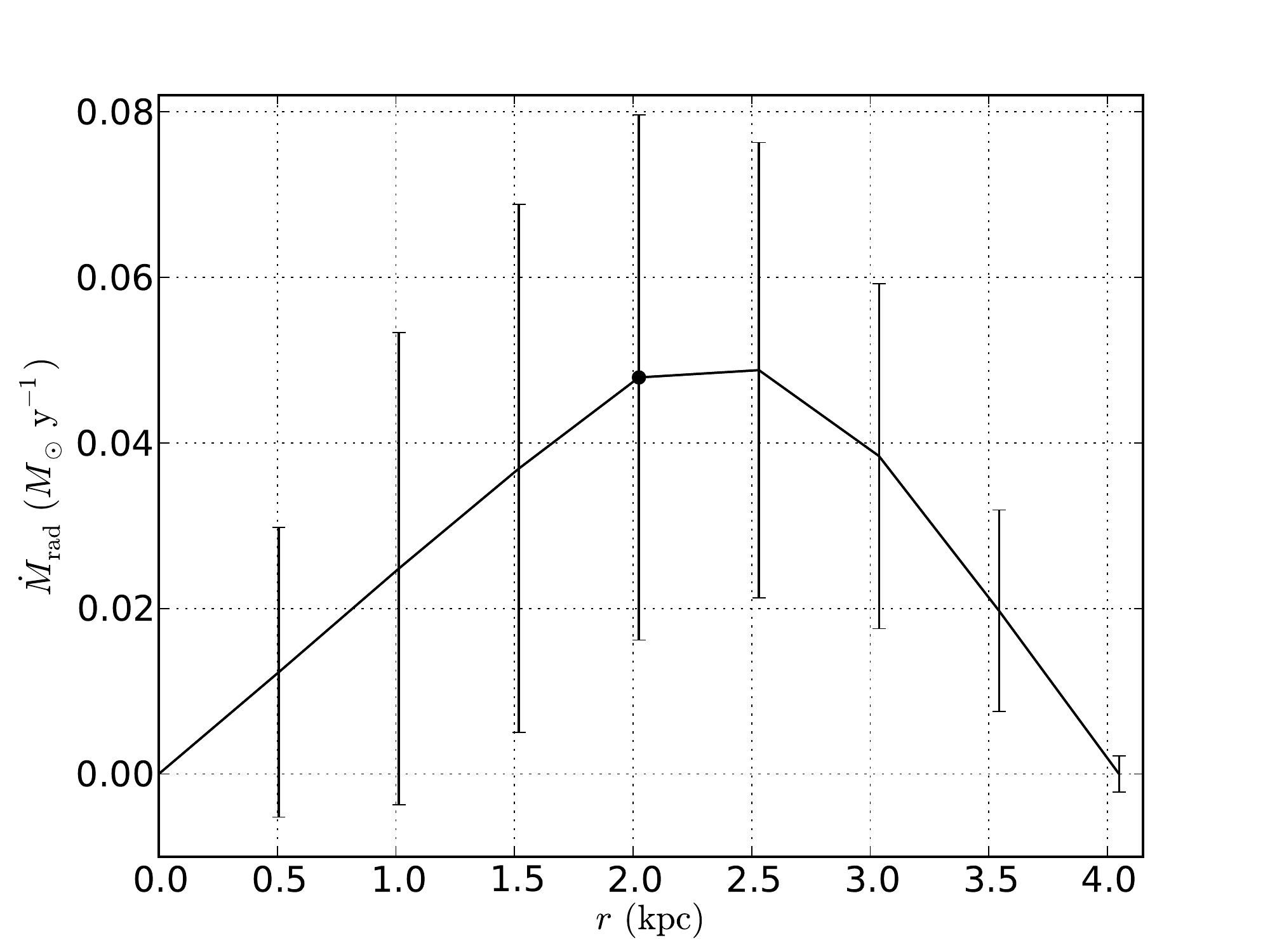}
 \caption{UGCA~105: radial $\ion{H}{i}$ inflow rate as derived from the tilted-ring model. The errors are formal errors. The most significant data point at a radius of $2.0\,{\rm kpc}$ is highlighted (see text).}
\label{fig:j1}
\end{figure}

The most striking feature of the model is the inwards motion above the plane, which basically appears to be inconsistent with a steady-state assumption for UGCA~105. This gas must be the result of an extreme fountain flow or is newly accreted gas. We calculated the radial inflow rate at a certain radius as the flow through a cylinder at radius r

\begin{equation}
\label{eq:j1}
 \dot{M}_{\rm rad} = 2\int\limits_{z_{\rm i}}^\infty \,2\pi r\,\varv_{\rm rad}(z)\,\sigma(r)\,h(z)\,\rm{d}z\;,
\end{equation}
where $z$ denotes the height above the plane, $\varv_{\rm rad}$ the radial velocity, $\sigma$ the surface density, $h$ the vertical profile ($sech^2$), and $z_{\rm i}$ the onset height above the plane. For $\varv_{\rm rad}$ we used
\begin{equation}
\label{eq:j2}
 \varv_{\rm rad} = 
  \left\{
  \begin{array}{l l}
    \Delta\varv_{\rm rad}(z-z_i) & \quad z \geq z_{\rm i}\\
                     0 & \quad z < z_{\rm i}
  \end{array}
\qquad
\right.
\end{equation}
according to our kinematical model, where $\Delta\varv_{\rm rad}$ is the vertical gradient in radial velocity. In observable units as used by TiRiFiC, we derived
\begin{equation}
\label{eq:j3}
 \dot{M}_{\rm rad}(r) = 6.3\cdot 10^{-2} \Delta\varv_{\rm rad}\,\Sigma\,r\,z_0\left[ln(2)+ln\left[cosh\left(z_i\,z_0^{-1}\right)\right]-z_i\,z_0^{-1}\right]\; {\rm ,}
\end{equation}
where $\dot{M}_{\rm rad}(r)$ is measured in $M_\odot {\rm yr^{-1}}$, $\Delta\varv_{\rm rad}$ in $\rm km \, s^{-1}\, arcsec^{-1}$, the {azimuthally averaged} surface brightness $\Sigma$ in $\rm Jy\, km \, s^{-1}\, arcsec^{-2}$, and $r$, $z_0$, and $z_{\rm i}$ in arcsec. We show the results as calculated over the radial range within which we include a radial inflow in Fig.~\ref{fig:j1}. Because we used a model that is sampled on a coarse grid (which reduces the number of fit parameters), we identified the point at $r = 2.0\,\rm kpc$ as the most reliable one (radial motion is only fitted at r = 0, 2.0, and 4.0 kpc, see Fig.~\ref{fig:14}), and derived an inflow rate of $\dot{M}_{\rm rad} = 0.05 \pm 0.03\,\rm M_\odot \rm yr^{-1}$. Assuming an onset height $z_{\rm i} = 0$ would yield $\dot{M}_{\rm rad} = 0.1 \pm 0.06\,\rm M_\odot \rm yr^{-1}$, which is a substantial increase. We therefore point out that while introducing this onset for our model improves our fit, its choice has also a significant impact 
on the derived physical properties. Correcting 
our 
preferred inflow rate for helium abundance by multiplying with a factor 1.3, we derived a total cold gas inflow rate of $\dot{M}_{\rm rad, total} = 0.06 \pm 0.05\,\rm M_\odot \rm yr^{-1}$. 

The star formation rate of UGCA\,105 as derived from UV- and $\ion{H}{\alpha}$ measurements amounts to $0.07 \, \rm M_\odot {\rm yr^{-1}}$ (UV) and $0.05 \, \rm M_\odot {\rm yr^{-1}}$ ($\ion{H}{\alpha}$), respectively \citep{Lee09}. This means that if we interpret the radial inflow, which in fact is modelled only near the star-forming disk, as gas accretion, the star formation in UGCA~105 can roughly be maintained by that inflow, even neglecting direct stellar feedback to the ISM. Even though the derived gradients appear to be extreme, the radial inflow rate is hence modest and in principle compatible with the star formation properties of UGCA~105. {We modelled the inflow assuming a symmetric surface density. Indeed, due to projection effects, a connection of the inflow to specific areas of star formation cannot be made on the basis of our data.}

{We point out that the gas fraction that starts to counterrotate (because at a certain height above the plane the sign of the rotation reverses in our simple model) is negligibly small according to our purely kinematical model (0.4 percent of the total gas mass). Comparing the inflow velocity at the same height above the plane, it is also similar to the asymptotic rotation velocity of UGCA~105. We therefore conclude that our description does not contradict simple physical considerations.}

The total mass of the anomalous gas component cannot be determined in a straightforward manner, because we modelled the $\ion{H}{i}$ in UGCA~105 as {one} continuous structure, in contrast to kinematical models used to explain the structure of NGC~2403 \citep{Fraternali02} or NGC~891 \citep{Oosterloo07}, where an additional thick disk accounts for the anomalous gas. In external galaxies, including NGC~891, it appears generally not straightforward to separate two \ion{H}{i} components \citep{Marinacci10}. Depending on the model used to explain the occurrence of anomalous gas, this is an expected result \citep{Fraternali06, Marinacci10}. We estimate the gas mass of the $\ion{H}{i}$ component with a rotational velocity of less than 90\% of the rotation velocity of the midplane gas to amount to $3\cdot10^8\,\rm M_\odot$, which is 40\% of the total gas mass. For NGC~891, \citet{Oosterloo07} estimated based on a kinematical model that 30\% of the total gas mass has anomalous kinematics (\citet{Fraternali06} 
estimated 
15\%), which means that we derive similar numbers{, albeit UGCA 105 would exhibit a slightly more extreme mass of its extraplanar component}. {Hence, if} our simple kinematical model does indeed reflect the physical status of the $\ion{H}{i}$ in UGCA~105, this galaxy would {contain the most massive, regular extraplanar \ion{H}{i} component described in a galaxy so far}.

Additional studies, which are beyond {the scope of} this work, are required to investigate under which circumstances we might witness the observed morphology and kinematics of UGCA~105. In particular, it remains to be seen whether the observed kinematics is compatible with a closed-box scenario, in which the observed anomalous component constitutes gas that was formerly expelled from the galaxy via star formation and now returns to the galaxy. \citet{Fraternali08, Fraternali06} argued for NGC~2403, the only other galaxy which to our knowledge has been observed to exhibit an inwards-flowing extraplanar gas component, that an inwards fountain flow is not expected, and therefore the observations hint towards a mixing of low-angular momentum gas{, external to the main disk or the fountain flow,} with the fountain gas. Within the framework of their ballistic model, 
Fraternali \& Binney found a possibility to generate a net inwards motion by taking into account a phase change and letting the fountain material appear at later times in their excursion from the plane. In that case, a net inwards motion can be observed without adding external material to the fountain flow. However, for their specific model, the signatures of inwards motion appear only at a weaker level than observed, and only in conjunction with a strong vertical velocity component that is not observed for NGC~2403. For UGCA~105 as well, a vertical velocity component is at most at a rather low level compared to the radial motion. \cite{Fraternali08} found for NGC~2403 that the required inward motion can be achieved by adding a small fraction of gas at low angular momentum. \cite{Marinacci11} {showed that for a Milky-Way-type galaxy this low-angular momentum material might be provided by a slowly rotating hot corona that slows down the fountain flow and condenses in the wake of the fountain 
flow} \citep{Marinacci10}.

We note that considered naively, our findings would be compatible with such a hypothesis. Apparently, at a large height above the plane, rotation becomes negligible, but a strong inwards gas motion is present at a radius where star formation takes place. Assuming this to be external accreting gas, it is expected that it mixes with the {cold} ISM and, moving further toward the disk, will be dragged along with the main {cold,} gaseous body of the galaxy. A possible vertical drizzle (which cannot be excluded at a low level) may then feed the star formation at the centre of the galaxy. {A caveat against indirect accretion (first onto the corona and from there onto the cold disk) comes from the mass of the galaxy. At low masses, accretion is expected to occur in a cold mode and not by shock-heating and accretion onto a hot corona} \citep[e.g. ][]{Birnboim03,Keres05}, {such that the mechanism originally proposed by} \citet{Fraternali08}, {the mixing of directly accreted, cold gas from the IGM,
 might be preferable for this galaxy.}

{Hence,} a direct comparison with the case study of NGC~2403 \citep{Fraternali08, Fraternali06} suggests that there is some chance that we witness a (very regular) gas accretion event for the dwarf galaxy UGCA~105. This hint can only be substantiated by performing dedicated simulations tailored to the case of UGCA~105, which is beyond the scope of this work.

{It is remarkable that we found kinematically anomalous gas in a galaxy that appears to be quiescent in the optical, the star formation rate of the galaxy being low. One of the distinguishing properties of UGCA 105 -- if any -- is its large \ion{H}{i} mass with respect to its optical luminosity} ($M_{\rm \ion{H}{i}}L_{\rm B}^{-1} = 5.5 \rm M_\odot\,L_\odot$), {which is rarely found in other galaxies. The other dwarf galaxy that has recently been analysed without a clear sign of anomalous gas, UGC~1281,} \citep[][ see introduction]{Kamphuis11b} has $M_{\rm \ion{H}{i}}L_{\rm B}^{-1} = 0.15 \rm M_\odot\,L_\odot$. {One might therefore be inclined to speculate whether a high \ion{H}{i} mass might be one of the prerequisites to find anomalous gas in galaxies.}

\section{Summary}

We have presented the analysis of WSRT $\ion{H}{i}$ observations of the nearby dwarf galaxy UGCA\,105, which was chosen from a sample of galaxies selected to have extended and massive $\ion{H}{i}$ disks with respect to their stellar counterparts. The data cubes and total intensity maps showed that the structure of its $\ion{H}{i}$ disk is dominated by low column density holes and high density clumps and filaments. The velocity field of UGCA\,105 was shown to be more symmetric than expected for an irregular galaxy, or arguably even for a very late-type spiral. Differential rotation sets in at a relatively small galacto-centric distance. The galaxy also shows clear indications of the presence of anomalous gas.

Applying the tilted-ring modelling routine \texttt{TiRiFiC}, we derived the rotation curve of the galaxy from an $\ion{H}{i}$ data cube, along with orientational parameters of the disk, its scale height, and its surface density profile. 
Model cubes showed that UGCA\,105 has a flat rotation curve, a slightly warped, albeit very diffuse, outer $\ion{H}{i}$ disk, as well as an $\ion{H}{i}$ flare outside of the optical radius.
Most notably, {searching a wide range of parameter combinations} to reproduce the morphological and kinematical structure of this galaxy, {we found that a basic tilted-ring model is not sufficient.} {We were able to reflect the major kinematical features of the galaxy by} {introducing} gradients in rotation ($60\,\rm km\,s^{-1}\,kpc^{-1}$) and radial velocity ($70\,\rm km\,s^{-1}\,kpc^{-1}$), {the latter} within the range of the stellar disk, to reproduce the data with the tilted-ring model. According to our model, UGCA~105 has a strongly lagging extraplanar $\ion{H}{i}$ component and strong inwards motion of the gas within the radius of the stellar disk, increasing with height above the plane{. While we cannot exclude that other kinematical models may match the data equally well, we interpret these findings as an indication for an extraplanar, symmetric, inward flow of neutral gas. This, in turn,} might be indicative of smooth accretion of 
material{, either directly from the IGM or from an unseen, hot component in the galaxy}. Vertical motion is not required to reproduce the data, but would be compatible with the observations.

We derived the radial gas inflow rate of UGCA~105 to be $\dot{M}_{\rm rad, total} = 0.06 \pm 0.05\,\rm M_\odot \rm yr^{-1}$, which is of the order of the star formation rate of the galaxy. The hypothesis that UGCA~105 is undergoing an accretion event is consequently in principle compatible with the observations. {An inflow was detected only above the bright stellar disk. Whether this indicates a direct connection of the inflow and a fountain flow, which has been proposed as a mechanism for high-mass galaxies, cannot be answered within the scope of this paper. At this stage, we consider the same mechanism of sweeping up gas from the surrounding medium of the disk via fountain flows a possible mechanism. Because UGCA~105 has a relatively low mass, the surrounding material may be directly inflowing cold gas and not a hot corona. So far, the theoretical focus in the context of cold gas accretion have been high-mass galaxies.} If future studies {of fountain mechanisms in the context of gas 
accretion at lower masses} are able to physically motivate our kinematical model (i.e. to reproduce our observations), UGCA~105 may be the most extreme case of smooth gas accretion known so far, and, to the best of our knowledge, it is the third galaxy so far ({apart from NGC~2403 and the Milky Way}), where a net radial motion of extraplanar gas is detected.

\begin{acknowledgements}
  We thank T.B. Georgiev for sending us an electronic version of his article.
  We thank the referee, T. Oosterloo, for his comments to improve the paper.
  Part of this work was supported by the German
  \textit{Deut\-sche For\-schungs\-ge\-mein\-schaft, DFG\/} project
  number KL 533/11-1.\\\\
TiRiFiC is publicly available from http://www.astron.nl/\textasciitilde jozsa/tirific/ .
\end{acknowledgements}

\bibliographystyle{aa} 
\bibliography{dwarfs} 

\begin{thebibliography}{56}
\expandafter\ifx\csname natexlab\endcsname\relax\def\natexlab#1{#1}\fi

\bibitem[{{Barbieri} {et~al.}(2005){Barbieri}, {Fraternali}, {Oosterloo},
  {Bertin}, {Boomsma}, \& {Sancisi}}]{Barbieri05}
{Barbieri}, C.~V., {Fraternali}, F., {Oosterloo}, T., {et~al.} 2005, \aap, 439,
  947

\bibitem[{{Barnab{\`e}} {et~al.}(2006){Barnab{\`e}}, {Ciotti}, {Fraternali}, \&
  {Sancisi}}]{Barnabe06}
{Barnab{\`e}}, M., {Ciotti}, L., {Fraternali}, F., \& {Sancisi}, R. 2006, \aap,
  446, 61

\bibitem[{{Begum} {et~al.}(2005){Begum}, {Chengalur}, \&
  {Karachentsev}}]{begum05}
{Begum}, A., {Chengalur}, J.~N., \& {Karachentsev}, I.~D. 2005, \aap, 433, L1

\bibitem[{{Benjamin}(2002)}]{Benjamin02}
{Benjamin}, R.~A. 2002, in Astronomical Society of the Pacific Conference
  Series, Vol. 276, Seeing Through the Dust: The Detection of HI and the
  Exploration of the ISM in Galaxies, ed. A.~R. {Taylor}, T.~L. {Landecker}, \&
  A.~G. {Willis}, 201

\bibitem[{{Birnboim} \& {Dekel}(2003)}]{Birnboim03}
{Birnboim}, Y. \& {Dekel}, A. 2003, \mnras, 345, 349

\bibitem[{{Boomsma} {et~al.}(2008){Boomsma}, {Oosterloo}, {Fraternali}, {van
  der Hulst}, \& {Sancisi}}]{Boomsma08}
{Boomsma}, R., {Oosterloo}, T.~A., {Fraternali}, F., {van der Hulst}, J.~M., \&
  {Sancisi}, R. 2008, \aap, 490, 555

\bibitem[{{Bregman}(1980)}]{Bregman80}
{Bregman}, J.~N. 1980, \apj, 236, 577

\bibitem[{{Buta} \& {McCall}(1999)}]{buta99}
{Buta}, R.~J. \& {McCall}, M.~L. 1999, \apjs, 124, 33

\bibitem[{{Carignan} \& {Purton}(1998)}]{carignan98}
{Carignan}, C. \& {Purton}, C. 1998, \apj, 506, 125

\bibitem[{{Collins} {et~al.}(2002){Collins}, {Benjamin}, \& {Rand}}]{Collins02}
{Collins}, J.~A., {Benjamin}, R.~A., \& {Rand}, R.~J. 2002, \apj, 578, 98

\bibitem[{{Franx} {et~al.}(1994){Franx}, {van Gorkom}, \& {de Zeeuw}}]{Franx94}
{Franx}, M., {van Gorkom}, J.~H., \& {de Zeeuw}, T. 1994, \apj, 436, 642

\bibitem[{{Fraternali} \& {Binney}(2006)}]{Fraternali06}
{Fraternali}, F. \& {Binney}, J.~J. 2006, \mnras, 366, 449

\bibitem[{{Fraternali} \& {Binney}(2008)}]{Fraternali08}
{Fraternali}, F. \& {Binney}, J.~J. 2008, \mnras, 386, 935

\bibitem[{{Fraternali} \& {Tomassetti}(2012)}]{Fraternali12}
{Fraternali}, F. \& {Tomassetti}, M. 2012, \mnras, 426, 2166

\bibitem[{{Fraternali} {et~al.}(2002){Fraternali}, {van Moorsel}, {Sancisi}, \&
  {Oosterloo}}]{Fraternali02}
{Fraternali}, F., {van Moorsel}, G., {Sancisi}, R., \& {Oosterloo}, T. 2002,
  \aj, 123, 3124

\bibitem[{{Gentile} {et~al.}(2013){Gentile}, {J{\'o}zsa}, {Serra}, {Heald}, {de
  Blok}, {Fraternali}, {Patterson}, {Walterbos}, \& {Oosterloo}}]{Gentile13}
{Gentile}, G., {J{\'o}zsa}, G.~I.~G., {Serra}, P., {et~al.} 2013, \aap, 554,
  A125

\bibitem[{{Gentile} {et~al.}(2007){Gentile}, {Salucci}, {Klein}, \&
  {Granato}}]{gentile07}
{Gentile}, G., {Salucci}, P., {Klein}, U., \& {Granato}, G.~L. 2007, \mnras,
  375, 199

\bibitem[{{Georgiev} {et~al.}(2005){Georgiev}, {Georgiev}, {Koleva},
  {Nedialkov}, \& {Stanchev}}]{georgiev04}
{Georgiev}, T.~B., {Georgiev}, I.~Y., {Koleva}, N.~A., {Nedialkov}, P.~L., \&
  {Stanchev}, O.~I. 2005, Aerospace Research in Bulgaria, 20, 138

\bibitem[{{Heald} {et~al.}(2011){Heald}, {J{\'o}zsa}, {Serra}, {Zschaechner},
  {Rand}, {Fraternali}, {Oosterloo}, {Walterbos}, {J{\"u}tte}, \&
  {Gentile}}]{Heald11}
{Heald}, G., {J{\'o}zsa}, G., {Serra}, P., {et~al.} 2011, \aap, 526, A118

\bibitem[{{Heald} {et~al.}(2007){Heald}, {Rand}, {Benjamin}, \&
  {Bershady}}]{Heald07}
{Heald}, G.~H., {Rand}, R.~J., {Benjamin}, R.~A., \& {Bershady}, M.~A. 2007,
  \apj, 663, 933

\bibitem[{{Hoeft} \& {Gottl{\"o}ber}(2010)}]{Hoeft10}
{Hoeft}, M. \& {Gottl{\"o}ber}, S. 2010, Advances in Astronomy, 2010

\bibitem[{{Hoffman} {et~al.}(2001){Hoffman}, {Salpeter}, \&
  {Carle}}]{hoffman01}
{Hoffman}, G.~L., {Salpeter}, E.~E., \& {Carle}, N.~J. 2001, \aj, 122, 2428

\bibitem[{{Jacobs} {et~al.}(2009){Jacobs}, {Rizzi}, {Tully}, {Shaya},
  {Makarov}, \& {Makarova}}]{jacobs09}
{Jacobs}, B.~A., {Rizzi}, L., {Tully}, R.~B., {et~al.} 2009, \aj, 138, 332

\bibitem[{{J{\'o}zsa}(2007)}]{Jozsa07b}
{J{\'o}zsa}, G.~I.~G. 2007, \aap, 468, 903

\bibitem[{{J{\'o}zsa} {et~al.}(2007){J{\'o}zsa}, {Kenn}, {Klein}, \&
  {Oosterloo}}]{Jozsa07a}
{J{\'o}zsa}, G.~I.~G., {Kenn}, F., {Klein}, U., \& {Oosterloo}, T.~A. 2007,
  \aap, 468, 731

\bibitem[{{Kamphuis} {et~al.}(2011){Kamphuis}, {Peletier}, {van der Kruit}, \&
  {Heald}}]{Kamphuis11b}
{Kamphuis}, P., {Peletier}, R.~F., {van der Kruit}, P.~C., \& {Heald}, G.~H.
  2011, \mnras, 414, 3444

\bibitem[{{Kennicutt} {et~al.}(2008){Kennicutt}, {Lee}, {Funes}, {Sakai}, \&
  {Akiyama}}]{Kennicutt08}
{Kennicutt}, Jr., R.~C., {Lee}, J.~C., {Funes}, Jos{\'e}~G., S.~J., {Sakai},
  S., \& {Akiyama}, S. 2008, \apjs, 178, 247

\bibitem[{{Kere{\v s}} {et~al.}(2005){Kere{\v s}}, {Katz}, {Weinberg}, \&
  {Dav{\'e}}}]{Keres05}
{Kere{\v s}}, D., {Katz}, N., {Weinberg}, D.~H., \& {Dav{\'e}}, R. 2005,
  \mnras, 363, 2

\bibitem[{{Kingsburgh} \& {McCall}(1998)}]{kingsburgh98}
{Kingsburgh}, R.~L. \& {McCall}, M.~L. 1998, \aj, 116, 2246

\bibitem[{{Larson} {et~al.}(1980){Larson}, {Tinsley}, \& {Caldwell}}]{Larson80}
{Larson}, R.~B., {Tinsley}, B.~M., \& {Caldwell}, C.~N. 1980, \apj, 237, 692

\bibitem[{{Lee} {et~al.}(2009){Lee}, {Gil de Paz}, {Tremonti}, {Kennicutt},
  {Salim}, {Bothwell}, {Calzetti}, {Dalcanton}, {Dale}, {Engelbracht}, {Funes},
  {Johnson}, {Sakai}, {Skillman}, {van Zee}, {Walter}, \& {Weisz}}]{Lee09}
{Lee}, J.~C., {Gil de Paz}, A., {Tremonti}, C., {et~al.} 2009, \apj, 706, 599

\bibitem[{{Lee} {et~al.}(2001){Lee}, {Irwin}, {Dettmar}, {Cunningham}, {Golla},
  \& {Wang}}]{Lee01}
{Lee}, S.-W., {Irwin}, J.~A., {Dettmar}, R.-J., {et~al.} 2001, \aap, 377, 759

\bibitem[{{Marinacci} {et~al.}(2010){Marinacci}, {Binney}, {Fraternali},
  {Nipoti}, {Ciotti}, \& {Londrillo}}]{Marinacci10}
{Marinacci}, F., {Binney}, J., {Fraternali}, F., {et~al.} 2010, \mnras, 404,
  1464

\bibitem[{{Marinacci} {et~al.}(2011){Marinacci}, {Fraternali}, {Nipoti},
  {Binney}, {Ciotti}, \& {Londrillo}}]{Marinacci11}
{Marinacci}, F., {Fraternali}, F., {Nipoti}, C., {et~al.} 2011, \mnras, 415,
  1534

\bibitem[{{Norman} \& {Ikeuchi}(1989)}]{Norman89}
{Norman}, C.~A. \& {Ikeuchi}, S. 1989, \apj, 345, 372

\bibitem[{{Oh} {et~al.}(2008){Oh}, {de Blok}, {Walter}, {Brinks}, \&
  {Kennicutt}}]{oh08}
{Oh}, S.-H., {de Blok}, W.~J.~G., {Walter}, F., {Brinks}, E., \& {Kennicutt},
  Jr., R.~C. 2008, \aj, 136, 2761

\bibitem[{{Oosterloo} {et~al.}(2007){Oosterloo}, {Fraternali}, \&
  {Sancisi}}]{Oosterloo07}
{Oosterloo}, T., {Fraternali}, F., \& {Sancisi}, R. 2007, \aj, 134, 1019

\bibitem[{{Pagel} \& {Patchett}(1975)}]{Pagel75}
{Pagel}, B.~E.~J. \& {Patchett}, B.~E. 1975, \mnras, 172, 13

\bibitem[{{Paturel} {et~al.}(2003){Paturel}, {Petit}, {Prugniel}, {Theureau},
  {Rousseau}, {Brouty}, {Dubois}, \& {Cambr{\'e}sy}}]{paturel03}
{Paturel}, G., {Petit}, C., {Prugniel}, P., {et~al.} 2003, \aap, 412, 45

\bibitem[{{Richter}(2012)}]{Richter12}
{Richter}, P. 2012, \apj, 750, 165

\bibitem[{{Rogstad} {et~al.}(1974){Rogstad}, {Lockhart}, \&
  {Wright}}]{rogstad74}
{Rogstad}, D.~H., {Lockhart}, I.~A., \& {Wright}, M.~C.~H. 1974, \apj, 193, 309

\bibitem[{{Sancisi} {et~al.}(2008){Sancisi}, {Fraternali}, {Oosterloo}, \& {van
  der Hulst}}]{Sancisi08}
{Sancisi}, R., {Fraternali}, F., {Oosterloo}, T., \& {van der Hulst}, T. 2008,
  \aapr, 15, 189

\bibitem[{{Sault} {et~al.}(1995){Sault}, {Teuben}, \& {Wright}}]{sault95}
{Sault}, R.~J., {Teuben}, P.~J., \& {Wright}, M.~C.~H. 1995, in Astronomical
  Society of the Pacific Conference Series, Vol.~77, Astronomical Data Analysis
  Software and Systems IV, ed. {R.~A.~Shaw, H.~E.~Payne, \& J.~J.~E.~Hayes},
  433--+

\bibitem[{{Schaap} {et~al.}(2000){Schaap}, {Sancisi}, \& {Swaters}}]{Schaap00}
{Schaap}, W.~E., {Sancisi}, R., \& {Swaters}, R.~A. 2000, \aap, 356, L49

\bibitem[{{Schlegel} {et~al.}(1998){Schlegel}, {Finkbeiner}, \&
  {Davis}}]{schlegel98}
{Schlegel}, D.~J., {Finkbeiner}, D.~P., \& {Davis}, M. 1998, \apj, 500, 525

\bibitem[{{Schoenmakers} {et~al.}(1997){Schoenmakers}, {Franx}, \& {de
  Zeeuw}}]{Schoenmakers97}
{Schoenmakers}, R.~H.~M., {Franx}, M., \& {de Zeeuw}, P.~T. 1997, \mnras, 292,
  349

\bibitem[{{Searle} \& {Sargent}(1972)}]{Sargent72}
{Searle}, L. \& {Sargent}, W.~L.~W. 1972, Comments on Astrophysics and Space
  Physics, 4, 59

\bibitem[{{Shapiro} \& {Field}(1976)}]{Shapiro76}
{Shapiro}, P.~R. \& {Field}, G.~B. 1976, \apj, 205, 762

\bibitem[{{Spekkens} \& {Sellwood}(2007)}]{Spekkens07}
{Spekkens}, K. \& {Sellwood}, J.~A. 2007, \apj, 664, 204

\bibitem[{{Swaters} {et~al.}(1997){Swaters}, {Sancisi}, \& {van der
  Hulst}}]{Swaters97}
{Swaters}, R.~A., {Sancisi}, R., \& {van der Hulst}, J.~M. 1997, \apj, 491, 140

\bibitem[{{van den Bergh}(1962)}]{Bergh62}
{van den Bergh}, S. 1962, \aj, 67, 486

\bibitem[{{van Woerden} \& {Wakker}(2004)}]{Woerden04}
{van Woerden}, H. \& {Wakker}, B.~P. 2004, in Astrophysics and Space Science
  Library, Vol. 312, High Velocity Clouds, ed. H.~{van Woerden}, B.~P.
  {Wakker}, U.~J. {Schwarz}, \& K.~S. {de Boer}, 195

\bibitem[{{Walter} \& {Brinks}(1999)}]{Walter99}
{Walter}, F. \& {Brinks}, E. 1999, \aj, 118, 273

\bibitem[{{Westmeier} {et~al.}(2005){Westmeier}, {Braun}, \&
  {Thilker}}]{westmeier05}
{Westmeier}, T., {Braun}, R., \& {Thilker}, D. 2005, \aap, 436, 101

\bibitem[{{Zschaechner} {et~al.}(2012){Zschaechner}, {Rand}, {Heald},
  {Gentile}, \& {J{\'o}zsa}}]{Zschaechner12}
{Zschaechner}, L.~K., {Rand}, R.~J., {Heald}, G.~H., {Gentile}, G., \&
  {J{\'o}zsa}, G. 2012, \apj, 760, 37

\bibitem[{{Zschaechner} {et~al.}(2011){Zschaechner}, {Rand}, {Heald},
  {Gentile}, \& {Kamphuis}}]{Zschaechner11}
{Zschaechner}, L.~K., {Rand}, R.~J., {Heald}, G.~H., {Gentile}, G., \&
  {Kamphuis}, P. 2011, \apj, 740, 35

\end{thebibliography}

\end{document}